\pdfoutput=1
\documentclass[10pt,letterpaper,final,journal,compsoc]{IEEEtran}

\usepackage{amssymb}
\usepackage{graphicx}
\usepackage{subfigure}
\usepackage{float}
\usepackage{amsmath}
\usepackage{booktabs}

\usepackage{balance}

\usepackage{ifpdf}
\ifCLASSOPTIONcompsoc
\usepackage[nocompress]{cite}
\else  
\usepackage{cite}  
\fi

\usepackage{rawfonts}
\usepackage{upgreek}
\usepackage{pifont}

\usepackage{url}
\usepackage{amssymb}
\usepackage{graphicx}

\usepackage[    
protrusion=true, 
expansion=true,
tracking=true,
kerning=true,
spacing=true,
letterspace=40,
shrink=40,      
factor=1000
]    
{microtype} 

\usepackage{ragged2e} 
\usepackage{psfrag} 
\renewcommand{\subsubsection}[1]{\medskip\noindent\textbf{{{#1}:\,}}} 
\def \paperTitle{Estimation of Graphlet Statistics}

\ifCLASSINFOpdf
\else
\fi

\ifCLASSOPTIONcaptionsoff
\usepackage[nomarkers]{endfloat}
\let\MYoriglatexcaption\caption
\renewcommand{\caption}[2][\relax]{\MYoriglatexcaption[#2]{#2}}
\fi

\newcommand\MYhyperrefoptions{bookmarks=false,bookmarksnumbered=false,
pdfpagemode={UseOutlines},plainpages=false,pdfpagelabels=true,
colorlinks=true,linkcolor={theblue},citecolor={theblue},urlcolor={theblue},
pdftitle={\paperTitle},
}

\ifCLASSINFOpdf
	\usepackage[\MYhyperrefoptions,pdftex]{hyperref}
\else
	\usepackage[\MYhyperrefoptions,breaklinks=true,dvips]{hyperref}
	\usepackage{breakurl}
\fi

\usepackage[notheorems]{rr-math}

\usepackage{verbatim}
\usepackage{wrapfig}
\usepackage[T1]{fontenc} 

\usepackage{booktabs}
\newcommand{\ra}[1]{\renewcommand{\arraystretch}{#1}}

\usepackage{rccol} 
\usepackage{graphicx}
\usepackage{multirow}
\usepackage{rotate}
\usepackage{subfigure}
\usepackage{epstopdf}
\usepackage{paralist}
\usepackage{url}
\usepackage{graphicx}
\usepackage{wrapfig}
\usepackage{xspace}
\usepackage[table,usenames,dvipsnames]{xcolor}
\usepackage{listings}

\usepackage{graphicx}

\usepackage{amssymb}
\usepackage{amsmath}
\usepackage{amsfonts}

\newcommand{\todo}[1]{}

\newcommand{\incomplete}[1]{}

\usepackage{tabularx}
\usepackage{booktabs}

\usepackage{relsize}
\usepackage{algorithm}
\usepackage{algorithmicx}
\usepackage{algpseudocode}

\algblockdefx[parallel]{ParFor}{EndPar}[1][]{$\textbf{parallel for}$ #1 $\textbf{do}$}{$\textbf{end parallel}$}
\algrenewcommand{\alglinenumber}[1]{\fontsize{6.5}{7}\selectfont#1}
\algtext*{EndPar}

\algblockdefx[parallel]{parfor}{endpar}[1][]{$\textbf{parallel for}$ #1 $\textbf{do}$}{$\textbf{end parallel}$}

\algblockdefx[parallel]{ParallelFor}{EndParallel}[1][]{$\textbf{parallel for}$ #1 $\textbf{do}$}{}
\providecommand{\multiline}[1]{\State \parbox[t]{\dimexpr\linewidth-\algorithmicindent}{#1\strut}}

\newcommand{\pluseq}{\;\ensuremath{\mathsmaller{\mathsmaller{\mathsmaller{\mathsmaller{\overset{\tiny +}{=}}}}}}\;}

\let\oldpluseq\pluseq
\renewcommand{\pluseq}[1][-2pt]{
\mathrel{\raisebox{#1}{$\oldpluseq$}}
}

\usepackage{nicefrac}

\algnotext{EndFor}
\algnotext{EndIf}
\algnotext{EndProcedure}
\algnotext{EndParallel}

\ifpdf
\usepackage{epstopdf}
\fi
\graphicspath{{./}{./figures/}}

\graphicspath{{graphics/}{../graphics/}{./}}

\usepackage{array}
\newcolumntype{P}[1]{>{\centering\arraybackslash}p{#1}}
\newcolumntype{M}[1]{>{\centering\arraybackslash}m{#1}}

\usepackage{listings}

\newcommand{\eol}{\end{enumerate}\setlength{\itemsep}{-\parsep}}

\relax

\let\hat\widehat

\def\Pr{{\mathbb P}}

\usepackage{wrapfig}
\usepackage{xspace}
\usepackage[table,usenames,dvipsnames]{xcolor}

\newcommand{\etal}{\emph{et al.}\xspace}
\newcommand{\cmark}{\ding{51}}

\newlength{\commentWidth}
\newcommand{\bspacing}{\begin{spacing}{1.4}}
\newcommand{\espacing}{\end{spacing}}

\definecolor{plotblue}{RGB}	{30,144,255}
\definecolor{plotgreen}{RGB}	{50,205,50}
\definecolor{plotred}{RGB}	{220,20,60}

\definecolor{myyellow}{RGB}{255,255,204}
\definecolor{myred}{RGB}{255,204,204}
\definecolor{myblue}{RGB}{204, 255, 255}
\definecolor{mygreen}{RGB}{204, 255, 204}

\definecolor{gray}{RGB}{150,150,150}
\definecolor{theblue}{RGB}{0,0,180}

\usepackage{textcase}
\usepackage{soul}

\newcommand*\hrulefillvar[1][0.4pt]{\leavevmode\leaders\hrule height#1\hfill\kern0pt}

\newcommand{\be}{\begin{equation}}
\newcommand{\ee}{\end{equation}}
\newcommand{\bea}{\begin{eqnarray}}
\newcommand{\eea}{\end{eqnarray}}
\newcommand{\bit}{\begin{itemize}}
\newcommand{\eit}{\end{itemize}}

\definecolor{lightgray}{rgb}{0.93,0.93,0.93}

\definecolor{lightblue}{rgb}{0.5,0.90,1.0}
\definecolor{lightgreen}{rgb}{0.5,0.92,0.5}
\definecolor{lightred}{rgb}{0.98,0.5,0.5}
\definecolor{lightyellow}{rgb}{1,0.90,0.40}

\newcommand{\nr}{\ensuremath{\textsc{nr}}}
\newcommand{\pgd}{\ensuremath{\textsc{pgd}}}

\newcommand{\gfd}{\ensuremath{\textsc{gfd}}}

\providecommand{\lge}{\ensuremath{\textsc{lge}}}

\newcommand\TTT{\rule{0pt}{3.2ex}}
\newcommand\BBB{\rule[-1.4ex]{0pt}{0pt}}

\newcommand\TT{\rule{0pt}{2.3ex}}
\newcommand\BB{\rule[-1.0ex]{0pt}{0pt}}

\usepackage{morefloats}
\usepackage{booktabs}
\usepackage{comment}

\newcommand{\eat}[1]{}

\definecolor{myyellow}{RGB}{255,255,204}
\definecolor{myred}{RGB}{255,204,204}
\definecolor{myblue}{RGB}{0,200,255}
\definecolor{mygreen}{RGB}{80,220,80}

\newcommand{\eg}{\emph{e.g.}}
\newcommand{\ie}{\emph{i.e.}}

\newcommand{\wrt}{\emph{w.r.t.}\ }

\newcommand{\rbr}[1]{\left(#1\right)}
\newcommand{\cbr}[1]{\left\{#1\right\}}

\newcommand{\ceil}[1]{\left\lceil #1 \right\rceil}

\algblockdefx[parallel]{parallelfor}{parallelend}
[1][]{\textbf{parallel for} #1}
{\textbf{end parallel}}

\RequirePackage{bm}

\RequirePackage{mathtools}
\RequirePackage{color}
\RequirePackage{xcolor}
\RequirePackage{listings}

\newcommand{\ds}\displaystyle
\newcommand{\mbb}\mathbb
\newcommand{\mc}\mathcal
\newcommand{\del}\nabla

\newcommand{\beqstar}{\begin{eqnarray*}}
\newcommand{\eeqstar}{\end{eqnarray*}}

\definecolor{verylightgreen}{RGB}	{204,255,204}
\definecolor{verylightred}{RGB}		{255,204,204}
\definecolor{verylightyellow}{RGB}		{255,255,204}

\definecolor{thegreen}{rgb}{0,.5,0}
\definecolor{idea}{rgb}{0,.6,0.1}
\definecolor{problem}{rgb}{0.7,0,0.1}
\definecolor{comment-green}{rgb}{0,.3,0}
\definecolor{theblue}{rgb}{0,0,.8}
\definecolor{light-gray}{gray}{0.98}
\definecolor{comment-color}{rgb}{0,0,.8}
\definecolor{string-color}{rgb}{0,.75,0}
\definecolor{border-blue}{rgb}{0,0,.6}

\usepackage{epsfig}

\definecolor{verylightgreen}{RGB}	{204,255,204}
\definecolor{verylightred}{RGB}		{255,204,204}
\definecolor{verylightyellow}{RGB}		{255,255,204}

\usepackage{ragged2e}
\usepackage{setspace}

\graphicspath{{./}{./images/}}
\newcolumntype{H}{>{\setbox0=\hbox\bgroup}c<{\egroup}@{}}

\definecolor{orange}{rgb}{1,0.5,0}
\definecolor{gray}{RGB}{20,20,20}
\definecolor{greencm}{RGB}{0,153,0}
\definecolor{thegreen}{RGB}{0,153,0}
\definecolor{thered}{RGB}{255, 0, 63}
\definecolor{Crimson}{RGB}{255, 0, 63}
\definecolor{crimson}{RGB}{255, 0, 63}

\newcommand{\cm}{ {\color{greencm}\cmark}}

\usepackage{amssymb}
\usepackage{pifont}
\renewcommand{\cmark}{\ding{51}}

\usepackage[T1]{fontenc}
\usepackage{pifont}
\usepackage{upgreek}

\usepackage{ifpdf}
\usepackage{xcolor}
\usepackage{paralist}
\usepackage{tabularx}
\usepackage{booktabs}
\usepackage{multirow}
\usepackage{graphicx}
\usepackage{sidecap}
\usepackage{mathdots}
\usepackage{lscape}
\makeatletter
\def\vcdots{\vbox{\baselineskip4\p@ \lineskiplimit\z@
\kern3\p@\hbox{.}\hbox{.}\hbox{.}\kern3\p@}}
\makeatother
\usepackage[font=scriptsize,labelfont=sf,textfont=sf]{caption}
\usepackage{subfigure}
\usepackage{array}
\usepackage{url}
\usepackage{float}

\usepackage{relsize}

\providecommand{\multiline}[1]{\State \parbox[t]{\dimexpr\linewidth-\algorithmicindent}{#1\strut}}

\newcommand{\Exp}{\mathbb{E}} 
\renewcommand{\Pr}{\mathbb{P}} 
\renewcommand{\Var}{\mathbb{V}} 
\renewcommand{\Std}{\mathbb{S}} 
\newcommand{\bK}{\mathbb{K}} 
 
\newcommand{\bD}{\mathbb{D}} 
\newcommand{\bB}{\mathbb{B}}

\providecommand{\Fdist}{\ensuremath{\mathsf{F}}}

\newcommand{\G}{\ensuremath{{\mathcal{G}}}}
\newcommand{\g}{\ensuremath{{G}}}

\newcommand{\mychoose}[2]{\left( \begin{smallmatrix} #1 \\ #2 \end{smallmatrix} \right)}

\newcommand{\T}{\ensuremath{{T}}} 
\renewcommand{\E}{\ensuremath{E}}

\newcommand{\e}{\ensuremath{e}}  		
\newcommand{\Es}{\ensuremath{J}} 		
\newcommand{\K}{\ensuremath{K}}  		
\newcommand{\trials}{\ensuremath{S}}  	
\providecommand{\trials}{\ensuremath{{\tau}}}	
\newcommand{\B}{\ensuremath{{\beta}}}			
\newcommand{\lb}{\ensuremath{{\B_{\text{lb}}}}}
\newcommand{\ub}{\ensuremath{{\B_{\text{ub}}}}}

\newcommand{\C}{\ensuremath{C}} 
\newcommand{\Xs}{\ensuremath{X}} 
\newcommand{\X}{\ensuremath{X}}  
\newcommand{\Y}{\ensuremath{Y}} 
\newcommand{\bC}{\ensuremath{\mC}} 
\newcommand{\bX}{\ensuremath{\mX}}  
\newcommand{\bY}{\ensuremath{\mY}}

\providecommand{\pinv}{\ensuremath{\sigma}} 
\providecommand{\p}{\ensuremath{p}} 
\providecommand{\pr}{\ensuremath{p}} 
\renewcommand{\prob}{\ensuremath{\phi}} 
\providecommand{\pe}{\ensuremath{\p}}

\providecommand{\lb}{\ensuremath{\Omega_{\rm lb}}}
\providecommand{\ub}{\ensuremath{\Omega_{\rm ub}}}

\newcommand{\Error}[2]{\ensuremath{{\mathbb{D}\,\bigr( \; {#1 \; \| \; #2} \; \bigl) }}}

\newcommand{\D}[2]{\ensuremath{{\mathbb{D}\,\bigr( \; {#1 \; \| \; #2} \; \bigl) }}}

\newcommand{\dmax}{\ensuremath{\Delta}}

\renewcommand{\p}{\ensuremath{p}} 
 
\newcommand{\hash}{\ensuremath{{\boldsymbol\Psi}}}

\newcommand{\perc}{\ensuremath{\%}}

\newcommand{\clique}{\ensuremath{{K}}}
\renewcommand{\S}{\ensuremath{{S}}} 
\newcommand{\cycle}{\ensuremath{{C}}}
 
\newcommand{\tri}{\ensuremath{{T}}}

\renewcommand{\O}{\ensuremath{\mathcal{O}}}
\renewcommand{\bigO}[1]{\ensuremath{\mathcal{O}(#1)}}

\newcommand{\data}[2]{{\mathsf{#1}\text{--}{\mathbf{\tt #2}}}} 
\newcommand{\datasm}[2]{{\fontsize{6}{7.5}\selectfont \mathsf{#1}\text{--}\mathsf{#2}}}

\newcommand{\opt}{\ensuremath{\star}}

\newcommand{\m}{\ensuremath{M}}
\newcommand{\n}{\ensuremath{N}}

\renewcommand{\d}{\ensuremath{d}}

\newcommand{\N}{\ensuremath{\Gamma}} 
\newcommand{\Ng}{\ensuremath{\boldsymbol \Gamma}} 
\newcommand{\Ne}{\ensuremath{\boldsymbol \Gamma}}

\newcommand{\loss}{\ensuremath{\mathcal{L}}}

\newcommand{\s}{\ensuremath{\tau}} 
 
\newcommand{\I}{\ensuremath{\mathcal{I}}}

\begin{document}
\title{\paperTitle}
\author{
Ryan~A.~Rossi,
Rong~Zhou,
and~Nesreen~K.~Ahmed
\IEEEcompsocitemizethanks{
\IEEEcompsocthanksitem 
R. A. Rossi and R. Zhou are with Palo Alto Research Center (Xerox PARC), 
3333 Coyote Hill Rd, Palo Alto, CA 94304\protect\\
Email: \{rrossi, rzhou\}@parc.com
\IEEEcompsocthanksitem
N. K. Ahmed is with Intel Labs, 3065 Bowers Ave, Santa Clara, CA 95052\protect\\
Email: Nesreen.K.Ahmed@intel.com
}
}

\markboth{}
{Rossi \MakeLowercase{\textit{et al.}}: \paperTitle}

\maketitle

\begin{abstract}
Graphlets are induced subgraphs of a large network and are important for understanding and modeling complex networks. 
Despite their practical importance, graphlets have been severely limited to applications and domains with relatively small graphs. 
Most previous work has focused on \emph{exact algorithms}, however, it is often too expensive to compute graphlets exactly in massive networks with billions of edges, and finding an approximate count is usually sufficient for many applications.
In this work, we propose an \emph{unbiased graphlet estimation framework} that is 
(\emph{a}) fast with significant speedups compared to the state-of-the-art,
(\emph{b}) parallel with nearly linear-speedups, 
(\emph{c}) accurate with <$1\%$ relative error, 
(\emph{d}) scalable and space-efficient for massive networks with billions of edges, and 
(\emph{e}) flexible for a variety of real-world settings, as well as estimating macro and micro-level graphlet statistics (\eg, counts) of both connected \emph{and} disconnected graphlets.
In addition, an adaptive approach is introduced that finds the smallest sample size required to obtain estimates within a given user-defined error bound.
On 300 networks from 20 domains, we obtain <$1\%$ relative error for all graphlets.
This is significantly more accurate than existing methods while using less data.
Moreover, it takes a few seconds on billion edge graphs (as opposed to days/weeks).
These are by far the largest graphlet computations to date.
\end{abstract}


\IEEEpeerreviewmaketitle

\section{Introduction}
\label{sec:intro}

\IEEEPARstart{G}{raphlets}
are \emph{induced subgraphs}\footnote{The terms graphlet and induced subgraph are interchangeable.} and are important for many predictive and descriptive modeling tasks~\cite{prvzulj2004modeling,milenkoviae2008uncovering,hayes2013graphlet}.
More recently, graphlets have been used to solve important and challenging problems in a variety of disciplines including 
image processing and computer vision~\cite{zhang-image-categorization-via-graphlets,zhang2013probabilistic}, 
bioinformatics~\cite{vishwanathan2010graph,shervashidze2009efficient}, 
and cheminformatics~\cite{rupp2010graph}.
Unfortunately, the application and general use of graphlets remains severely limited to a few specialized problems/domains where the networks are small enough to avoid the scalability and performance limitations of existing methods.
For instance, Shervashidze~\etal~\cite{shervashidze2009efficient} takes hours to count motifs on small biological networks (\ie, few hundreds/thousands of nodes/edges) and uses such counts as features for graph classification~\cite{vishwanathan2010graph}.
Thus, this work provides a foundation for using graphlets to solve countless other important and unsolved problems, especially those with data that is large, massive, or streaming, as well as those with space- or time-constraints (real-time settings, interactive queries).

In many applications, finding an `approximate' answer is usually sufficient where the extra cost and time in finding the exact answer is often not worth the extra accuracy. 
The recent rise of Big Data~\cite{boyd2012critical} has made approximation methods even more important and critical~\cite{zaslavsky2013sensing}, especially for many practical applications~\cite{fischer2015approximation,badoiu2002approximate,henzinger2015almost,pfeffer2012k,stutzbach2006usul}.
More recently, approximation methods have been proposed for numerous important problems including triangle counting~\cite{Ahmed-gSH,lim2015mascot,tsourakakis2009doulion,pagh2012colorful,rahman2013approximate}, 
shortest path problems~\cite{henzinger2015almost,roditty2012dynamic}, 
finding max cliques~\cite{pmc}, 
and many others.

This work aims to overcome the above limitations to make graphlets more accessible to other applications/domains with much larger graphs.
In particular, this work proposes a general estimation framework for computing unbiased estimates of graphlet statistics (\eg, frequency of an arbitrary k-vertex induced subgraph) from a small set of edge-induced neighborhoods. 
The graphlet estimators provide accurate and fast approximations of a variety of macro and micro-level graphlet statistics for both connected and disconnected graphlets.
Moreover, the estimation framework is also scalable to massive networks with billions of edges and nodes.
We also propose an approach for automatically determining the appropriate sample size for estimating various graphlet statistics within a given error bound.
Parallel methods are introduced for each of the proposed techniques.
Furthermore, a number of important machine learning tasks are likely to benefit from the proposed methods, including graph anomaly detection~\cite{noble2003graph,akoglu2014graph}, entity resolution~\cite{bhattacharya2006entity}, as well as features for improving community detection~\cite{schaeffer2007graph}, role discovery~\cite{rossi2014roles}, and relational classification~\cite{getoor2007introduction}.

{
\smallskip
\noindent
\textbf{Summary of contributions.}
The key contributions of this work are as follows:
\vspace{0mm}
\begin{itemize}[\small$\bullet$\leftmargin=0em]
\item \textbf{Novel graphlet estimation framework and algorithms:}
A general unbiased edge-centric estimation framework is proposed for approximating macro and micro graphlet counts in massive networks with billions of edges.
The framework is shown to be accurate, fast, and scalable for \emph{both} dense and sparse networks of arbitrary size.

\item \textbf{Efficient:}
The proposed estimation algorithms are orders of magnitude faster than the recent state-of-the-art algorithm and take a few seconds as opposed to days/months.

\item \textbf{Accurate:} 
For all graphlets and data (300 graphs from 20 domains), the methods are more accurate than existing state-of-the-art methods (<$1\%$ relative error) while using only a small fraction of the data.
Provable error bounds are also derived and shown to be tight (see Section~\ref{sec:exp-confidence-bounds}).

\item \textbf{Parallel methods:}
This work proposes parallel graphlet estimation methods for shared and distributed-memory architectures.
Strong scaling results with nearly linear speedups are observed across a wide variety of graphs from 20 domains.

\item \textbf{Adaptive estimation:}
While existing work requires the number (or proportion) of samples to be given as input, 
we instead introduce an approach that automatically determines the number of samples required to obtain estimates within a given error bound. 
Thus, this approach effectively balances the trade-offs between accuracy, time, and space.

\item \textbf{Full spectrum of graphlets and novel sufficient statistics:}
Our algorithms provide efficient computation of the full spectrum of graphlets including both connected and disconnected graphlets. 
Existing work has mainly focused on connected graphlets~\cite{guise,graft,path-sampling,rage}, despite the importance of disconnected graphlets.
For instance, Shervashidze~\etal~\cite{shervashidze2009efficient} 
found that disconnected graphlets are \emph{essential} for correct classification on some datasets (See~\cite{shervashidze2009efficient} pp. 495 where disconnected graphlets lead to a $10\%$ improvement in accuracy).

\item \textbf{Largest investigation and graphlet computations:}
To the best of our knowledge, this work provides the 
(i) largest graphlet computations to date \emph{and} the 
(ii) largest empirical investigation using $300$+ networks from $20$+ domains.

\end{itemize}
\smallskip
}

\noindent
The proposed \emph{localized graphlet estimation} ($\lge$) framework is flexible and gives rise to many important estimation methods for approximating a wide range of 
graphlet statistics (\eg, frequency of all k-vertex induced subgraphs) and 
distributions including 
(i) macro-level graphlet statistics for the graph $G$ as well as 
(ii) micro-level statistics for individual edges.
Furthermore, we also propose estimators for both connected \emph{and} disconnected graphlet counts (as opposed to only connected graphlet counts).
The framework naturally allows for both uniform and weighted sampling designs, and has many other interchangeable components as well.

\section{Localized Estimation Framework}
\label{sec:framework}
\noindent
In this section, we propose a new family of graphlet estimation methods based on selecting a set of localized edge-centric neighborhoods $\{\Ne(\e_1), \dotsc, \Ne(\e_{\K})\}$.
This gives rise to the \emph{localized graphlet estimation framework} ($\lge$) which serves as a basis for deriving unbiased and consistent estimators that are fast, accurate, and scalable for massive networks.
Moreover, the $\lge$ framework is also flexible with many interchangeable components. 
As shown later in Section~\ref{sec:exp}, $\lge$ is useful for a wide variety of networks, applications, and domains (\eg, biological, social, and infrastructure/physical networks), which have fundamentally different structural properties. 

{
\setlength{\tabcolsep}{2.0pt}
\begin{table}[h!]
\vspace{-0mm}
\centering
\caption{Summary of graphlet properties \emph{and} notation}
\label{table:graphlet_notation}
\vspace*{-2mm}
\scriptsize
\scalebox{1.00}{
\begin{tabularx}{1.0\linewidth}{
M{0.2cm} 
M{0.4cm}
M{0.6cm} 
l
c
c
XX
r
c
XXX
HH
}
\toprule
\multicolumn{15}
{p{1.0\linewidth}}
{
\footnotesize
Summary of the notation and properties for graphlets of size $k = \{2,3,4\}$.
Note that $\rho$ denotes density, $\Delta$ and $\bar{d}$ denote the max and mean degree, whereas assortativity is denoted by $r$.
Also, $|T|$ is the total number of triangles, $\bK$ is the max k-core number, $\chi$ denotes the Chromatic number, whereas ${\bD}$ denotes the diameter.
}
\\
\midrule
&
\multicolumn{2}{c}
{}
& \textbf{Description} & \textbf{Comp.} & $\;\;\;\; \rho$ & $\Delta$ & ${\rm \bar{d}}$ & $r$ & $|T|$ & $\bK$ & 
${\rm \chi}$ & $ \bD$ & $\bB$ & $|{\rm C}|$\\ 
\midrule 
\multirow{8}{*}{\rotatebox{90}{\mbox{}
}} 
&  \includegraphics[scale=0.04]{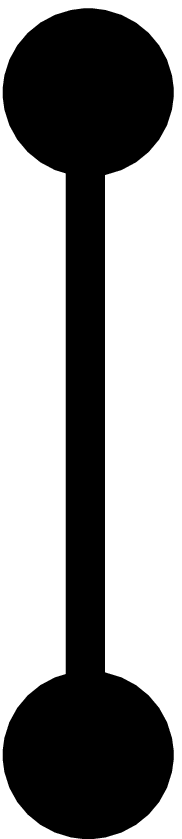} 
& $\g_1$  & edge 
&  \includegraphics[scale=0.04]{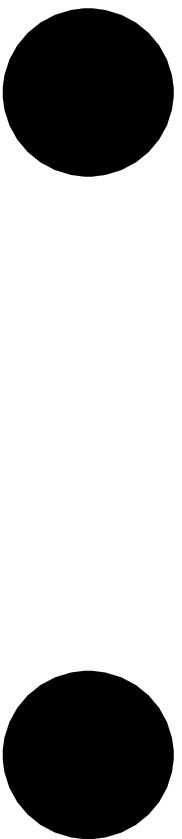} 
& 1.00 & 1 & 1.0 & 1.00  & 0 & 1 & 2 & 1 & 0 & 1\\ 

&  \includegraphics[scale=0.04]{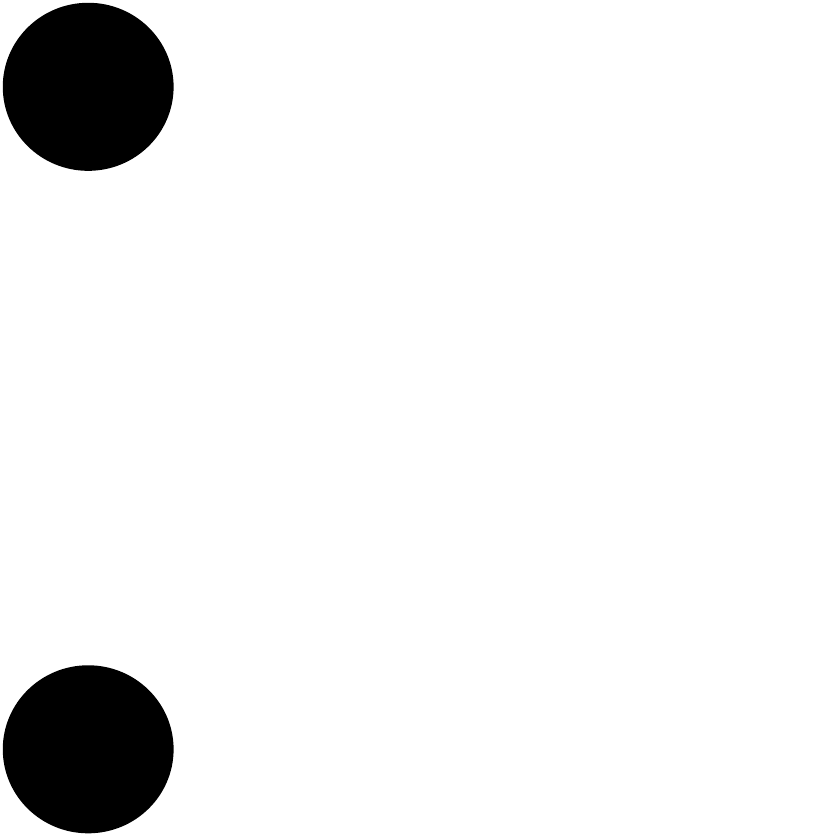} 
& $\g_2$  & 2-node-independent 
&  \includegraphics[scale=0.04]{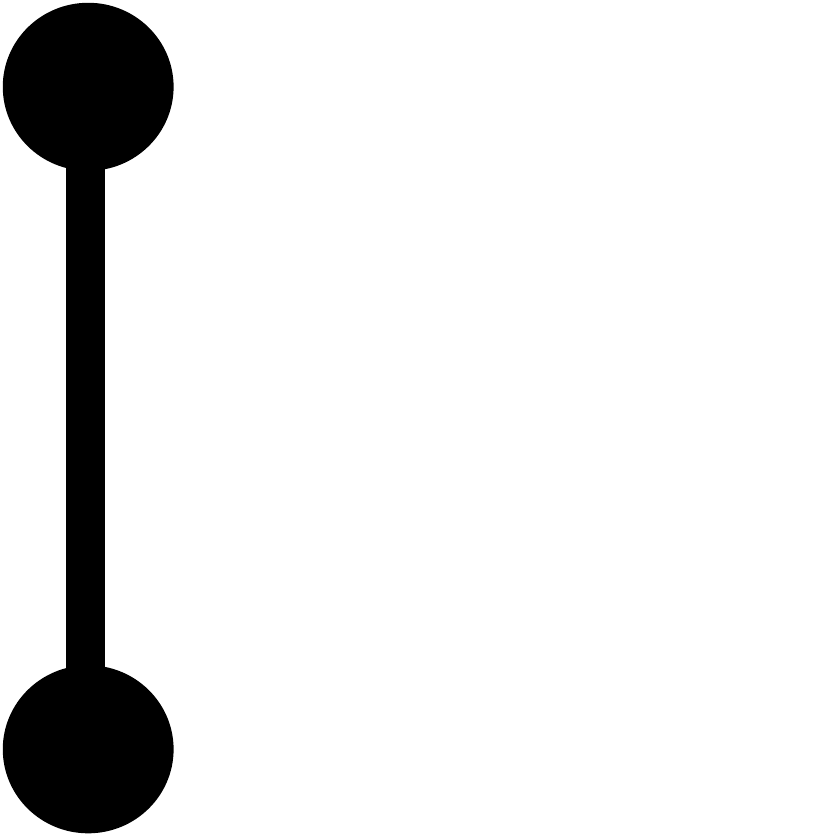} 
& 0.00 & 0 & 0.0 & 0.00  & 0 & 0 & 1 & $\infty$ & 0 & 2\\ 

\midrule 
\multirow{8}{*}{\rotatebox{90}{\mbox{}
}} 
&  \includegraphics[scale=0.04]{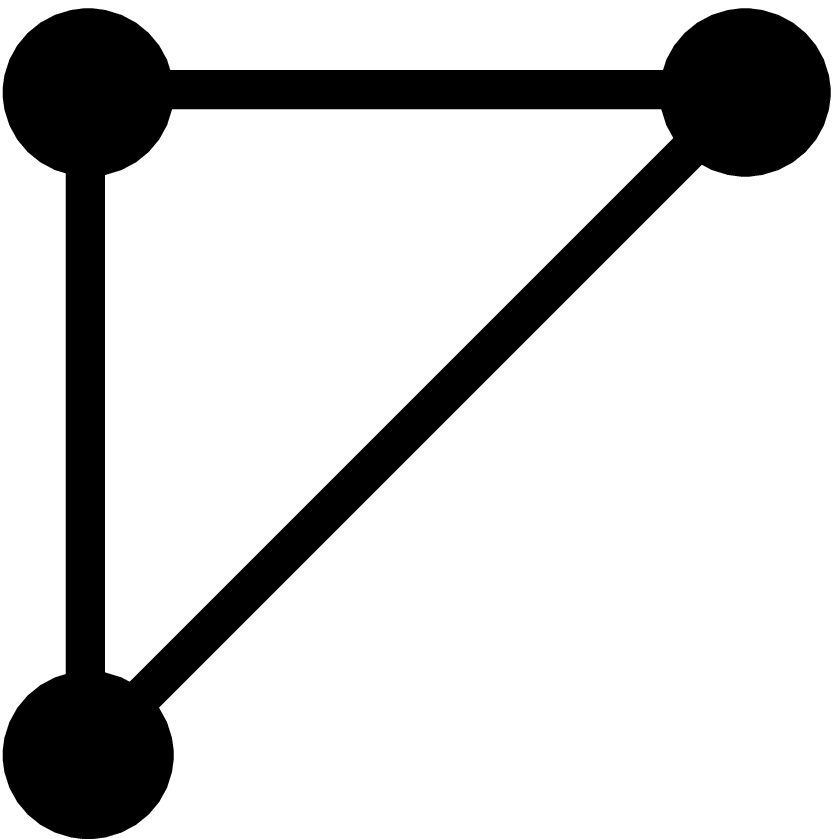} 
& $\g_3$  & triangle 
&  \includegraphics[scale=0.04]{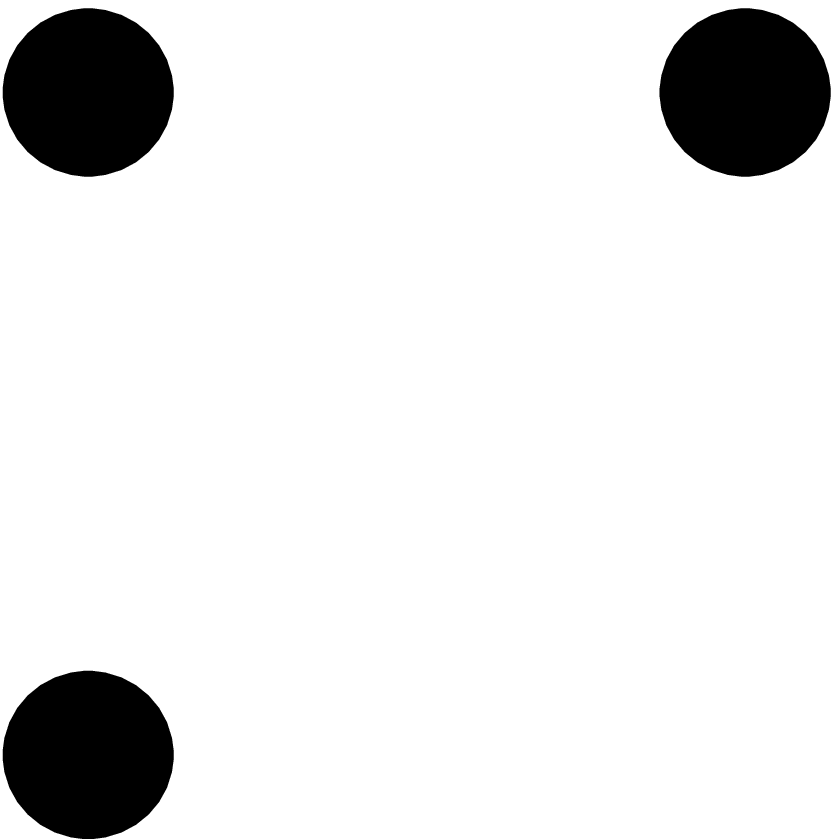} 
& 1.00 & 2 & 2.0 & 1.00 & 1 & 2 & 3 & 1 & 0 & 1\\ 
\TTT\BBB
&  \includegraphics[scale=0.04]{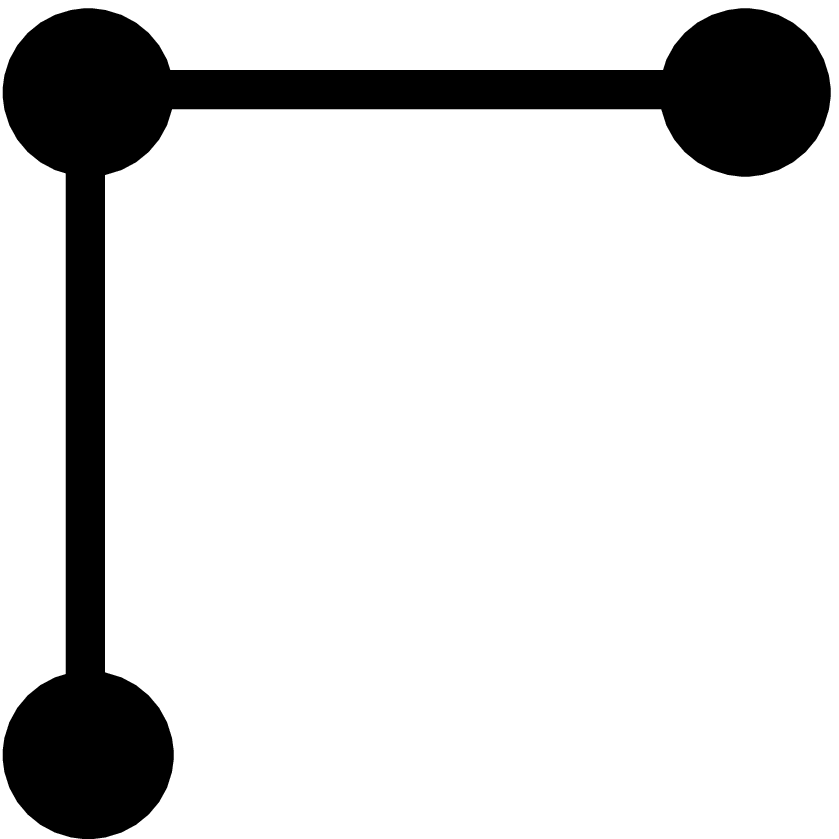} 
& $\g_4$  & 2-star 
&  \includegraphics[scale=0.04]{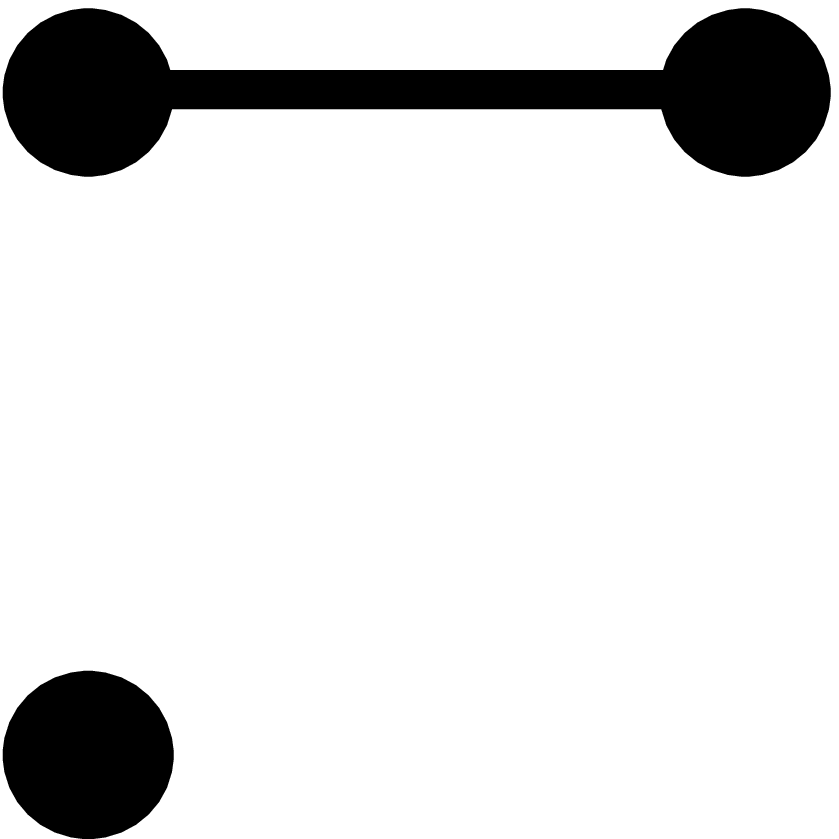} 
& 0.67 & 2 & 1.33 & -1.00  & 0 & 1 & 2 & 2 & 1 & 1\\ 

\TTT\BBB
&  \includegraphics[scale=0.04]{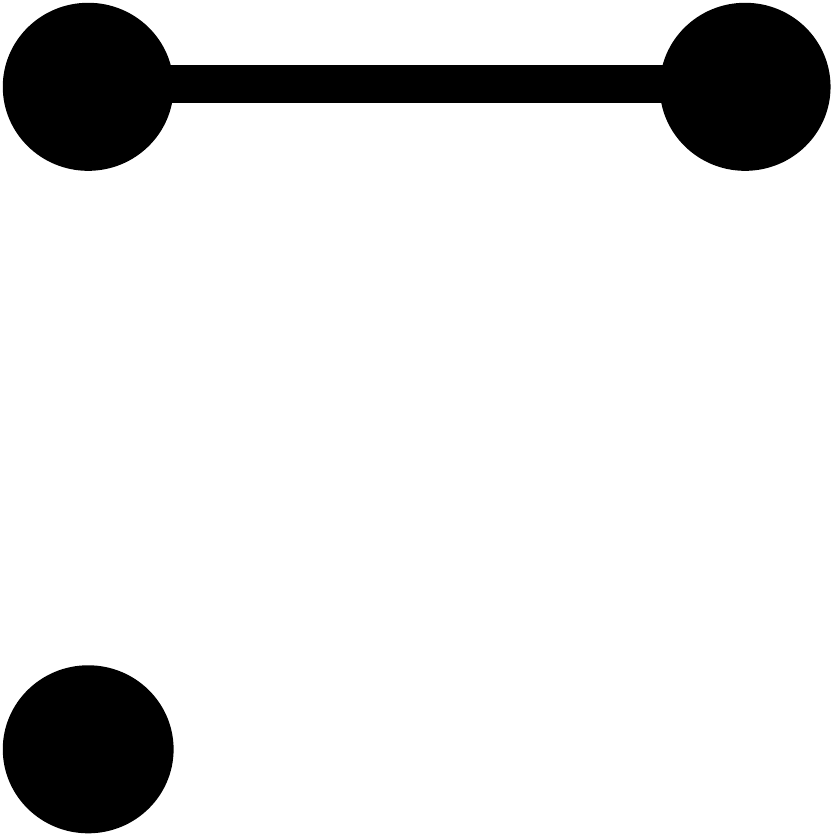} 
& $\g_5$  & 3-node-1-edge 
&  \includegraphics[scale=0.04]{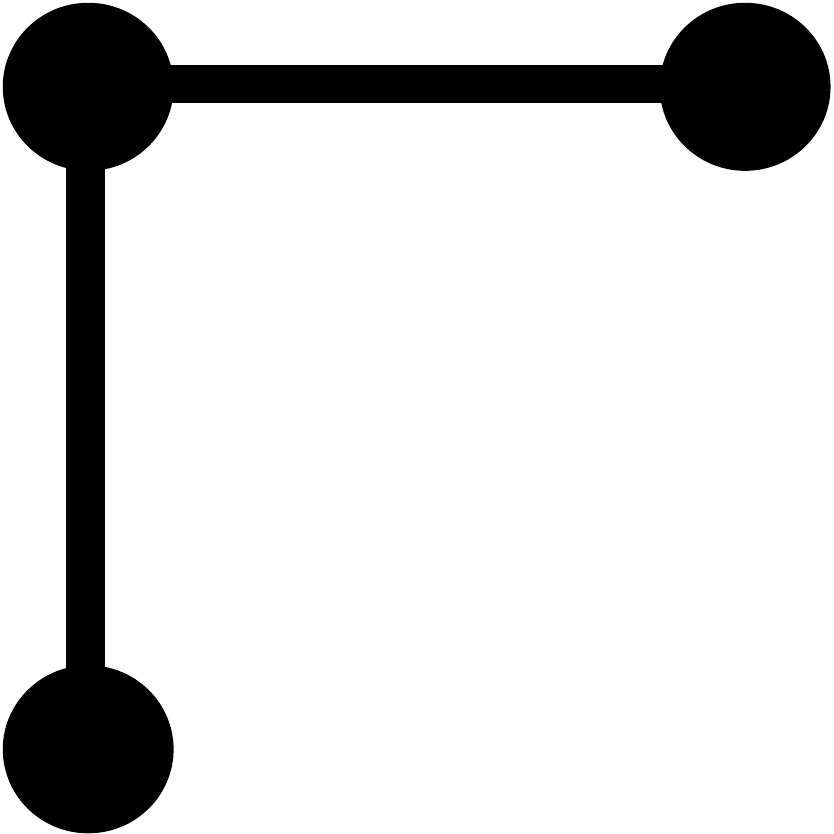} 
& 0.33 & 1 & 0.67 & 1.00  & 0 & 1 & 2 & 1 & 0 & 2\\ 
\TTT\BBB
&  \includegraphics[scale=0.04]{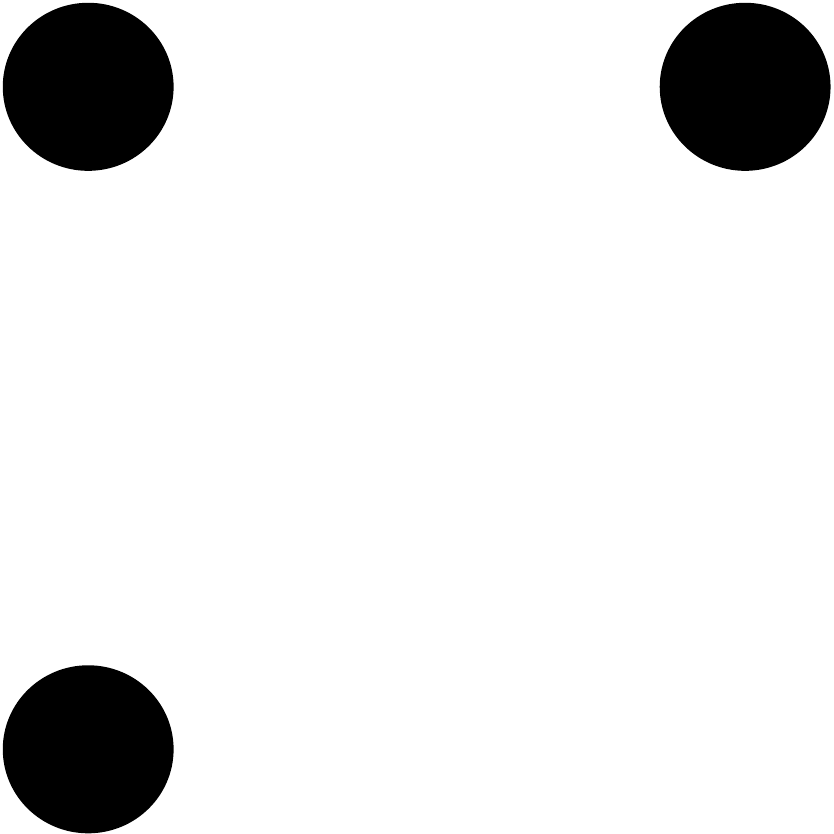} 
& $\g_6$  & 3-node-independent 
&  \includegraphics[scale=0.04]{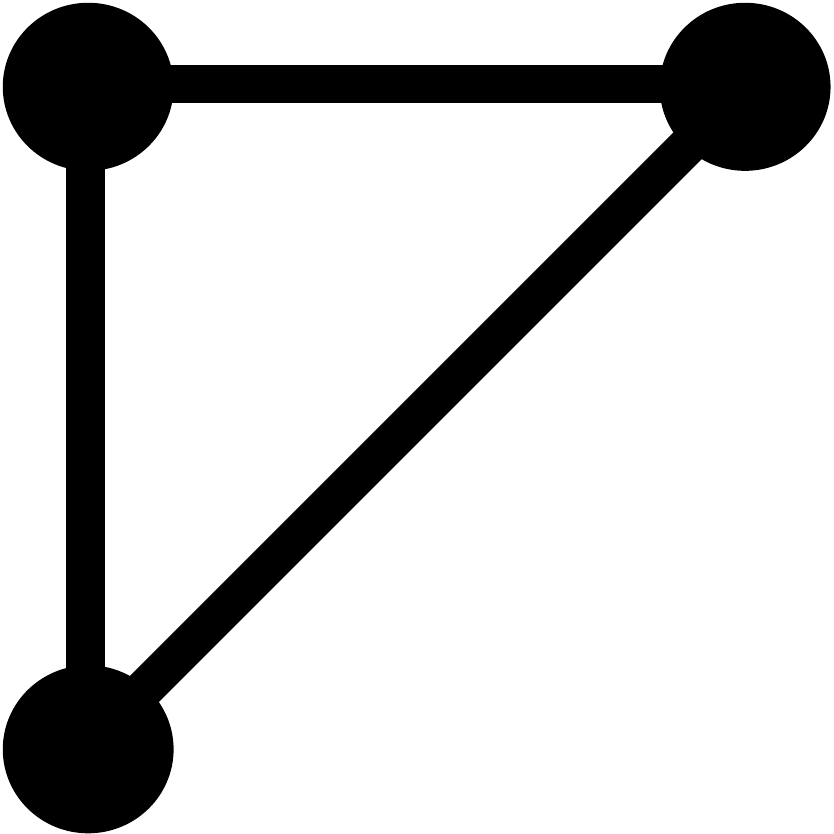} 
& 0.00 & 0 & 0.00 & 0.00  & 0 & 0 & 1 & $\infty$ & 0 & 3\\ 

\midrule 
\multirow{10}{*}{\rotatebox{90}{\mbox{\textsc{Connected}}}}
&  \includegraphics[scale=0.04]{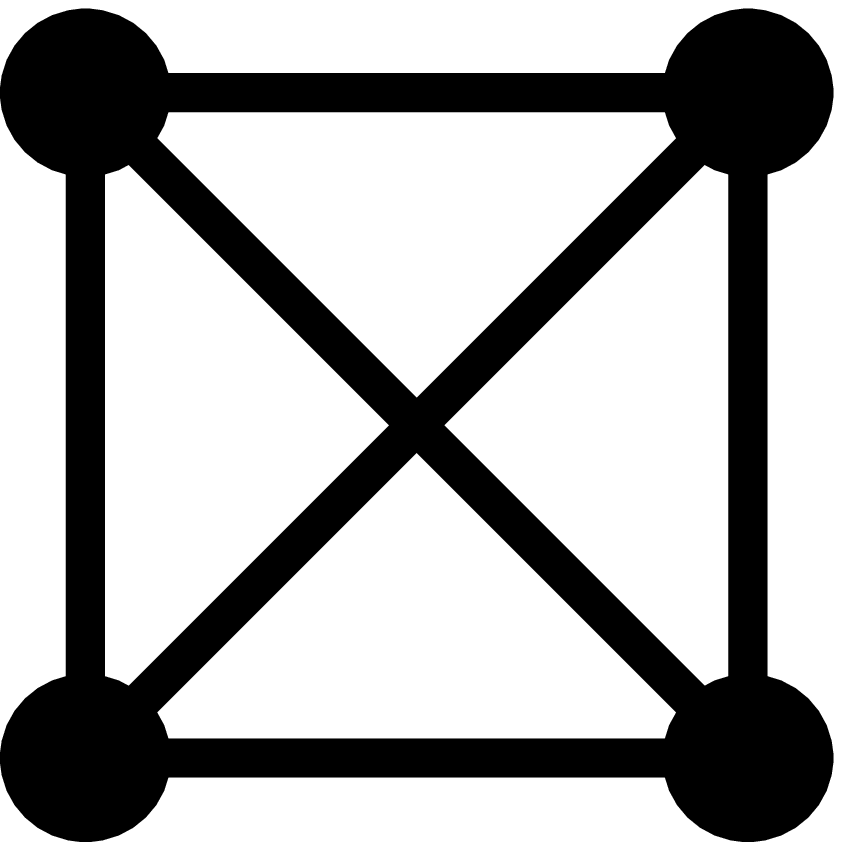} 
& $\g_7$  & 4-clique 
&  \includegraphics[scale=0.04]{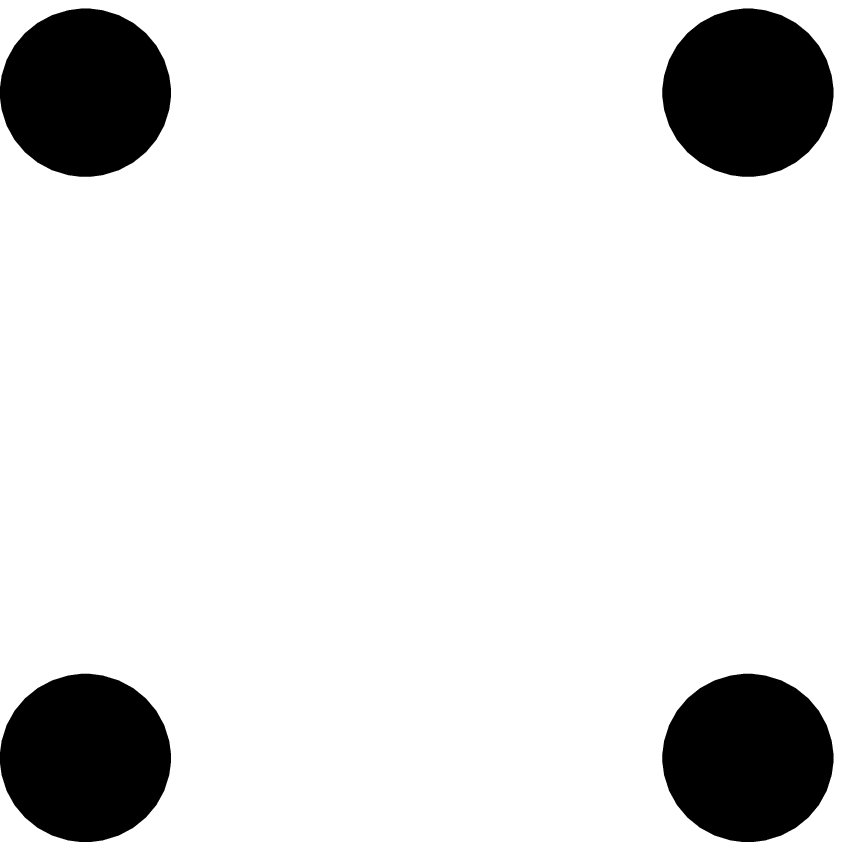} 
& 1.00 & 3 & 3.0 & 1.00 & 4 & 3 & 4 & 1 & 0 & 1\\ 
\TTT\BBB
&  \includegraphics[scale=0.04]{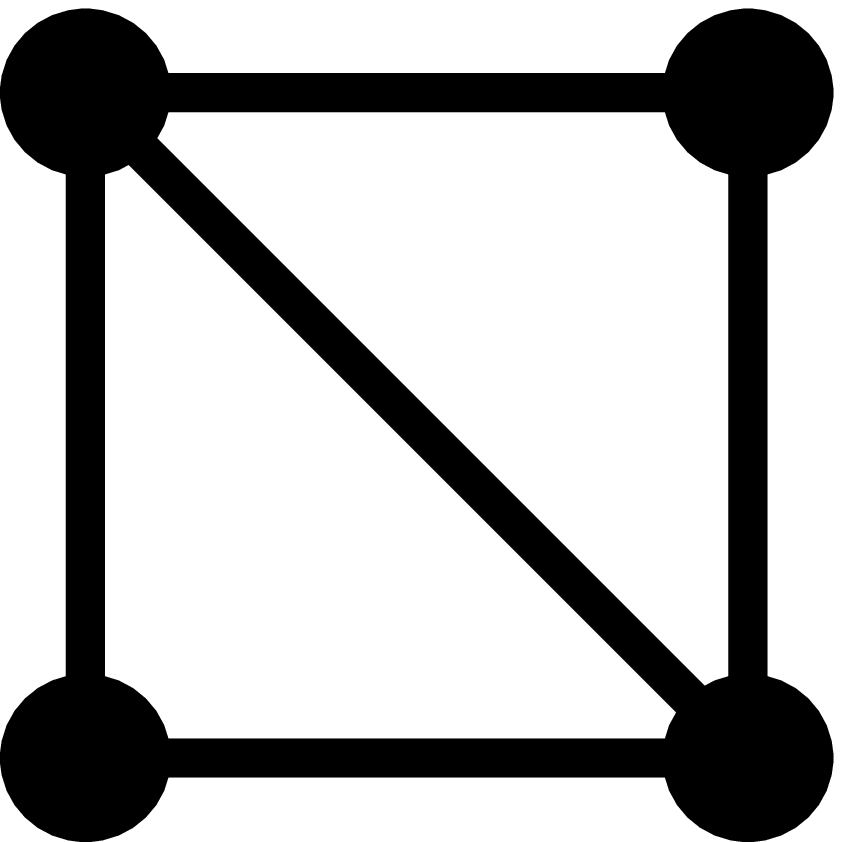} 
& $\g_8$  & 
chordal-cycle
&  \includegraphics[scale=0.04]{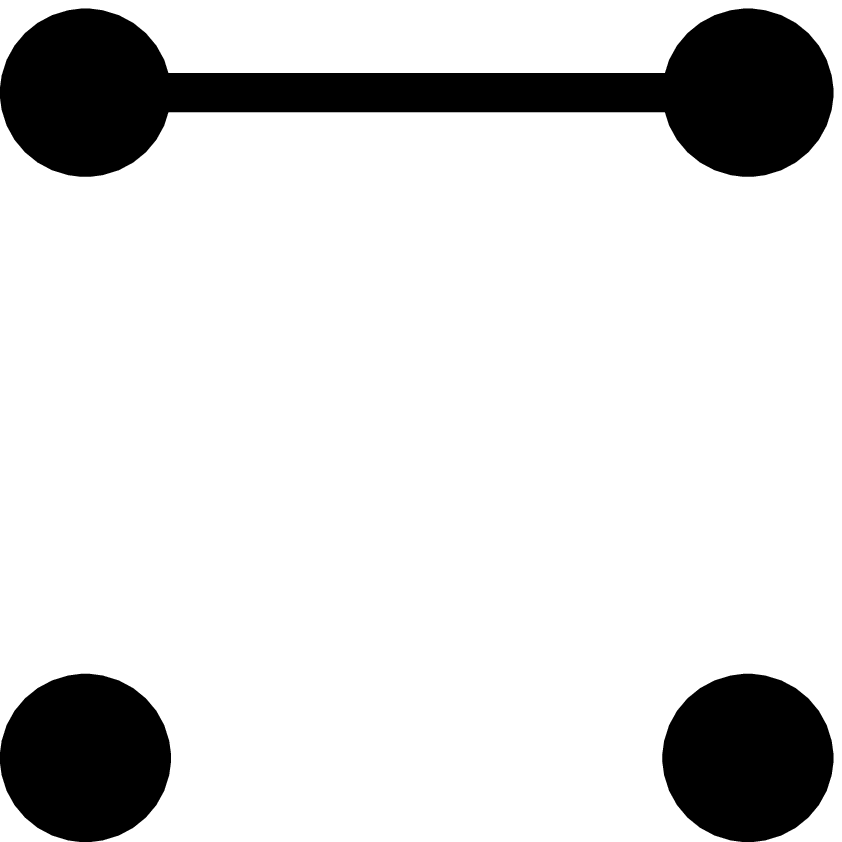} 
& 0.83 & 3 & 2.5 & -0.66 & 2 & 2 & 3 & 2 & 1 & 1 \\ 
\TTT\BBB
&  \includegraphics[scale=0.04]{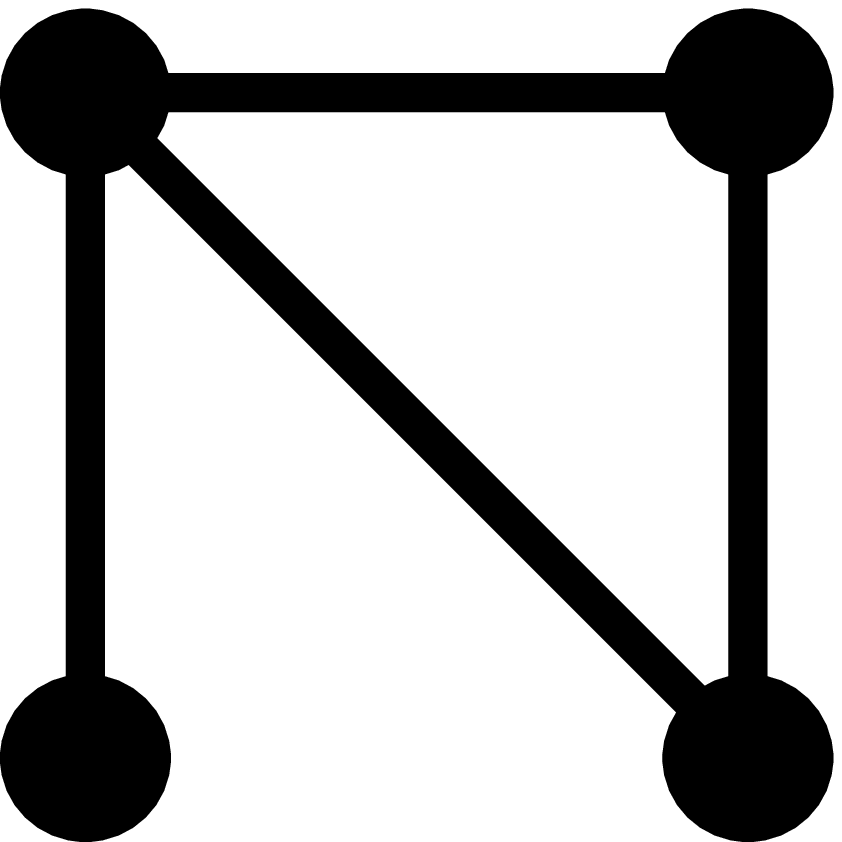} 
& $\g_9$  & 
tailed-triangle
&  \includegraphics[scale=0.04]{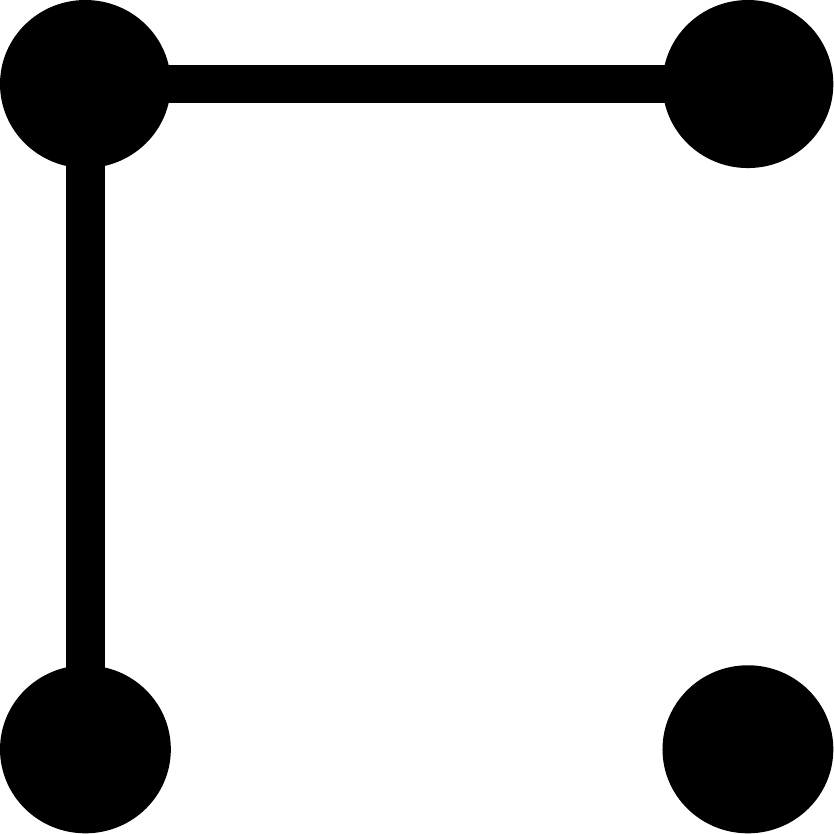} 
& 0.67 & 3 & 2.0 & -0.71 & 1 & 2 & 3 & 2 & 2 & 1\\
\TTT\BBB
&  \includegraphics[scale=0.04]{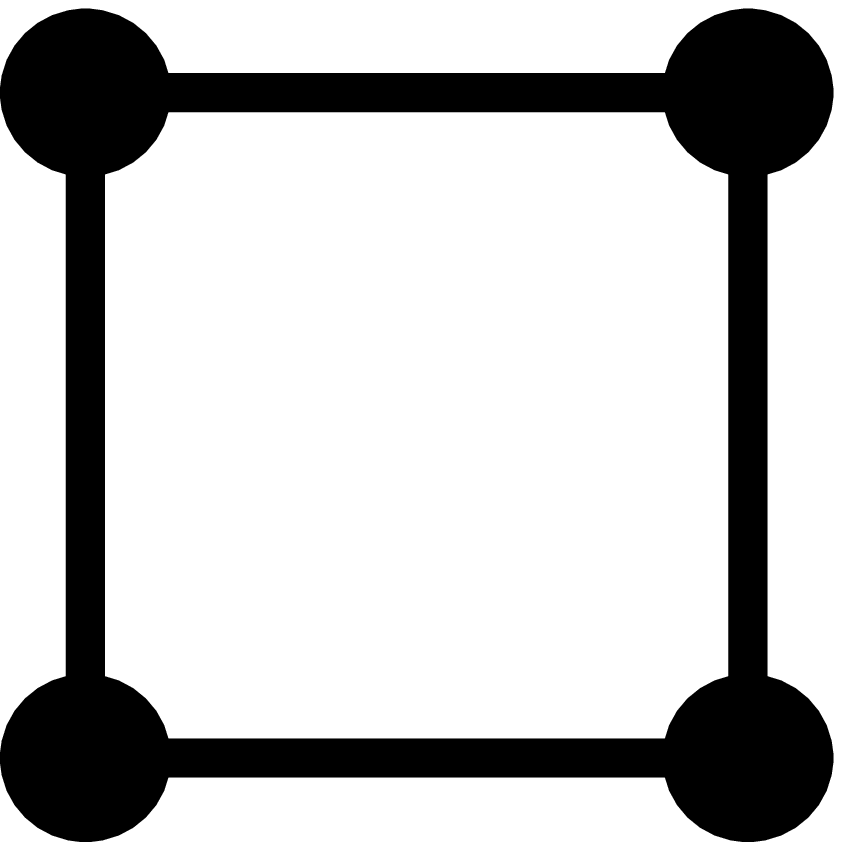} 
& $\g_{10}$  & 4-cycle 
&  \includegraphics[scale=0.04]{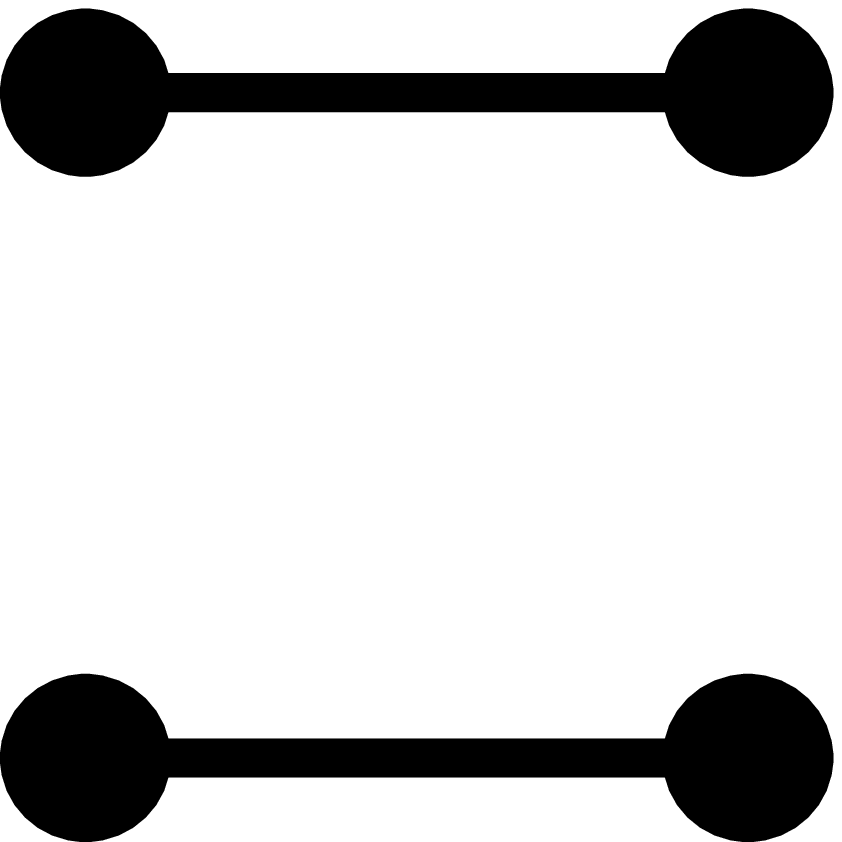} 
& 0.67 & 2 & 2.0 & 1.00 & 0 & 2 & 2 & 2 & 1 & 1\\
\TTT\BBB
&  \includegraphics[scale=0.04]{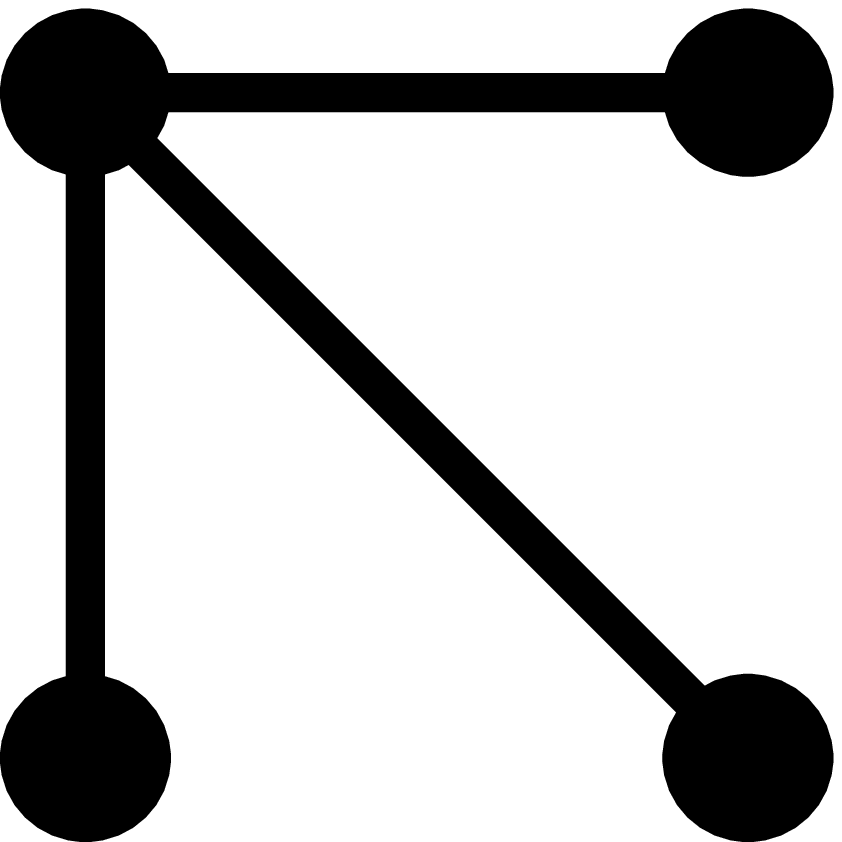} 
& $\g_{11}$  & 3-star 
&  \includegraphics[scale=0.04]{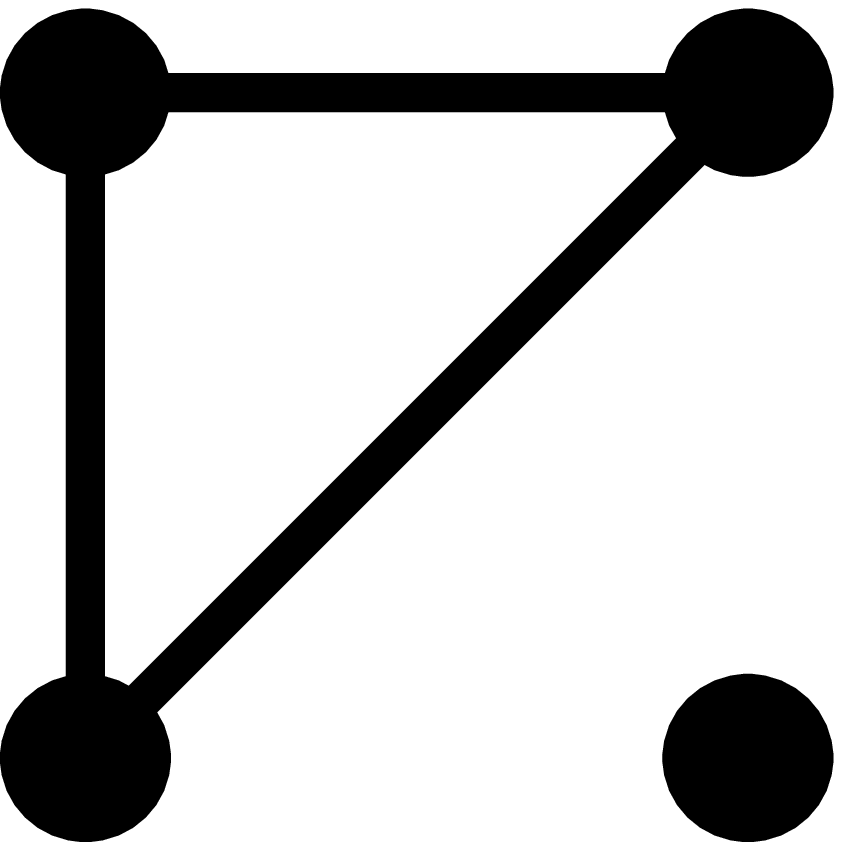} 
& 0.50 & 3 & 1.5 & -1.00 & 0 & 1 & 2 & 2 & 3 & 1\\
\TTT\BBB
&  \includegraphics[scale=0.04]{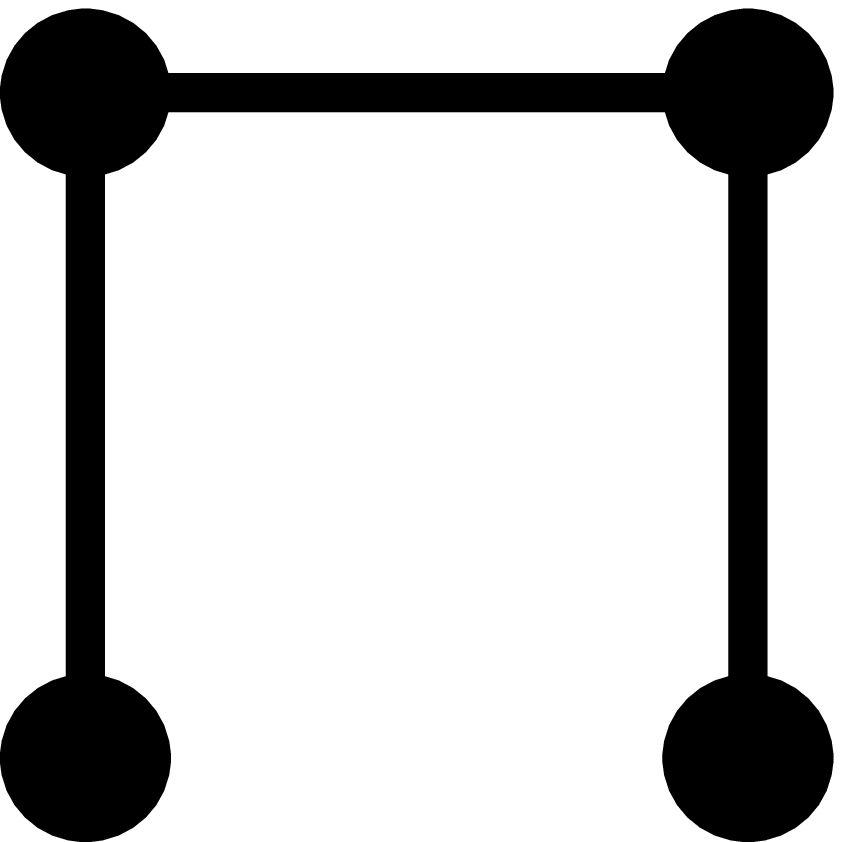} 
& $\g_{12}$  & \cellcolor{white}4-path
&  \includegraphics[scale=0.04]{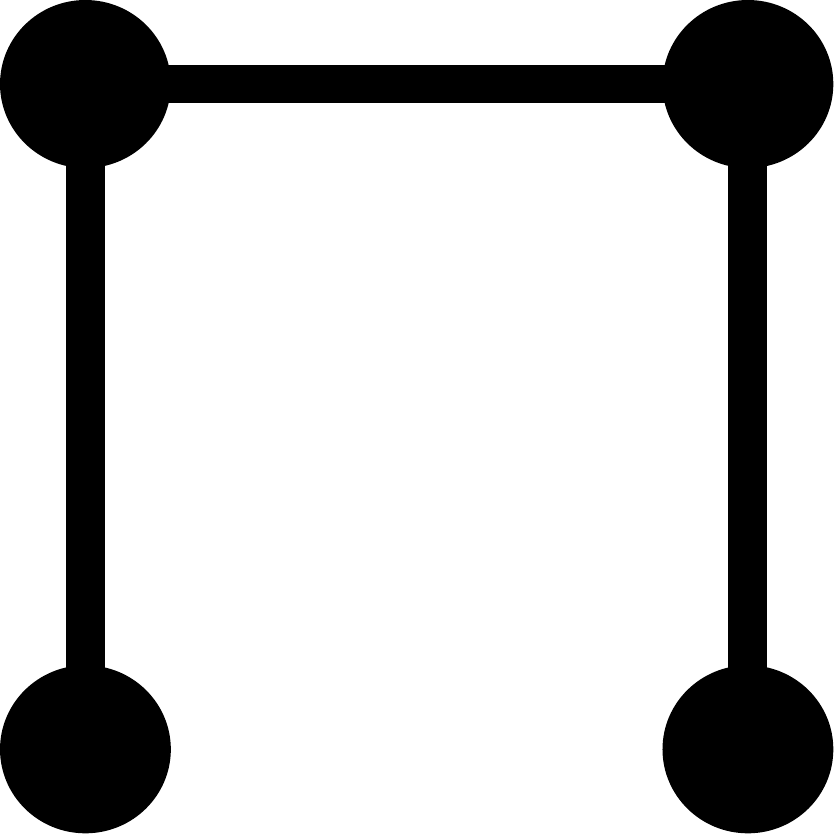} 
& 0.50 & 
2 & 
1.5 &
-0.50 &
0 &
1 &
2 &
3 &
2 &
1\\
\TTT\BBB
\multirow{8}{*}{\rotatebox{90}{\mbox{\textsc{Disconnected}}}} 
&  \includegraphics[scale=0.04]{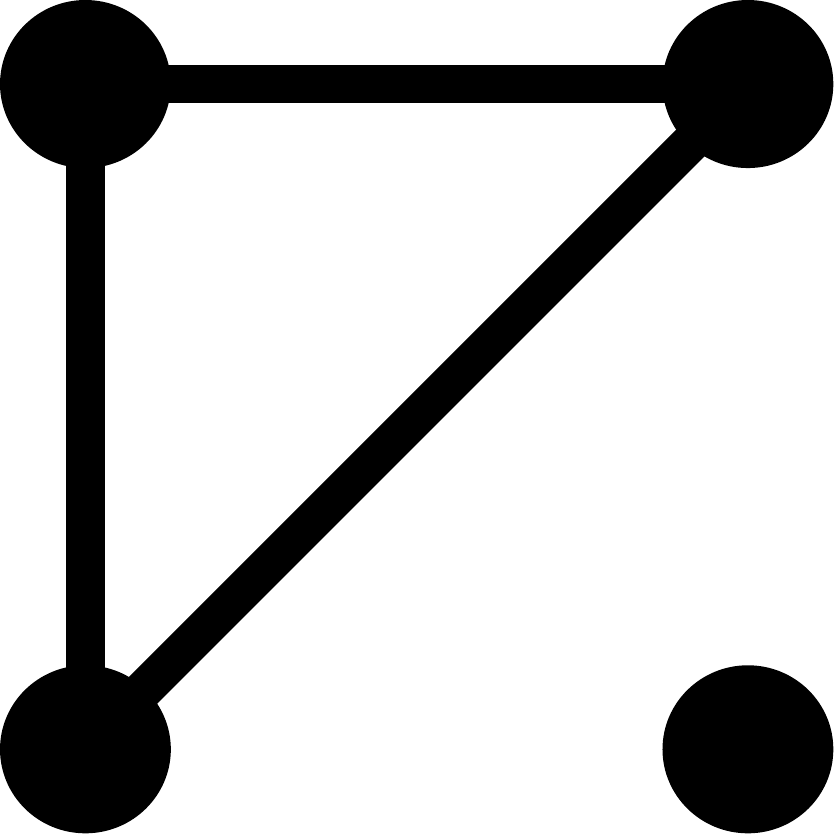} 
& $\g_{13}$  & 4-node-1-triangle 
&  \includegraphics[scale=0.04]{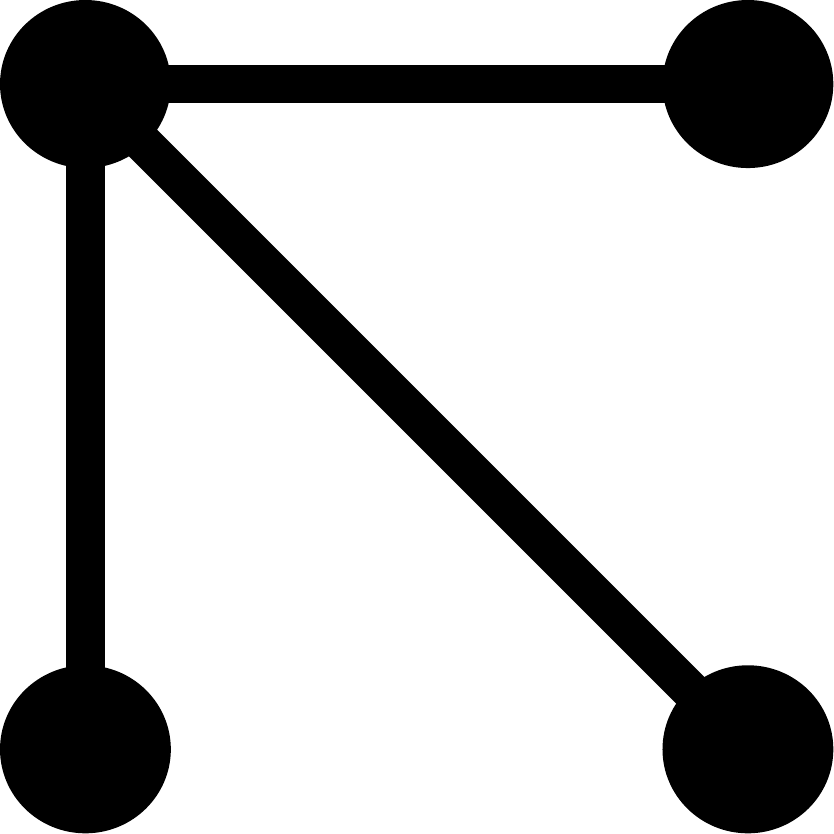} 
& 0.50 & 2 & 1.5 & 1.00 & 1 & 2 & 3 & 1 & 0 & 2\\
\TTT\BBB
&  \includegraphics[scale=0.04]{graphlets/4-node-star} 
& $\g_{14}$  & 4-node-2-star 
&  \includegraphics[scale=0.04]{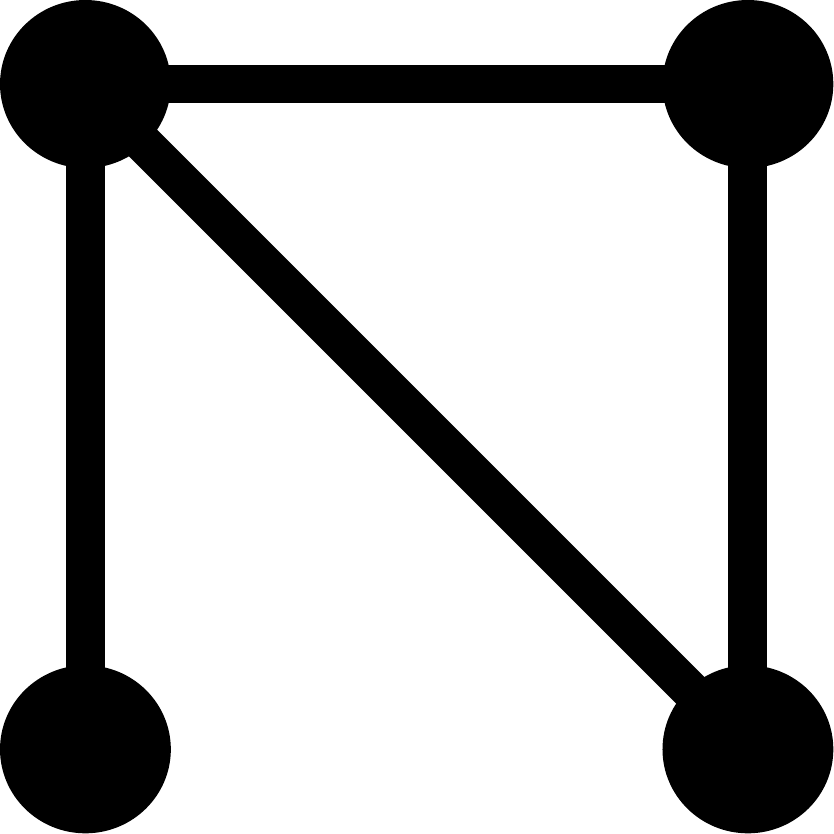} 
& 0.33 & 2 & 1.0 & -1.00 & 0 & 1 & 2 & 2 & 1 & 2\\
\TTT\BBB
&  \includegraphics[scale=0.04]{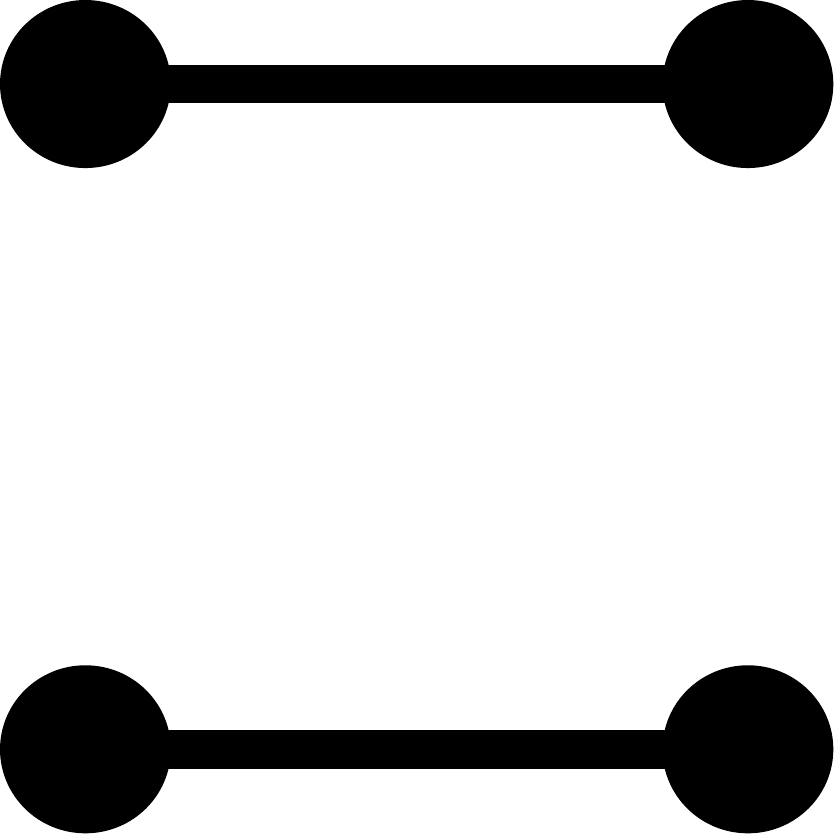} 
& $\g_{15}$  & 4-node-2-edge 
&  \includegraphics[scale=0.04]{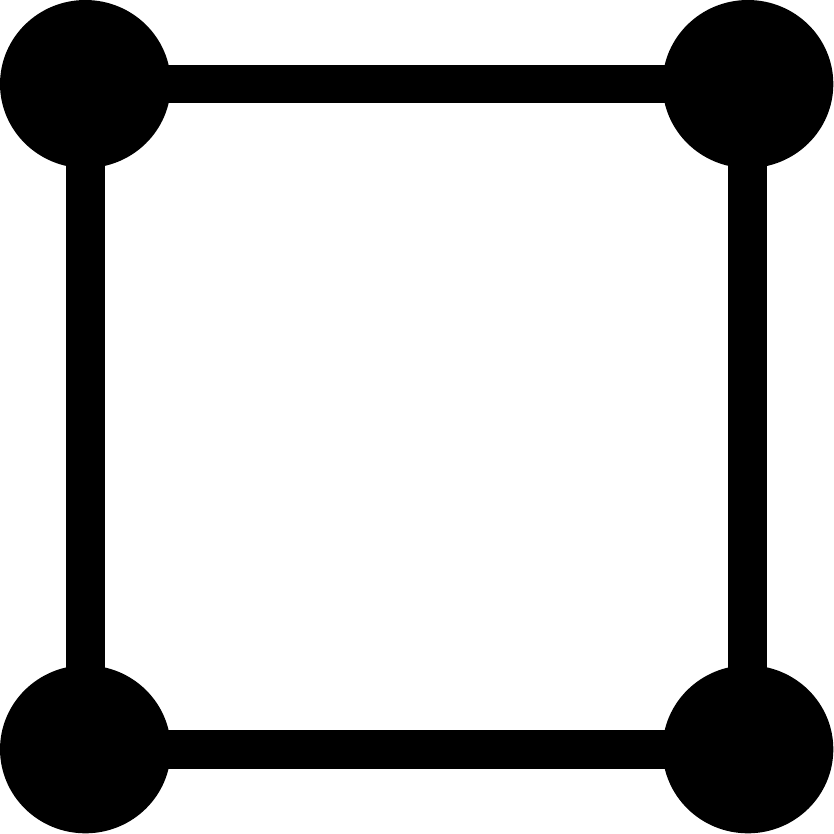} 
& 0.33 & 1 & 1.0 & 1.00 & 0 & 1 & 2 & 1 & 0 & 2\\
\TTT\BBB
&  \includegraphics[scale=0.04]{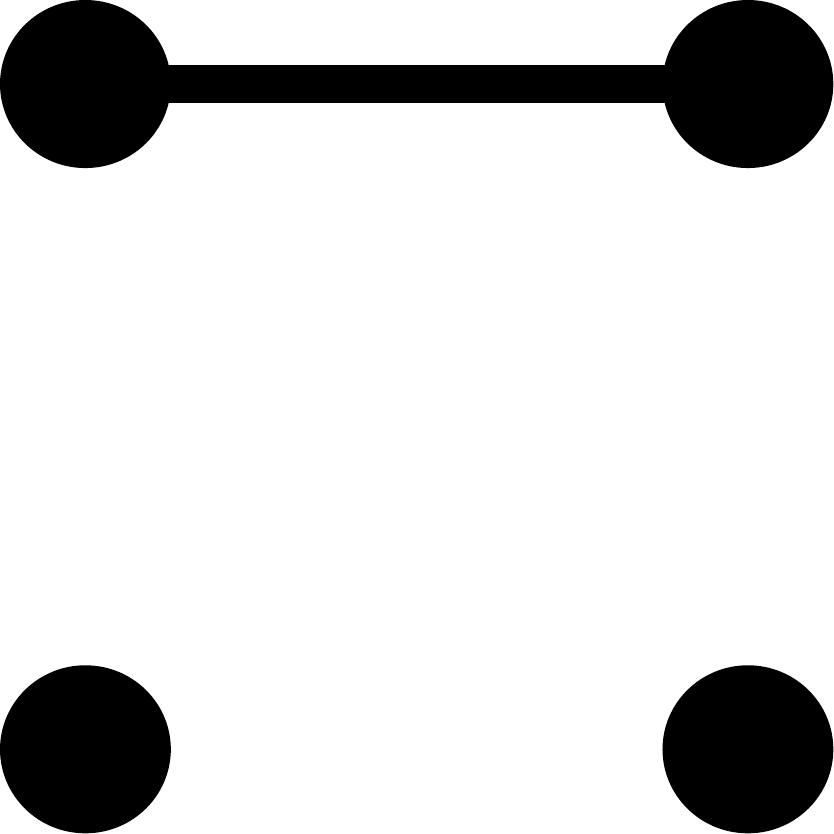} 
& $\g_{16}$  & 4-node-1-edge 
&  \includegraphics[scale=0.04]{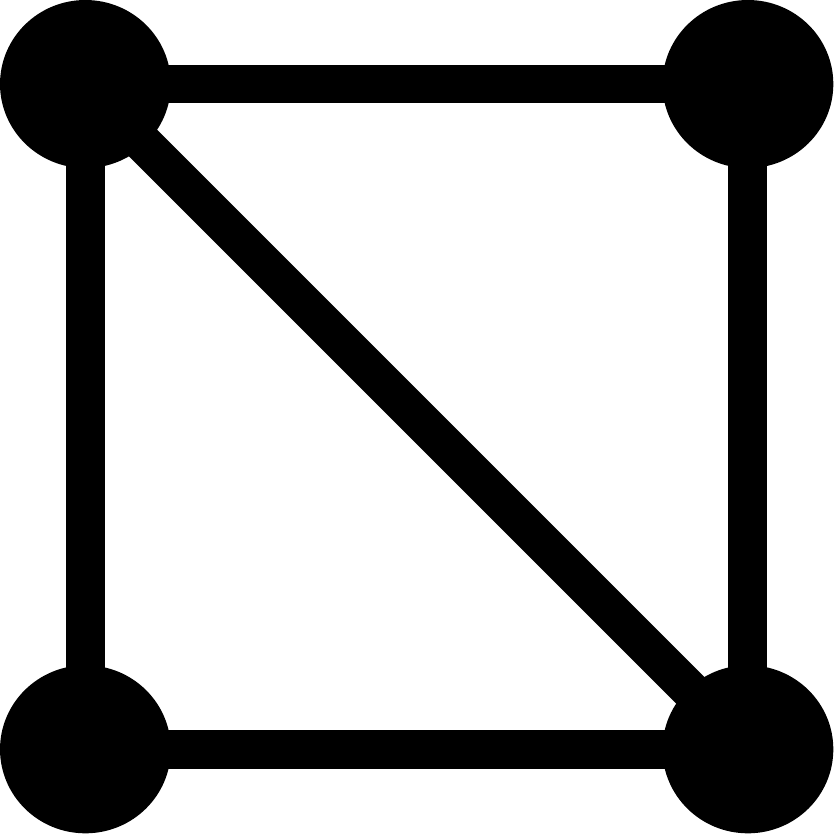}
& 0.17 & 1 & 0.5 & 1.00 & 0 & 1 & 2 & 1 & 0 & 3\\
\TTT\BBB
&  \includegraphics[scale=0.04]{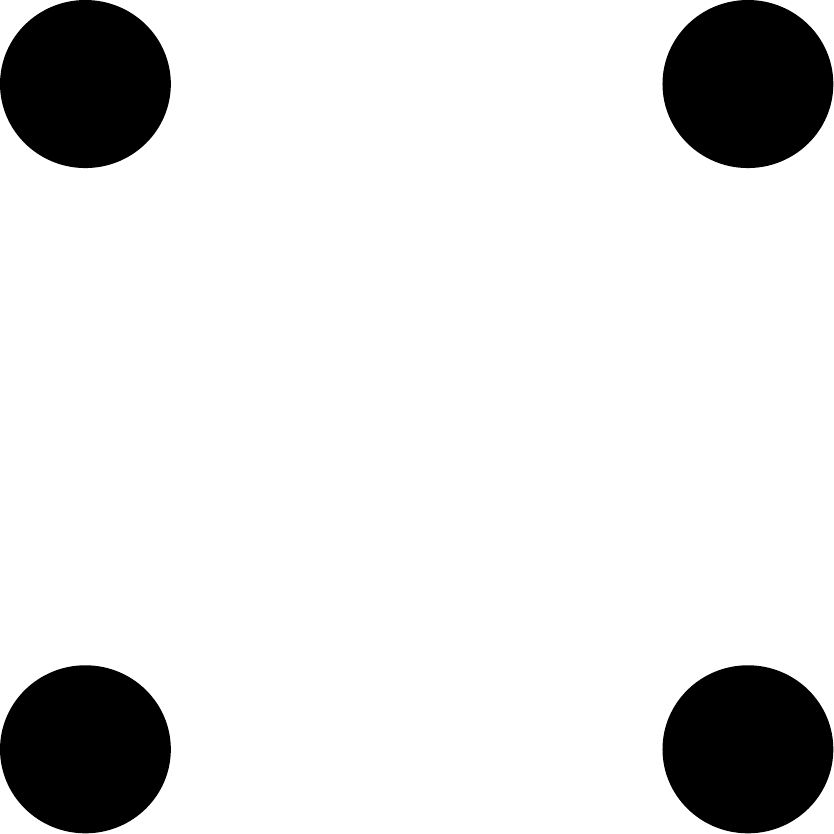} 
& $\g_{17}$  & 4-node-independent 
&  \includegraphics[scale=0.04]{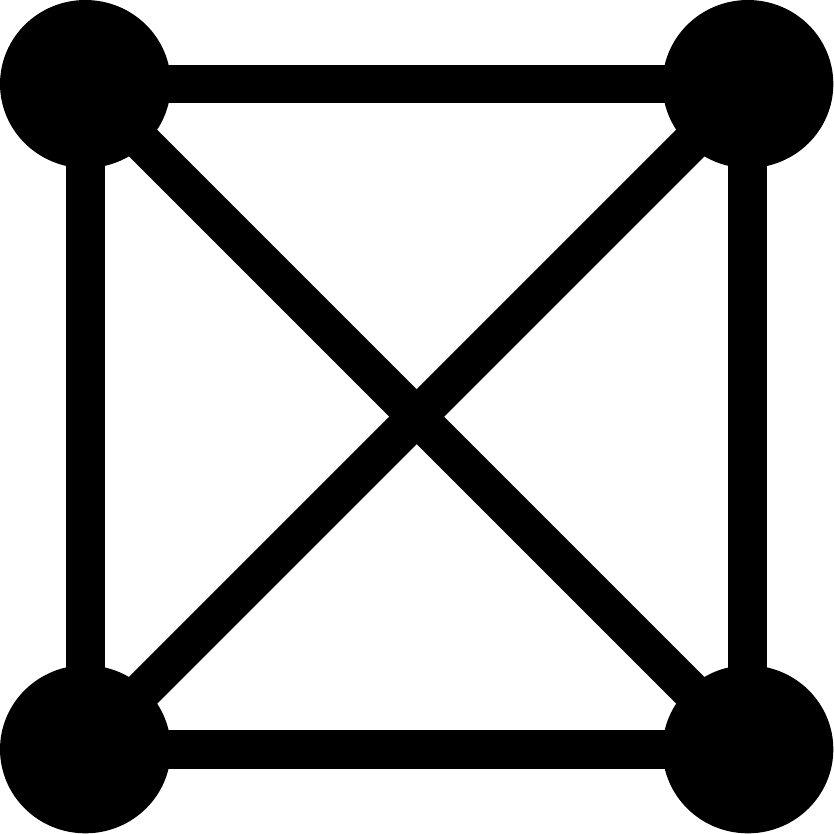} 
& 0.00 & 0 & 0.0 & 0.00 & 0 & 0 & 1 & $\infty$ & 0 & 4\\
\bottomrule
\end{tabularx}
}
\vspace{-0mm}
\end{table}
}

\subsection{Preliminaries}
\noindent
Let $G=(V,\E)$ be an undirected simple graph with $\n = |V|$ vertices and $\m=|\E|$ edges.
Sets are \textit{ordered}.
Given a vertex $v \in V$, let $\N(v) = \{w \, | \, (v,w) \in E\}$ be the set of vertices adjacent to $v$ in $G$. 
We also define $\d_v$ as the degree of $v \in V$, where the degree $\d_v$ of $v \in V$ is defined as the size of the neighborhood of $v$, \ie, $\d_v = |\N(v)|$.
Further, let $\dmax(G)$ be the maximum vertex degree in $G$.
Let $\G^{(k)}$ denote the set of $k$-vertex subgraphs and $\G = \G^{(1)} \cup \G^{(2)} \cup \cdots \cup \G^{(k)}$.
Given a set $U=\{u_1,...,u_k\} \subset V$ of $k$ vertices, 
we define a $k$-graphlet as any k-vertex induced subgraph $\g_i=(U, E[U])$ where $\g_i \subset \G^{(k)}$.
Note that $E[U]$ is the set of edges between the vertices in $U$.
This work focuses on estimating statistics of these induced subgraphs called graphlets.

It is important to distinguish between \emph{connected} and \emph{disconnected} graphlets (see Table~\ref{table:graphlet_notation}). 
A graphlet is connected if there is a path from any node to any other node in the graphlet, otherwise it is disconnected.
Table~\ref{table:graphlet_notation} summarizes the important and fundamental properties of all graphlets of size $k \in \{2,3,4\}$.

\subsection{Objective}
\noindent
The goal of this work is to obtain fast and accurate estimates of a variety of 
\emph{macro} and 
\emph{micro} graphlet properties (including both single-valued network statistics as well as distributions, that is, multi-valued network statistics) for both connected and disconnected graphlets (See Table~\ref{table:graphlet_notation}) that include: 
(a) frequency of graphlets $\g_i \in \G$, for all $i=1,2,...|\G|$ or frequency of a specific graphlet $\g_i \in \G$,
(b) graphlet frequency distributions ($\gfd$) including the connected, disconnected, and combined $\gfd$ containing both.
(c) univariate statistics such as mean, median, min, max, variance, Q1, Q3, etc.
(d) probability distribution (\textsc{pdf}, \textsc{cdf}, \textsc{ccdf}) of a particular graphlet $\g_i$. 

\begin{figure}[t!]
\vspace{-2mm}
\centering
\includegraphics[width=0.7\linewidth]{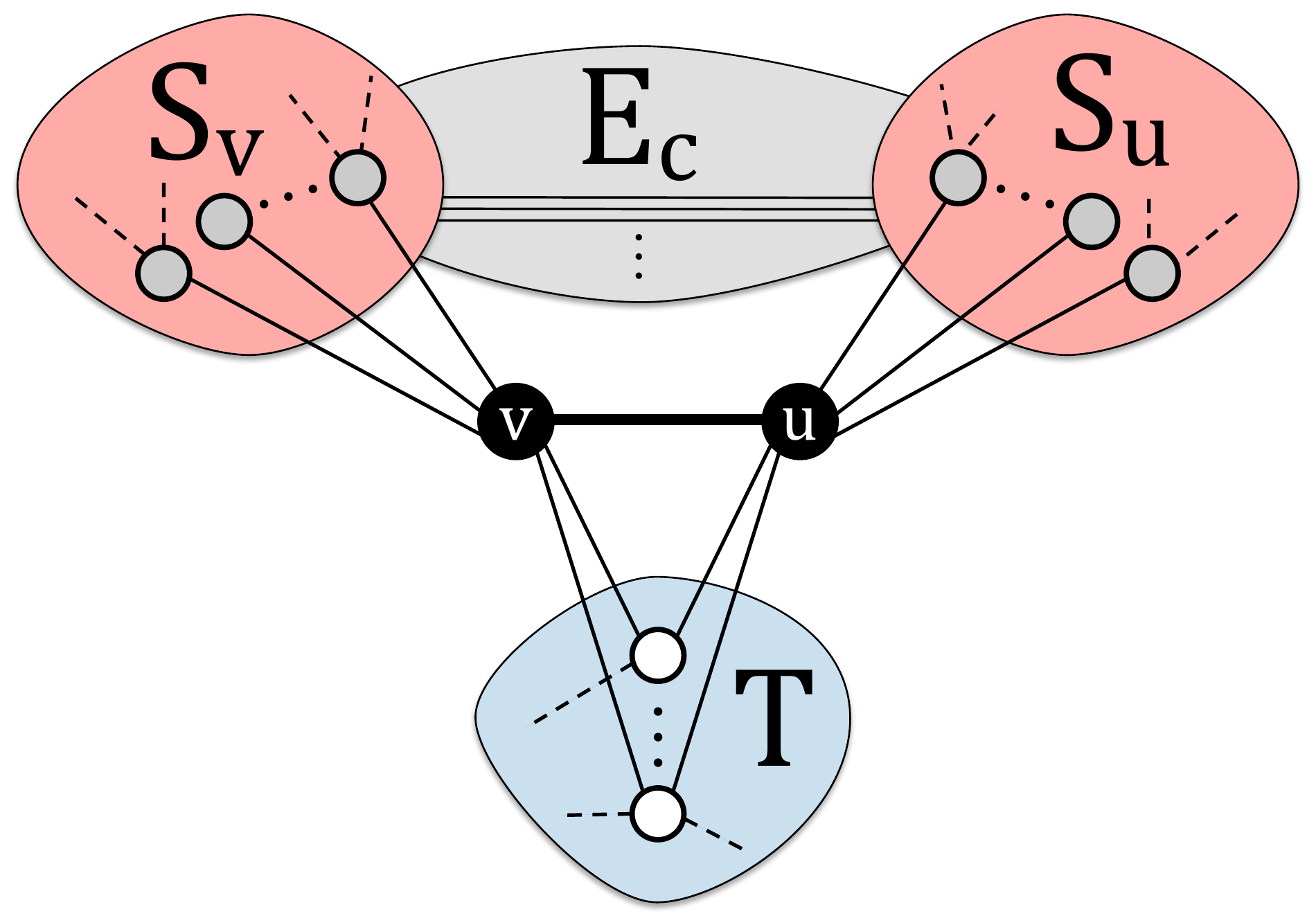}
\vspace{-1mm}
\caption{
Let $\tri$ be the set of nodes completing a triangle with the edge $(v,u) \in \E$, and let $\S_v$ and $\S_u$ be the set of nodes that form a 2-star with $v$ and $u$, respectively.
Note that $\S_u \cap \S_v = \emptyset$ by construction and $|\S_u \cup \S_v| = |\S_u|+|\S_v|$.
Further, let $\E_c$ be the set of edges that complete a cycle (of size 4) $\wrt$ the edge $\e=(v,u)$ where for each edge $(r,s) \in \E_c$ such that $r \in \S_v$ and $s \in \S_u$ and both $(r \cap \S_u) \cup (s \cap \S_v) = \emptyset$, that is, $r$ is not adjacent to $u$ ($r \not\in \N(u)$) and $s$ is not adjacent to $v$ ($s \not\in \N(v)$).
}
\label{fig:graphlets-intuition}
\vspace{-4mm}
\end{figure}

Despite the practical importance of these graphlet properties, this work is the first to propose and investigate many of these novel problem variants.
In addition, this paper proposes and investigates methods for estimating not only connected graphlet counts, but also disconnected graphlet counts, as well as a variety of other important and novel graphlet properties beyond simple counts.
Disconnected graphlets are vital for many problems, including graph and node classification, graph kernels, among others~\cite{shervashidze2009efficient,pgd}.
For instance, Shervashidze~\etal~\cite{shervashidze2009efficient} find that disconnected graphlets are essential for correct classification on some datasets (See~\cite{shervashidze2009efficient} pp. 495).
Nevertheless, this leads us to define and investigate many novel problem variants.
For instance, this work estimates the three possible $\gfd$ variations (connected, disconnected, and a $\gfd$ containing both).

{
\algrenewcommand{\alglinenumber}[1]{\fontsize{6.5}{7}\selectfont#1}
\begin{figure}[b!]
\vspace{-3mm}
\begin{center}
\begin{algorithm}[H]
\caption{\,\small
Edge-centric graphlet estimators
}
\label{alg:graphlet-est}
{
\begin{spacing}{1.3}
\fontsize{8}{9}\selectfont
\begin{algorithmic}[1]
\vspace{-1.3mm}
\Require ~\newline
\quad\quad a graph $G = (V,\E)$ 
\newline
\quad\quad a sample size $\K$, or sample probability $\pr$
\smallskip
\parfor[$j = 1,2,...,\K$] \label{algline:sample-a-graphlet}
	\State Select $\e$ via an arbitrary (weighted/uniform) distribution $\Fdist$ \label{algline:select-edge-neighborhood-via-F} \label{algline:sample-an-edge-uniformly}
	\State Set $\Es \leftarrow \Es \cup \{\e_{}\}$ \label{algline:add-edge-to-sample}
\endpar
\State Obtain estimated graphlet counts $\mX$ for $\Es$ via Alg~\ref{alg:pgd} \label{algline:obtain-est-graphlet-counts}
\State \textbf{return} $\bX$ -- the estimated graphlet counts \label{algline:return-estimated-graphlet-counts}
\end{algorithmic}
\end{spacing}
}
\end{algorithm}
\end{center}
\vspace{-3mm}
\end{figure}
}

While previous methods have been proposed for computing counts of connected induced subgraphs, we instead propose a unifying unbiased estimation framework that is robust and generalizes for connected \emph{and} disconnected graphlets.
In addition to the above limitation, previous work has also been limited to simple count statistics (frequency or proportion of a graphlet).
This work introduces and investigates estimators for a number of important macro \emph{and} micro-level graphlet statistics (\eg, graphlet counts for individual edges, as well as the total frequency of a graphlet in $G$).
In particular, the framework gives rise to graphlet estimation methods that are fast and accurate for both 
(a) macro and 
(b) micro-level graphlet properties including 
(i) single-valued graphlet statistics and 
(ii) distributions (multiple-valued network statistics).
For each of these new problem variants, we introduce fast, parallel, and accurate techniques and demonstrate their effectiveness and utility on a variety of networks.

Table~\ref{table:taxonomy-and-overview} summarizes existing related methods as well as our proposed approach according to the 
types of graphlets computed (connected and/or disconnected), 
the macro and micro-level graphlet statistics estimated by each (including single-valued statistics and distributions), 
as well as computational and algorithmic aspects/features.

{
\providecommand{\rotateDeg}{90}
\setlength{\tabcolsep}{1.2pt}
\providecommand{\rotDeg}{70}
\definecolor{verylightgreennew}{RGB}	{220,255,220}
\definecolor{verylightrednew}{RGB}		{255, 230, 230}
\definecolor{verylightrednewlighter}{RGB}		{255, 229, 239}
\definecolor{lightgraynew}{rgb}{0.9,0.9,0.9}
\providecommand{\cellsz}{0.32cm}
\providecommand{\cellno}{\cellcolor{verylightrednew}}
\providecommand{\cellyes}{{\TTT \BBB  \cm \cellcolor{verylightgreennew}}}
\begin{table}[t!]
\vspace{-0mm}
\scriptsize
\tiny
\caption{
Qualitative and quantitative comparison of the two main classes of graphlet estimation methods, namely, \emph{direct graphlet sampling} and \emph{localized graphlet estimation} ($\lge$) methods.
Direct methods are those that sample $k$-vertices directly and retrieve the graphlet induced by that subset. 
This work proposes the family of \emph{localized graphlet estimation} ($\lge$) methods that select (sample) localized $\ell$-neighborhoods for estimation.
The first two columns refer to the type of graphlets estimated (connected and/or disconnected graphlets).
The next six columns refer to the \emph{macro} and \emph{micro} graphlet estimation problems.
In particular, columns 3-5 refer to the \emph{macro} graphlet statistics and distributions estimated by the methods (counts, $\gfd$, and others such as extremal stats./distributions, etc.),
whereas columns 6-8 refer to the \emph{micro} graphlet statistics and distributions.
``Parallel" refers to parallel estimation methods. 
``Space efficient" holds true if the space requirements of the algorithm are sublinear (preferably poly-logarithmic in the size of the input).
``Massive 1B+" holds true if the methods is capable of handling massive graphs of 1 billion or more edges.
``Streaming" holds true if the method is amenable to streaming implementation. 
``Position-aware" is true if the algorithm supports position-aware graphlets (orbits).
``Sparse \& dense" is true if the method has limited assumptions, and designed/capable of handling both sparse and dense graphs.
``Parameter-free" methods are those that do not expect any user-specified input parameters (though they can be set, but is not required).
``All graphlets" holds true if the method computes graphlet statistics and distributions for all graphlets up to size $k$.
}
\label{table:taxonomy-and-overview}
\centering\scriptsize
\vspace{-1mm}
\begin{tabular}
{
lr c !{\vrule width 0.4pt} 
P{\cellsz} P{\cellsz} !{\vrule width 0.4pt} 
P{\cellsz} P{\cellsz} P{\cellsz} !{\vrule width 0.4pt} 
P{\cellsz} P{\cellsz} P{\cellsz} !{\vrule width 0.4pt} 
ccHcc cccc  !{\vrule width 0.8pt}}

\multicolumn{3}{c !{\vrule width 0.4pt} }{} &
\multicolumn{2}{c !{\vrule width 0.4pt} }{} &
\multicolumn{3}{c !{\vrule width 0.4pt} }{\textsc{Macro}} &
\multicolumn{3}{c !{\vrule width 0.4pt} }{\textsc{Micro}} &
\multicolumn{9}{c !{\vrule width 0.7pt} }{\textsc{Computational}} 
\\

& \textbf{Method} \TTT \BBB 
& 
&
\rotatebox{\rotateDeg}{\textbf{Connected}} &
\rotatebox{\rotateDeg}{\textbf{Disconnected}} &
\rotatebox{\rotateDeg}{\textbf{Counts}} &
\rotatebox{\rotateDeg}{\textbf{\gfd}} &
\rotatebox{\rotateDeg}{\textbf{Others}} & 
\rotatebox{\rotateDeg}{\textbf{Counts}} &
\rotatebox{\rotateDeg}{\textbf{\gfd}} &
\rotatebox{\rotateDeg}{\textbf{Others}} &
\rotatebox{\rotateDeg}{\textbf{Parallel}} &
\rotatebox{\rotateDeg}{\textbf{Space efficient}} &
 &
\rotatebox{\rotateDeg}{\textbf{Massive 1B+}} &
\rotatebox{\rotateDeg}{\textbf{Streaming}} &
\rotatebox{\rotateDeg}{\textbf{Position-aware}} &
\rotatebox{\rotateDeg}{\textbf{Sparse \& dense}} &
\rotatebox{\rotateDeg}{\textbf{Parameter-free}} &
\rotatebox{\rotateDeg}{\textbf{All graphlets}} 
\\
\hline

\multirow{6}{*}{\rotatebox[origin=c]{90}{\mbox{
\normalsize
\small \bf \sc
$\mathsf{DIRECT}$}}}
& \textsc{Sherv.}~\etal~\cite{shervashidze2009efficient}  
&
& \cellyes 
& \cellyes 
& \cellno
& \cellyes 
& \cellno 
& \cellno  
& \cellno 
& \cellno
& \cellno  
& \cellyes 
& \cellno  
& \cellno  
& \cellyes 
& \cellno  
& \cellno  
& \cellno  
& \cellno  
\\

& \textsc{guise}~\cite{guise}   
&
& \cellyes 
& \cellno  
& \cellno
& \cellyes 
& \cellno 
& \cellno  
& \cellno 
& \cellno   
& \cellno  
& \cellno 
& \cellno  
& \cellno 
& \cellno  
& \cellno 
& \cellno  
& \cellno   
& \cellno
\\

& \textsc{graft}~\cite{graft}      
&
& \cellyes 
& \cellno  
& \cellno
& \cellyes 
& \cellno 
& \cellno  
& \cellno 
& \cellno
& \cellno  
& \cellno 
& \cellno  
& \cellno 
& \cellno  
& \cellno 
& \cellno  
& \cellno   
& \cellno
\\

& \textsc{3-path samp.}~\cite{path-sampling} 
&    
& \cellyes 
& \cellno
& \cellyes 
& \cellno
& \cellno 
& \cellno  
& \cellno 
& \cellno
& \cellno  
& \cellno 
& \cellno  
& \cellno 
& \cellno  
& \cellno 
& \cellno  
& \cellno 
& \cellno
\\
\hline

\multirow{2}{*}{\rotatebox{90}{\mbox{
\small
$\mathsf{LGE}$}}}
& \textsc{uniform}   
&
& \cellyes 
& \cellyes 
& \cellyes 
& \cellyes 
& \cellyes 
& \cellyes 
& \cellyes 
& \cellyes
& \cellyes 
& \cellyes 
& \cellyes 
& \cellyes 
& \cellyes 
& \cellyes 
& \cellyes 
& \cellyes 
& \cellyes 
\\

& \textsc{kcore}       
&
& \cellyes 
& \cellyes 
& \cellyes 
& \cellyes 
& \cellyes 
& \cellyes 
& \cellyes 
& \cellyes
& \cellyes  
& \cellno   
& \cellyes  
& \cellyes  
& \cellno   
& \cellyes  
& \cellno   
& \cellyes  
& \cellyes  
\\
\noalign{\hrule height 0.7pt}
\end{tabular}
\vspace{-0mm}
\end{table}
}

\subsection{Class of Localized Graphlet Estimation Algorithms} \label{sec:lge-class-of-algorithms}
\noindent
The general unbiased graphlet estimation framework is based on sampling (or selecting) edge-induced neighborhoods $\N(\e)$.
Given an edge $\e=(u,v) \in E$, 
let $\N(\e)$ denote the edge neighborhood of $\e$ defined as:
\begin{equation}\label{eq:edge-neighborhood}
\N(e) = \N(u,v) = \N(u) \cup \N(v) \setminus \{u,v\},
\end{equation} 
\noindent
where $\N(u) \textrm{ and } \N(v)$ are the neighbors of $u \textrm{ and } v$, respectively.
The (explicit) edge-induced neighborhood is $\Ng_{\e} = G(\{\N(v) - u\} \cup \{\N(u) - v\})$.
The subgraph $\Ng(\e)$ consists of the set of vertices adjacent to $v$ or $u$ (non-inclusive) and all edges between that set.

In particular, given an edge-centric neighborhood $\N(\e)$, we compute all graphlets $\g_i$ that include $\e$.
Intuitively, an edge neighborhood $\N(\e)$ is sampled with some probability from the set of all edge-induced neighborhoods (See Alg~\ref{alg:graphlet-est}).
Using the edge neighborhood $\N(\e)$ centered at $\e\in E$ as a basis, we compute the frequency of each graphlet $\g_i \in \G$, for $i=1,...,|\G|$.
Let us note that edge neighborhoods may be selected uniformly at random or by an arbitrary weighted distribution $\Fdist$ (as shown in Alg~\ref{alg:graphlet-est}).
For instance, edge neighborhoods may be sampled uniformly at random or according to an arbitrary weight/property such as k-core numbers, degrees, or any attribute of interest.
Further, an edge neighborhood may be selected with replacement or without.
Selecting an edge neighborhood with replacement allows each edge neighborhood $\N(\e)$ to be used multiple times, whereas sampling without replacement ensures that each edge neighborhood included in the sample is unique (by label) and never repeated.
We have experimented with both on a few networks and there was no significant difference (using a fixed sampling probability $\p$ and number of trials $\trials$). 
Edge-centric graphlet decomposition algorithms also lend themselves for (parallel) implementation on both shared-memory and distributed-memory architectures (see Section~\ref{sec:parallel}).

{
\algblockdefx[parallel]{parfor}{endpar}[1][]{$\textbf{parallel for}$ #1 $\textbf{do}$}{$\textbf{end parallel}$}
\algrenewcommand{\alglinenumber}[1]{\fontsize{6.5}{7}\selectfont#1}
\begin{figure}[t!]
\vspace{-2.mm}
\begin{center}
\begin{minipage}{1.0\linewidth}
\begin{algorithm}[H]
\caption{\,
\small
Family of edge-centric parallel \emph{localized graphlet estimation} ($\lge$) algorithms
}
\label{alg:pgd}
\begin{spacing}{1.3}
\fontsize{8}{9}\selectfont
\begin{algorithmic}[1]
\vspace{-1.6mm}
\Procedure {LocalizedGraphletEst}
{$G$, $\Es$}
\parfor[{\bf each} $\e=(v,u) \in \Es$ in order] \label{algline:graphlet-for}
	\State Reset $\tri_e = \emptyset$ and $\S_u = \emptyset$ \label{algline:graphlet-init}
	\For {$w \in \N(v)$} \label{algline:graphlet-create-hash}
		\If{$w \not= u$} $\hash(w) = \lambda_1$ \label{algline:graphlet-mark-neigh-of-v}
		\EndIf
	\EndFor
	\For {$w \in \N(u)$} \label{algline:triangles-and-wedges}
		\If{$w = v$} \textbf{continue} \EndIf \label{algline:graphlet-tri-w=v}
		\If{$\hash(w) = \lambda_1$} \label{algline:triangle} 
			\State $\tri_e \leftarrow \tri_e \cup \{w\}$ and set $\hash(w)=\lambda_3$ \label{algline:triangle-found} \Comment{triangle} 
		\Else \, $\S_u \leftarrow \S_u \cup \{w\}$ and set $\hash(w)=\lambda_2$ \label{algline:wedge-found} \Comment{wedge} 
		\EndIf
	\EndFor
	\multiline{Update unrestricted connected counts via Eq.~\ref{eq:non-induced-conn-n1}--\ref{eq:non-induced-conn-n4} and 
		\text{unrestricted \;\;\;} disconnected counts via Eq.~\ref{eq:non-induced-disconn-n1}--\ref{eq:non-induced-disconn-n4}} \label{algline:pgd-est-update-non-induced-counts}
		
	\State $\C_{3}\,\,\pluseq\,|\tri_e|$ \label{algline:graphlet-C-tri} \Comment{Note $\pluseq$ is the addition sum $\C_{3} = \C_{3} + |\tri_e|$}
	\State $\C_{4}\,\,\pluseq\,\C_{4}(\e) = (\d_u + \d_v -2) - 2|\tri_e|$ \label{algline:graphlet-C-2star} \Comment{equiv. $|S_u|+|S_u|$}
	\State $\C_{5}\,\,\pluseq\,\C_{5}(\e) = n - \C_{4}(\e) + |\tri_e| - 2$ \label{algline:graphlet-C-3-node-1-edge} 
	\State $\C_{7}\,\,\pluseq\,\C_{7}(\e) =$ \textsc{Clique}$(\hash,\tri_e)$ \Comment{$\mathsf{in \; parallel}$} \label{algline:pgd-clique}
	\State $\C_{10}\pluseq\C_{10}(\e) =$ \textsc{Cycle}$(\hash,\S_u)$ \Comment{$\mathsf{in \; parallel}$} \label{algline:pgd-cycle}
\endpar
\State Compute estimated graphlet counts $\bX$ via Eq.~\ref{eq:graphlet-X-start}-\ref{eq:graphlet-X-end} \label{algline:graphlet-est}
\State {\bf return} $\bX$, where $\X_i$ is the estimate for graphlet $\g_i$
\EndProcedure
\smallskip
\end{algorithmic}
\end{spacing}
\vspace{-1.mm}
\end{algorithm}
\end{minipage}
\end{center}
\vspace{-4.mm}
\end{figure}
}

Given the sampled set of edge-centric neighborhood, we show how to compute the estimated graphlet counts in Alg~\ref{alg:pgd}. 
More formally, let $\tri_{e} = \N(u) \cap \N(v)$ be the set of nodes that complete triangles with $\e(v,u) \in \Es$.
Likewise, $\S_u = \{ w \in \N(u) \setminus \{v\} | w \notin \N(v) \}$ and $\S_v = \{ w \in \N(v) \setminus \{u\} | w \notin \N(u) \}$, and thus $|\S_v|$ and $|\S_u|$ are the number of 2-stars centered at $v$ and $u$, respectively.
Note that $\S_u \cap \S_v = \emptyset$ by construction and 
\[
\S_u \cup \S_v = \{\underbrace{w_1,\ldots,w_{i}}_{\S_u}, \underbrace{w_{i+1},\ldots,w_{n}}_{\S_v}\}.
\] 
Thus, $|\S_u \cup \S_v| = |\S_u|+|\S_v|$.
These quantities are computed in Line~\ref{algline:triangles-and-wedges}-\ref{algline:wedge-found} of Alg~\ref{alg:pgd}.
For further intuition, see Figure~\ref{fig:graphlets-intuition}.
Let us also note that $\hash(\cdot)$ is a hash table for checking edge existence in $o(1)$ time (see Alg~\ref{alg:pgd}).
As an aside, this is an implementation detail and $\hash(\cdot)$ can easily be replaced with another data structure (bloom filters, etc) or even removed entirely in favor of binary search (which may be favorable in situations where memory is limited). These possibilities are discussed in detail later.
Notice that $\hash(\cdot)$ is also used as a way to encode the different types of nodes.
Thus, nodes are hashed using $\lambda_1$, $\lambda_2$, and $\lambda_3$, which may be defined as any unique symbol.
In our implementation, we avoid the cost of resetting by ensuring that each $\lambda_i$ is unique for each edge-centric neighborhood.
In Alg~\ref{alg:pgd} Line~\ref{algline:graphlet-mark-neigh-of-v}, we mark the neighbors $\N(v)$ of $v$ as $\lambda_1$.
Later in Line~\ref{algline:triangle-found} a triangle is marked with $\lambda_3$, whereas Line~\ref{algline:wedge-found} encodes a wedge as $\lambda_2$.

Moreover, given that all count variables are initialized to zero, Alg~\ref{alg:pgd} maintains the \emph{unrestricted connected graphlet counts}\footnote{Note $\pluseq$ is the addition assignment operator.} (Eq.~\ref{eq:non-induced-conn-n1}-\ref{eq:non-induced-conn-n4}):
{
\vspace{-3mm}
\begin{align}
\C_{8} 
&\pluseq {|\tri_e| \choose 2}  \label{eq:non-induced-conn-n1}
\quad\quad\quad\quad\quad\quad
\quad\quad\quad
\\
\C_{9} & \pluseq |\tri_e| \cdot \Big(|\S_u|+|\S_v|\Big) \label{eq:non-induced-conn-n2} \\
\C_{11} & \pluseq {|\S_u| \choose 2} + {|\S_v| \choose 2}  \label{eq:non-induced-conn-n3} \\
\C_{12} & \pluseq |\S_u| \cdot |\S_v| \label{eq:non-induced-conn-n4}
\end{align}}
\noindent where $\C_{8}$, $\C_{9}$, $\C_{11}$, and $\C_{12}$ are later used for computing chordal-cycles, tailed-triangles, 3-stars, and 4-paths in constant time, respectively. 
For clarity, $\C_i$ represents the unrestricted counts for graphlet $\g_i \in \G$\footnote{Recall the graphlet notation summarized in Table~\ref{table:graphlet_notation})}.
Let us note that $\C_{7}$ and $\C_{10}$ (4-cliques and 4-cycles, respectively) are computed in Alg~\ref{alg:pgd} Line~\ref{algline:pgd-clique}-\ref{algline:pgd-cycle}.
Further, $\C_{3}$, $\C_{4}$, and $\C_{5}$ are computed in Line~\ref{algline:graphlet-C-tri}-\ref{algline:graphlet-C-3-node-1-edge} and represent the ($k=3$)-graphlets of triangles, 2-stars, and 3-node-1-edge, respectively.
Note that we refer to the unrestricted counts as the combinatorial counts that can be computed in constant time and using only the knowledge obtained from the quantities discussed above (\ie, triangle and $2$-star counts in the edge-centric neighborhood of an edge $\e$).

Similarly, we compute the \emph{unrestricted disconnected counts} (Eq.~\ref{eq:non-induced-disconn-n1}-\ref{eq:non-induced-disconn-n4}): 
{\begin{align}
\C_{13} 
&\pluseq \, |\S_u| \cdot  (\n - |\N(u) \cup \N(v)|) \,+  \label{eq:non-induced-disconn-n1} \\ \nonumber
& \quad\quad |\S_v| \cdot (\n - |\N(u) \cup \N(v)|) \\
\C_{14}  
&\pluseq \, |\tri_e| (\n - |\N(u) \cup \N(v)|) \label{eq:non-induced-disconn-n2} \\
\C_{15}  
&\pluseq \, {{\, \n - |\N(u) \cup \N(v)|} \choose 2 \,} \label{eq:non-induced-disconn-n3} \\
\C_{16}  
&\pluseq \, \m - |\N(u) \setminus \{v\}| - |\N(v) \setminus \{u\}| - 1 \label{eq:non-induced-disconn-n4} 
\end{align}
}
\noindent
In addition, we maintain $\C_3$, $\C_4$, $\C_5$, $\C_7$, and $\C_{10}$ (see Alg~\ref{alg:pgd}).
Note that the $(k\text{--}1)$-graphlets are used to compute the k-clique/k-cycle counts directly.
These quantities are computed for each edge-centric neighborhood in the sample, and then used for estimation.
In particular, the $3$-vertex graphlet counts are estimated from their counts via Eq.~\ref{eq:graphlet-X-start}-\ref{eq:graphlet-X3-end} as follows:
\begin{align}
\X_{3} &=  W_{3} \pinv_{3} \C_{3} \label{eq:graphlet-X-start} \\
\X_{4} &= W_{4} \pinv_{4}  \C_{4} \\ 
\X_{5} &= W_{5} \pinv_{5}  \C_{5} \\ 
\X_{6} &=  W_{6} \cdot \Bigg[ {n \choose 3} - \X_{3}-\X_{4}-\X_{5} \Bigg] \label{eq:graphlet-X3-end} 
\end{align}
\noindent\\
where $X_{3}, X_{4}, X_{5}, X_{6}$ are the estimated counts of the graphlets $\g_3, \g_4, \g_5, \g_6$ respectively, and $W$, $\pinv$ are the weights used to fix the sampling bias.

Similarly, the $4$-vertex \emph{connected graphlet} counts are estimated via Eq~\ref{eq:graphlet-X4-conn-start}-\ref{eq:graphlet-X4-conn-end} as follows:
\begin{align}
\X_{7} &= W_{7} \pinv_{7} \C_{7} 
\quad\quad \quad\quad \quad\quad \quad\quad 
\quad\quad \quad\quad 
\label{eq:graphlet-X4-conn-start} \\	
\X_{8} &= W_{8} \pinv_{8} (\C_{8} - \C_{7}) 		\\	
\X_{9} &=W_{9} (\pinv_{9} \C_{9} - 4\X_{8}) 		\\	
\X_{10} &= W_{10} \pinv_{10} \C_{10} 		\\ 
\X_{11} &= W_{11} (\pinv_{11}\C_{11} - \X_{9}) 		\\ 
\X_{12} &= W_{12} \pinv_{12} (\C_{12} - \C_{10}) 	\label{eq:graphlet-X4-conn-end}	 
\end{align}
\noindent
and the $4$-vertex \emph{disconnected graphlet} counts are estimated via Eq~\ref{eq:graphlet-X4-disconn-start}-\ref{eq:graphlet-X-end} as follows:
\begin{align}
\X_{13} &= W_{13}  (\pinv_{13} \C_{13} - \X_{9})  	\label{eq:graphlet-X4-disconn-start} \\	
\X_{14} &= W_{14}  (\pinv_{14} \C_{14} - 2\X_{12}) 	\\	
\X_{15} &= W_{15} (\pinv_{15} \C_{15} - 6\X_{7} - 4\X_{8} - 2\X_{9} -  \\ \nonumber
& \quad\;\;  4\X_{10} - 2\X_{12}) 		\\ 
\X_{16} &= W_{16} ( \pinv_{16} \C_{16} - 2\X_{15}) 	\\	
\X_{17} &= W_{17} \cdot \Bigg[ {n \choose 4} - \sum_{i=7}^{16} \X_{i} \Bigg] 	 \label{eq:graphlet-X-end} 
\end{align}
\noindent
where $X_{7}$--$X_{17}$ are the estimated counts of the graphlets $G_{7}$--$G_{17}$ respectively.
Further, $\mW \in \RR^{\kappa}$ is a vector of weights defined as:
\begin{equation}
\mW = \big[ \begin{smallmatrix} 
	1 					
& 	1 					
& 	\frac{1}{3} 		
& 	\frac{1}{2} 		
& 	1  					
& 	1  					
& 	\frac{1}{6}  		
& 	1  					
& 	\frac{1}{2}  		
& 	\frac{1}{4}  		
& 	\frac{1}{3}  		
& 	1  					
& 	\frac{1}{2}  		
& 	1  					
& 	\frac{1}{2}  		
& 	\frac{1}{3}  		
& 	1  					
\end{smallmatrix}\bigl]^\T
\label{eq:graphlet-W}
\end{equation}
\noindent
where each $W_i$ is a scalar that aims to correct the bias for the induced subgraph $\g_i$ (See Table~\ref{table:graphlet_notation} to determine the corresponding induced subgraph for each $\g_i \in \G$).
However, $\mW$ can be adapted to account for other known biases.
Further, $\vp \in \RR^{\kappa}$ is a vector of sampling probabilities for all graphlets where $\p_i$ is the sampling probability of graphlet $\g_i \in \G$.
Note that $\p_i$ can be proportional to any arbitrary function/weight computed on the graph $G$.
One possibility is to use uniform sampling probabilities such that each $\p_i$ is:
\[
\p_i = |\Es| / |\E|
\]
\noindent
where $\p_i$ is the fraction of edge neighborhoods selected thus far.
Results for both uniform and non-uniform sampling probabilities are discussed and investigated in Section~\ref{sec:exp}.
In addition, let $\pinv_i$ be defined as:
\[ \pinv_i = \frac{1}{\,\,\p_i\,\,} \]
\noindent
where $\pinv_i$ is the inverse sampling probability of graphlet $i$ used to correct the sampling bias.

{
\algblockdefx[parallel]{ParFor}{EndPar}[1][]{$\textbf{parallel for}$ #1 $\textbf{do}$}{$\textbf{end parallel}$}
\algrenewcommand{\alglinenumber}[1]{\fontsize{6.5}{7}\selectfont#1}
\algtext*{EndPar}
\begin{figure}[t!]
\vspace{-2.mm}
\begin{center}
\begin{minipage}{1.0\linewidth}
\begin{algorithm}[H]
\caption{\,{Clique counts via neigh-iter. 
}}
\label{alg:cliques}
\begin{spacing}{1.2}
\fontsize{8}{9}\selectfont
\begin{algorithmic}[1]
\vspace{-1.6mm}
\Procedure {Clique}{$\hash,\tri_e$} 
\State Set $\clique_e \leftarrow 0$ \label{algline:clique-init}
\ParFor[{\bf each} $w \in \tri_e$] \label{algline:clique-outer-parfor-T}
	\For{{\bf each} $r \in \N(w)$ {\bf where} $\hash(r) = \lambda_3$} \label{algline:clique-inner-for-neigh}
			set $\clique_e \leftarrow \clique_e + 1$ \label{algline:clique-detected}
	\EndFor
	\State Reset $\hash(w)$ to $0$  \label{algline:clique-reset-hash}	
\EndPar
\State \textbf{return} $\clique_e$ \label{algline:clique-return}
\EndProcedure
\smallskip
\end{algorithmic}
\end{spacing}
\vspace{-1.mm}
\end{algorithm}
\vspace{-3.1mm}
\vspace{-4.2mm}
\begin{algorithm}[H]
\caption{\,{Cycle counts via neigh-iter. 
}}
\label{alg:cycles}
\begin{spacing}{1.2}
\fontsize{8}{9}\selectfont
\begin{algorithmic}[1]
\vspace{-1.6mm}
\Procedure {Cycle}{$\hash, \S_u$} 
\State Set $\cycle_e \leftarrow 0$ \label{algline:cycle-init}
\ParFor[{\bf each} $w \in \S_u$] \label{algline:cycle-outer-parfor-Su}
	\For{{\bf each} $r \in \N(w)$ {\bf where} $\hash(r) = \lambda_2$} 	
		set $\cycle_e \leftarrow \cycle_e + 1$ \label{algline:cycle-detected}	
	\EndFor
	\State Reset $\hash(w)$ to $0$ \label{algline:cycle-reset-hash}
\EndPar
\State \textbf{return} $\cycle_e$  \label{algline:cycle-return}
\EndProcedure
\smallskip
\end{algorithmic}
\end{spacing}
\vspace{-1.mm}
\end{algorithm}
\end{minipage}
\end{center}
\vspace{-4.mm}
\end{figure}
}

Let us note that in Alg~\ref{alg:pgd}, the cliques and cycles are computed via Alg.~\ref{alg:cliques} and Alg.~\ref{alg:cycles} using information from the $(k\text{--}1)$-graphlets compute them directly.
However, in situations where memory is limited, then Alg.~\ref{alg:cliques-res} and Alg.~\ref{alg:cycles-res} should be used.
These methods search over the sets $\tri_{e}$, $\S_u$, $\S_v$ from the $(k\text{--}1)$-graphlets directly using binary search.
See Section~\ref{sec:complexity-edge-centric-graphlet-alg} for further details.

\subsection{Error Analysis} \label{sec:error-analysis}
\noindent
Let $\Y_i(\e)$ be the total count of an arbitrary induced subgraph $\g_i \in \G$ iff the subgraph is incident to $\e$, then $\Y_i = \sum_{\e \in \E} \Y_i(e)$.
Assume we sample a set of edge neighborhoods with probability $\prob$, then $\X = \sum_{\e \in \Es} \frac{\Y_i(e)}{\prob}$.
$\Exp[\X_i] = \Y$ is an unbiased estimate.
The proof is as follows.
\begin{align}
\Exp\bigl[\X_i\bigr] &= \Exp\Biggl[ \sum_{\e \in \Es} \frac{\X_i(\e)}{\prob} \Biggr] \, = \, 
\sum_{\e \in \Es} \Exp\Biggl[ \frac{\X_i(\e)}{\prob} \Biggr]  \\
\, &= \, \sum_{\e \in \E} \frac{\Exp\bigl[ \I_{\e} \bigr]}{\prob} \cdot \X_{i}(\e) 
\; = \; \sum_{\e \in \E} \frac{ \X_i(\e)}{\prob} \cdot \prob \; = \; \Y 
\end{align}
\noindent
\noindent
since $\I_{\e}$ is a Bernoulli r.v. that indicates whether $\e$ and its neighborhood is sampled.
Further, 
the mean squared error \textsc{mse}($\Xs_i$) is:
\begin{equation}
\Exp[(\Xs_i - \Y_i)^2] = \underbrace{\Var[\Xs_i]}_{\text{Variance}} + \underbrace{ (\Exp[\Xs_i] - \Y_i)^2 }_{\text{Bias}}
\end{equation}
\noindent
where $\Var[\Xs_i]$ is the variance component and $(\Exp[\Xs_i] - \Y_i)^2$ is the bias component of the estimator $\Xs_i$.
Therefore, $\text{MSE}(\Xs_i)=\Var[\Xs_i]$ since $\Xs_i$ is an \emph{unbiased estimator}. 

\subsection{Complexity} \label{sec:complexity-edge-centric-graphlet-alg}
\noindent 
Let $T_{\max}$ and $S_{\max}$ denote the maximum number of triangles and stars incident to a selected edge $\e \in \Es$.
Note that $S_{\max}$ in reality is significantly smaller since for each edge $\e=(v,u) \in \Es$, Alg.~\ref{alg:pgd} computes only $\S_u$\footnote{As opposed to both $\S_u$ and $\S_v$} such that $\d_u \leq \d_v$, and thus $|\S_u| \leq |\S_v|$.
For a single $\N(\e)$,
Alg~\ref{alg:pgd} counts $4$-cliques and $4$-cycles centered at $\e$ in $\O( \dmax T_{\max})$ and $\O(\dmax  S_{\max})$, respectively.
From either $4$-cliques/cycles, we derive all other graphlet counts in $o(1)$ using combinatorial relationships along with the $(k$--$1)$-graphlets.
Thus, Alg~\ref{alg:pgd} counts all graphlets for $\{\N(\e_1),...,\N(\e_K)\}$ up to $k=4$ in:
\begin{align} \label{eq:computational-complexity-pgd} \nonumber
\O\Bigr( \K \dmax T_{\max} + \K \dmax S_{\max} \Bigl) 
\; = \; \O\Bigr( \K \dmax \bigr( T_{\max} + S_{\max} \bigl) \Bigl)
\end{align}
\noindent
Using $K$ processing units (cores, workers), this reduces to $\O(\Delta (\tri_{\rm \max}+\S_{\max}) )$.

Space-efficient algorithms are crucial when dealing with such massive networks.
Thus, our method is designed to be space-efficient, and as we shall see requires significantly less space than existing approaches~\cite{orca,guise,graft,rage,fanmod}.
Space complexity of Alg~\ref{alg:pgd} is $\O(\n + 2\Delta - 1)=\O(\n)$ using a hash table $\Phi$ of size $n=|V|$.
However, this can be reduced to $\O(3\Delta - 1) = \O(\Delta)$ using binary search over $\tri$ or $\S_u$ and $\S_v$ directly, e.g., see Alg.~\ref{alg:cliques-res} and Alg.~\ref{alg:cycles-res}.
Note that Alg~\ref{alg:cliques}--\ref{alg:cycles} assumes a fast hash table-like data structure to efficiently check the existence of a neighbor.

{
\algrenewcommand{\alglinenumber}[1]{\fontsize{6.5}{7}\selectfont#1}
\begin{figure}[t!]
\vspace{-3.mm}
\begin{center}
\begin{minipage}{1.0\linewidth}
\begin{algorithm}[H]
\caption{\,{\small Clique counts restricted to searching $\tri_e$
}}
\label{alg:cliques-res}
\begin{spacing}{1.15}
\fontsize{7}{8}\selectfont
\begin{algorithmic}[1]
\vspace{-1.2mm}
\Procedure {CliqueRes}{$\hash$, $\tri_e$
}
\State Set $\clique_e \leftarrow 0$
\parfor[{\bf each} vertex $w_i$ in an ordering $w_1, w_2, \cdots$ of $\tri_e$] \label{algline:clique-res-wi}
	\ForAll{{\bf each} $w_j \in \{w_{i+1}, ..., w_{|\tri_e|} \}$ \textsf{in order}} \label{algline:clique-res-wj}
		\If{$w_i \in \N(w_j)$ via $\mathsf{binary}$ $\mathsf{search}$} \label{algline:clique-res-bsearch}
			$\clique_e \leftarrow \clique_e + 1$ \Comment{$4$-clique}
		\EndIf
	\EndFor
\endpar
\State \textbf{return} $\clique_e$
\EndProcedure
\smallskip
\end{algorithmic}
\end{spacing}
\end{algorithm}
\vspace{-3.25mm}
\algrenewcommand{\alglinenumber}[1]{\fontsize{6.5}{7}\selectfont#1}
\vspace{-4.mm}
\begin{algorithm}[H]
\caption{\,{\small Cycle counts restricted to $\S_u$ and $\S_v$
}}
\label{alg:cycles-res}
\begin{spacing}{1.15}
\fontsize{7}{8}\selectfont
\begin{algorithmic}[1]
\vspace{-1.2mm}
\Procedure {CycleRes}{$\hash$, $\S_u$, $\S_v$
}
\State Set $\cycle_e \leftarrow 0$
\parfor[{\bf each} $w \in \S_u$]		\label{algline:cycle-res-Su}
	\ForAll {$r \in \S_v$} 	\label{algline:cycle-res-Sv}
		\If{$r \in \N(w)$ via $\mathsf{binary}$ $\mathsf{search}$} \label{algline:cycle-res-bsearch}
			$\cycle_e \leftarrow \cycle_e + 1$ \Comment{$4$-cycle}
		\EndIf
	\EndFor
\endpar
\State \textbf{return} $\cycle_e$ 
\EndProcedure
\smallskip
\end{algorithmic}
\end{spacing}
\end{algorithm}
\end{minipage}
\end{center}
\vspace{-3.5mm}
\end{figure}
}

\subsection{Discussion} \label{sec:framework-discussion}
\noindent
The family of localized graphlet estimation methods easily generalize to
graphlets of arbitrary size by replacing the definition of an edge-centric neighborhood with the more general and suitable notion of an edge $\ell$-neighborhood:
\begin{equation}\label{eq:l-neighborhood}
\N_{\ell}(v,u) = \Big\{ w\in V \setminus \{v,u\} \, |\, D(v,w)\leq \, \ell \, \vee D(u,w)\leq\ell \Big\} \nonumber
\end{equation}
\noindent 
where $\N_{\ell}(v,u)$ represents the set of vertices with distance less than or equal to $\ell$ from $e=(v,u) \in E$.
Thus, we set $\ell=1$ for graphlets of size $k \leq 4$, and $\ell=2$ for graphlets of size $k=5$, and so on.
Note that if the total number of edges is unknown (due to streaming, problem constraints, or other issues), 
then Alg~\ref{alg:graphlet-est} is easily adapted, e.g., one may simply specify the number of graphlets to sample (instead of the fraction of graphlets to sample denoted by $\prob$ in Alg.~\ref{alg:graphlet-est}).
Unlike existing work, the proposed $\lge$ methods are naturally amenable to streaming graphs, and processing (for graphs to large to fit into memory).
For instance, we do not need to read the entire graph into memory, as long as there is an efficient way to obtain the $\ell$-neighborhood subgraph $\Ne(\e_i)$ required for estimation.

In the interest of space and to keep the presentation simple, we have left out several details on performance enhancement that we have in our implementation. 
To give a small example, we use an adjacency matrix structure for small graphs in order to facilitate $o(1)$ edge checks.
For larger graphs, we efficiently encode the neighbors of the top-k vertices with largest degree (and relabel to save space/time) for $o(1)$ graph ops.
We use a fast $O(d)$ neighborhood set intersection procedure, 
dynamically select local search procedures over $\tri_{e}$, $\S_u$, 
and have many other optimization's throughout the code (bit-vector graph representation, etc.).

{
\algrenewcommand{\alglinenumber}[1]{\fontsize{6.5}{7}\selectfont#1}
\algblockdefx[parallel]{parfor}{endpar}[1][]{$\textbf{parallel for}$ #1 $\textbf{do}$}{}
\algnotext{endpar}
\begin{figure}[b!]
\vspace{-4.mm}
\begin{center}
\begin{algorithm}[H]
\caption{\,{
\small
Micro-level graphlet estimation framework
}}
\label{alg:lge-local}
\vspace{-1mm}
\begin{spacing}{1.2}
\fontsize{7}{8}\selectfont
\algrenewcommand{\alglinenumber}[1]{\fontsize{6}{7}\selectfont#1}
\begin{algorithmic}[1]
\Procedure{MicroGraphletEstimation}{$\Ne(\e_k) \text{ or } G$, $\e_k$, $\pr_{\e}$}

\State Initialize variables
\parfor[{\bf each} $w \in \N(v)$] \label{algline:local-lge-graphlet-create-hash}
	\If{$w \not= u$} 
		$\S_v \leftarrow \S_v \cup \{w\}$ and $\hash(w) = \lambda_1$ \label{algline:local-lge-graphlet-mark-neigh-of-v}
	\EndIf
\endpar

\parfor[{\bf each} $w \in \N(u)$ {\bf and} $w \not= v$] \label{algline:local-lge-triangles-and-wedges}
	\If{$\hash(w) = \lambda_1$} \label{algline:local-lge-triangle} 
		\State \, $\tri_e \leftarrow \tri_e \cup \{w\}$ and set $\hash(w)=\lambda_3$ \label{algline:local-lge-triangle-found} \Comment{triangle} 
		\State \, $\S_v \leftarrow \S_v \setminus \{w\}$ 
	\Else \, $\S_u \leftarrow \S_u \cup \{w\}$ and set $\hash(w)=\lambda_2$ \label{algline:local-lge-wedge-found} \Comment{wedge} 
	\EndIf
\endpar

\State $x_3 = |\tri_e|$	\Comment{triangles/3-cliques} \label{algline:local-lge-3-graphlets-start}
\State $x_4 = \big( \d_u + \d_v - 2 \big) - 2|\tri_e|$ \Comment{2-stars}
\State $x_5 =  n - (|S_v| + |S_u| + |\tri_e| - 2)$ \Comment{3-node-1-edge}
\State $x_6 = \mychoose{\n}{3} - x_3 - x_4 - x_5$ \Comment{3-node-indep.} \label{algline:local-lge-3-graphlets-end}

\ParFor[{\bf each} $w \in \tri_e$] \label{algline:local-lge-clique-outer-parfor-T}
	\For{$j=1,...,\ceil{\d_w \cdot \pe_{\e}}$} \label{algline:local-lge-sample-T-for}
		\State Select a vertex $r \in \N(w)$ via an arbitrary distribution $\Fdist$ \label{algline:local-lge-F-dist-clique}
		\If{$\hash(r) = \lambda_3$} Set $x_7 \leftarrow x_7 + \big( \nicefrac{\d_w}{\ceil{\d_w \cdot \pe_{\e}}} \big)$ \Comment{$4$-clique} \label{algline:local-lge-clique-detected}
		\EndIf
	\EndFor
	
	\State Set $\hash(w)$ to $\lambda_4$  \label{algline:local-lge-clique-reset-hash} 
\EndPar
\State $x_8 = {|\tri_e| \choose 2} - x_7$ \Comment{chordal-cycles} \label{algline:local-lge-chordal-cycles}

\ParFor[{\bf each} $w \in \S_u$] \label{algline:local-lge-cycle-outer-parfor-Su}
	\For{$j=1,...,\ceil{\d_w \cdot \pe_{\e}}$} \label{algline:local-lge-sample-Su-for}
		\State Select a vertex $r \in \N(w)$ via an arbitrary distribution $\Fdist$ \label{algline:local-lge-F-dist-Su} 
		\If{$\hash(r) = \lambda_1$} set $x_{10} \leftarrow x_{10} + \big( \nicefrac{\d_w}{\ceil{\d_w \cdot \pe_{\e}}} \big)$ \Comment{$4$-cycle} \label{algline:local-lge-cycle-detected} 
		\EndIf
		\If{$\hash(r) = \lambda_2$} set $x_{9} \leftarrow x_{9} + \big( \nicefrac{\d_w}{\ceil{\d_w \cdot \pe_{\e}}} \big)$ \Comment{tailed-tri} \label{algline:local-lge-tailed-tri-est}
		\EndIf
		\If{$\hash(r) = \lambda_4$} set $\omega \leftarrow \omega + \big( \nicefrac{\d_w}{\ceil{\d_w \cdot \pe_{\e}}} \big)$ \label{algline:local-lge-ne-est} 
		
		\EndIf
	\EndFor
	\State Set $\hash(w)$ to $0$ \label{algline:local-lge-Su-reset-hash}
\EndPar

\ParFor[{\bf each} $w \in \S_v$] \label{algline:local-lge-cycle-outer-parfor-Sv}

	\For{$j=1,...,\ceil{\d_w \cdot \pe_{\e}}$} \label{algline:local-lge-sample-Sv-for}
		\State Select a vertex $r \in \N(w)$ via an arbitrary distribution $\Fdist$ \label{algline:local-lge-F-dist-Sv} 
	
		\If{$\hash(r) = \lambda_1$} set $x_{9} \leftarrow x_{9} + \big( \nicefrac{\d_w}{\ceil{\d_w \cdot \pe_{\e}}} \big)$ \Comment{tailed-tri} 
		\EndIf
		\If{$\hash(r) = \lambda_4$} set $\omega \leftarrow \omega + \big( \nicefrac{\d_w}{\ceil{\d_w \cdot \pe_{\e}}} \big)$ 
		\EndIf
	\EndFor
	\State Set $\hash(w)$ to $0$ \label{algline:local-lge-Sv-reset-hash}
\EndPar

\State $x_{11} = {|\S_v| \choose 2} + {|\S_v| \choose 2} - x_{9}$ \Comment{3-stars}  \label{algline:local-lge-x11}
\State $x_{12} = (|S_v| \cdot |S_u|) - x_{10}$ \Comment{4-paths}  \label{algline:local-lge-x12}
\State $x_{13} = |\tri_e| \cdot \big[n - (|\tri_e|+|S_u|+|S_v|+2) \big]$ \Comment{4-node-1-tri}  \label{algline:local-lge-x13}
\State $x_{14} = (|\S_u|+|\S_v|) \cdot \big[n - (|\tri_e|+|S_u|+|S_v|+2) \big]$ \Comment{4-node-2-star}  \label{algline:local-lge-x14}
\State $x_{15} = m - (|\tri_e|+\d_u + \d_v +1) - \omega$ \Comment{4-node-2-edge} \label{algline:local-lge-4-node-2-edge}
\State $x_{16} = \mychoose{n - \big[ |\tri_e|+|S_u|+|S_v|+2 \big]}{2}$ \Comment{4-node-1-edge}  \label{algline:local-lge-x16}
\State $x_{17} = \mychoose{n}{4} - \sum_{i=7}^{16} x_i$ \Comment{4-node-indep.}  \label{algline:local-lge-x17}
\State {\bf return} $\vx$, where $x_i$ is the estimate of graphlet $\g_i$ for $\e_k$ \label{algline:local-lge-return-x}
\EndProcedure
\smallskip
\end{algorithmic}
\end{spacing}
\end{algorithm}
\end{center}
\vspace{-3.mm}
\end{figure}
}

\section{Estimating Micro Graphlet Counts}
\label{sec:local-graphlet-est}
\noindent
This section formulates the micro-level graphlet estimation problem, then derives a flexible computational framework.
The experiments in Section~\ref{sec:exp-local-graphlet-est} demonstrate the effectiveness of these methods.
Computing \emph{micro-level graphlet statistics} $\vx_{i}$ for an individual edge $\e_{i} \in E$ (or node) in $G$ (as opposed to the global graph $G$) is important with numerous potential applications.
For instance, they can be used as powerful discriminative features $\{\vx_1, \vx_2, \ldots, \vx_{\m}\}$ 
for improving statistical relational learning (SRL) tasks~\cite{rossi12jair} such as 
relational classification~\cite{getoor2007introduction},  
link prediction and weighting tasks (e.g., recommending items, friends, web sites, music, events, etc.)~\cite{liu2013social},
detecting anomalies in graphs (e.g., detecting fraud, or attacks/malicious behavior in computer networks)~\cite{noble2003graph,akoglu2014graph},
among many others~\cite{bhattacharya2006entity,schaeffer2007graph,rossi2014roles}.

{
\smallskip 
\noindent
\textbf{Problem.}\; ({\sc Micro-level Graphlet Estimation})  
Given a graph $G=(V,\E)$ and an edge $e_i=(v,u) \in \E$, 
the \emph{micro graphlet estimation problem} is to find 
\[ 
\vx_{i} = \bmat{x_1 & x_2 & x_3 & \cdots & x_6 & x_7& \cdots & x_{17}}^{\T}
\]
\noindent
where $\vx_{i}$ is an approximation of the exact micro-level graphlet statistics denoted by $\vy_i$ for edge $e_i$ such that $\Error{\vx_i}{\vy_i}$ is minimized (\ie, ${\vx}_i \approx \vy_i$) as well as the computational cost associated with the estimation.
}
Note that $\Error{\vx_i}{\vy_i}$ can be any loss function.
The aim of the \emph{micro graphlet estimation problem} is to compute a fast approximation of the graphlet statistics (such as counts) centered at an individual edge.
Instead of approximating all graphs up to size $k$, one may relax the above problem to estimate a single graphlet pattern $\g_k \in \G$ of interest (e.g., 4-cliques).

A generalized and flexible framework for the \emph{micro graphlet estimation problem} is given in Alg~\ref{alg:lge-local}.
In particular, Alg.~\ref{alg:lge-local} takes as input an edge $\e_i$, a graph $G$ or $\Ne(\e_i)$ (neighborhood subgraph of $\e_i$), a sampling probability $\pr_{\e}$, and it returns the graphlet feature vector $\vx_i \in \RR^{\kappa}$ for $\e_i \in \E$ where $\kappa=|\G|$.
This generalization gives rise to a highly flexible and expressive unifying framework and serves as a basis for investigating this novel graphlet estimation problem.
Moreover, the class of micro graphlet approximation methods have many attractive properties such as unbiasedness, consistency, among others.
The algorithm estimates micro graphlet properties including micro single-valued statistics and multi-valued distributions (for a given edge or set of edges).

Alg.~\ref{alg:lge-local} shows how to efficiently count all graphlets of size $k \in \{2,3,4\}$ for an edge $\e_i \in E$.
First, we compute $\tri_{\e}$, $\S_u$, and $\S_v$ in Lines~\ref{algline:local-lge-graphlet-create-hash}-\ref{algline:local-lge-wedge-found}. 
Afterwards, Lines~\ref{algline:local-lge-3-graphlets-start}-\ref{algline:local-lge-3-graphlets-end} compute all graphlets of size $k=3$ exactly. 
Next, we compute $4$-cliques in Lines~\ref{algline:local-lge-clique-outer-parfor-T}-\ref{algline:local-lge-clique-reset-hash}.
In particular, Line~\ref{algline:local-lge-clique-outer-parfor-T} searches each vertex $w \in \tri_{\e}$ in parallel.
Given $w \in \tri_{\e}$, we select a neighbor $r \in \N(w)$ with probability $\pr_{\e}$ accordingly to an arbitrary weighted/uniform distribution $\Fdist$.
Then, we check if $r$ is of type $\lambda_{3}$ (from Line~\ref{algline:local-lge-triangle-found}), as this indicates that $r$ also participates in a triangle with $\e=(v,u)$, and since $r \in \N(w)$, then $\{v,u,w,r\}$ is a 4-clique.
Finally, Line~\ref{algline:local-lge-clique-reset-hash} ensures that the same 4-clique is not counted twice.
Chordal-cycles are derived in Line~\ref{algline:local-lge-chordal-cycles}.
Further, 4-cycles are computed in Lines~\ref{algline:local-lge-cycle-outer-parfor-Su}-\ref{algline:local-lge-Su-reset-hash} as well as a fraction of the tailed-triangles. 
The remaining tailed-triangles are computed in Lines~\ref{algline:local-lge-cycle-outer-parfor-Sv}-\ref{algline:local-lge-Sv-reset-hash}.
As an aside, $\omega$ is also computed (Lines~\ref{algline:local-lge-cycle-outer-parfor-Su}-\ref{algline:local-lge-Sv-reset-hash}) and used for estimating $\g_{15}$ (Line~\ref{algline:local-lge-4-node-2-edge}).
Finally, the remaining graphlets $\{x_{11}, \ldots, x_{17}\}$ are estimated in $o(1)$ time (Lines~\ref{algline:local-lge-x11}-\ref{algline:local-lge-x17}) using knowledge from the previous steps.
Notably, Alg.~\ref{alg:lge-local} gives rise to an efficient exact method, \eg, if $\pr_{\e}=1$ and selection is performed without replacement.

The computational complexity is summarized in Table~\ref{table:local-and-global-time-space-complexity}.
Note that just as before, we only need to compute a few graphlets and can directly obtain the others in constant time.
To compute all micro-level graphlet statistics for a given edge, it takes:
$
\O\Big( \dmax_{\rm ub} \big(|\S_u|+|\S_v|+|\tri_{\e}|\big) \Big)
$
\noindent
where $\dmax_{\rm ub}$ is the maximum degree from any vertex in $\S_v$, $\S_u$, and $\tri_{\e}$.
Alternatively, we can place an upper bound $\dmax_{\rm ub}$ on the number of neighbors searched from any vertex in $\S_v$, $\S_u$, and $\tri_{\e}$.
This can reduce the time quite significantly.
The intuition is that for vertices with large neighborhoods we only need to observe a relatively small (but representative) fraction of it to accurately extrapolate to the unobserved neighbors and their structure.

\begin{table}[b!]
\vspace{-1mm}
\centering
\scriptsize
\caption{Computational complexity}
\vspace{-2mm}
\label{table:local-and-global-time-space-complexity}
\begin{tabularx}{1.0\linewidth}{r ll H HHH}
\toprule
\textbf{Graphlet} & \textbf{Macro} & \textbf{Micro} & Graphlet & & & \\
\midrule
\textbf{4-clique} & $\O(\K \dmax \tri_{\max})$ & $\O(\dmax_{\rm ub} \cdot |\tri_{\e}| )$ & 4-clique & & \\
\textbf{4-cycle}  & \multirow{1}{*}{$\O(\K \dmax \S_{\max})$} & $\O(\dmax_{\rm ub} \cdot |\S_{u}| )$ & 4-cycle & & \\
\textbf{tailed-tri} & 	$\O(\K \dmax \S_{\max})$						& $\O\big( \dmax_{\rm ub} \cdot  (|\S_{u}|+|\S_{v}|) \big)$ & tailed-tri & & \\  
\textbf{all} & $\O\big( \K \dmax (\S_{\max}+\tri_{\max}) \big)$ & $\O\big( \dmax_{\rm ub} (|\S_{u}|+|\S_{v}|+|\tri_{\e}|) \big)$ & all & & \\ 
\bottomrule
\end{tabularx}
\vspace{-0mm}
\end{table}

\section{Adaptive Graphlet Estimation} 
\label{sec:est-opt}
In previous work, the user must specify the number (or proportion) of samples to use.
This is impractical in real-world settings, since the appropriate sample size is intrinsically tied to the required accuracy sufficient for a given problem or application.
Thus, we introduce an adaptive optimization scheme for graphlet estimation where the user can specify a bound on the accuracy and the technique automatically 
finds an approximation that is within the desired accuracy.
Thus, since the exact graphlet counts $\bY$ are unknown, we use the error between $\bX_{t-1}$ and $\bX_{t}$ as a proxy (See Alg~\ref{alg:graphlet-est-optimization}).
It is straightforward to see that $\Error{\bX_{t}}{\bX_{t-1}}$ decreases as $\K$ increases.
Therefore, $\D{\bX_{t-1}}{\bX_{t-2}} \geq \D{\bX_{t}}{\bX_{t-1}}$.
Intuitively, as $\K$ increases the variance $\bX_{t-1}$ and $\bX_{t}$ shrinks toward zero.

\subsubsection{Algorithmic Template} \label{sec:est-opt-template}
Alg~\ref{alg:graphlet-est-optimization} learns the appropriate sample size automatically given the desired error.
A parallel scheme to solve the graphlet optimization problem is also proposed (see Section~\ref{sec:parallel} for further details) and used in Section~\ref{sec:exp}.

\subsubsection{Objective Function} \label{sec:est-opt-obj}
The objective function aims to minimize an arbitrary loss (See Alg~\ref{alg:graphlet-est-optimization} Line~\ref{algline:for-each-graphlet-obj-func}--\ref{algline:est-opt-obj-func}).
For this, we use the maximum relative error.
\begin{equation} \label{eq:est-obj-max-rel-error}
\min_{\bX^{(t)}, \bX^{\opt}} \quad
\Biggl\{
\max_{\g_{i} \in \G^{(4)}} \quad
\frac{ |\X^{(t)}_{i} - \X_i^{\opt}| }{\X_i^{\opt}}
\Biggr\}
\end{equation}
\noindent
where $\X_i^{\star}$ is the best solution found thus far.
The inner part computes the maximum graphlet estimation error using relative error.
However, we also investigated KS-statistic, KL/Skew-divergence, and squared-loss.

{
\algrenewcommand{\alglinenumber}[1]{\fontsize{6.0}{7}\selectfont#1}
\begin{figure}[t!]
\vspace{-2mm}
\begin{center}
\begin{minipage}{1.0\linewidth}
\begin{algorithm}[H]
\caption{
\small
Adaptive graphlet estimation.
}
\label{alg:graphlet-est-optimization}
{
\begin{spacing}{1.2}
\fontsize{7}{8}\selectfont
\algrenewcommand{\alglinenumber}[1]{\fontsize{6.0}{6.5}\selectfont#1}
\begin{algorithmic}[1]
\vspace{-1mm}
\Require ~\newline
\quad\quad a graph $G = (V,\E)$ 
\newline
\quad\quad an arbitrary loss $\loss\rbr{\cdot}$ \Comment{for instance, max. relative error}
\newline
\quad\quad an error bound $\B$ such that $0 \leq \B \leq 1$
\newline
\quad\quad max number of iteration $t_{\max}$
\Ensure $\bX$, where $\X_i$ is the estimate for the graphlet $\g_i$
\smallskip
\State $\phi = \frac{1}{ \bigl( \nicefrac{1}{(\delta_{\text{err}}+\epsilon)} \bigr) \cdot \sqrt{m}}$ \Comment{Set $\phi$ if not specified by user}
\State Set $\Es\leftarrow \emptyset$,\quad $t \leftarrow 0$,\quad $\delta_{\rm err} \leftarrow 1$
\State Initialize $\bX$ uniformly at randomly
\While{$\delta_{\text{err}} - \epsilon > \B$ {\bf and} $t < t_{\max}$}

	\State $\K_t = \ceil{\phi \cdot (\m - |\Es|)}$ \Comment{Update sample size} \label{algline:num_samples_to_use_for_est}
	\State Set $\Es_t = \emptyset$
	\parfor[$\s = 1,2,...,\K_t$] \label{algline:est-opt-sample-edge-neighrborhoods}
		
		\State $\e \sim \text{UniformDiscrete} \cbr{1,2,...,\m}$ \label{algline:est-opt-sample-edge-neighborhoods}
		
		\If{{sampling without replacement}} \label{algline:est-opt-sampling-without-replacement}
			\While{$\hash(\e) > 0$} \Comment{edge has been sampled} 
				\State $\e \sim \text{UniformDiscrete} \cbr{1,2,...,\m}$ \label{algline:est-opt-sample-an-edge-uniformly}				
			\EndWhile
			\State Mark edge $\e$ in $\hash(\e)$

		\EndIf
		
		\State Set $\Es_{t} \leftarrow \Es_{t} \cup \{\e\}$ 
		\State Obtain $\bC_{}(\e)$ for $\e$ via Alg~\ref{alg:pgd}~Line~\ref{algline:graphlet-for}--\ref{algline:pgd-cycle}
		
		\State Set $\C_i^{(t)} \leftarrow \C_i^{(t)} + \C_{i}(\e)$, for all $i=1,2,...,|\G|$ \label{algline:est-opt-sum-uninduced-counts}
	\endpar
	\State Set $\C_i \leftarrow \C_i + \C_{i}^{(t)}, \;$ for all $i=1,2,...,|\G|$ {\bf in parallel}
	\State Obtain updated graphlet estimates $\bX^{(t)}$ using $\bC$ via Eq.~\ref{eq:graphlet-X-start}--\ref{eq:graphlet-X-end} 
	
	\State $\Es \leftarrow \Es \cup \Es_{t}$

	\parfor[$\g_i \in \G$] \label{algline:for-each-graphlet-obj-func}
		
		\State Compute loss $\sw_i \leftarrow \loss\rbr{ \X^{(t)}_{i} \; \| \; \X_{i} }$ \label{algline:est-opt-compute-loss}		
		\State Update $\delta_{\text{err}}$ via $\sw_i$ if required by obj. func. \label{algline:est-opt-obj-func}
	\endpar
	\State Set\; $\phi = \nicefrac{\phi}{2}$ \label{algline:est-opt-update-sampling-prob} \Comment{Update sampling probability}
	\State $\X \leftarrow \X^{(t)}$ \label{algline:est-opt-update-graphlet-est} \Comment{Update current graphlet estimates}
	\State $t \leftarrow t+1$ 
\EndWhile
\end{algorithmic}
\end{spacing}
\vspace{0.5mm}
}
\end{algorithm}
\end{minipage}
\end{center}
\vspace{-3mm}
\end{figure}
}

\begin{figure}[h!]
\vspace{-0mm}
\centering
\includegraphics[width=0.58\linewidth]{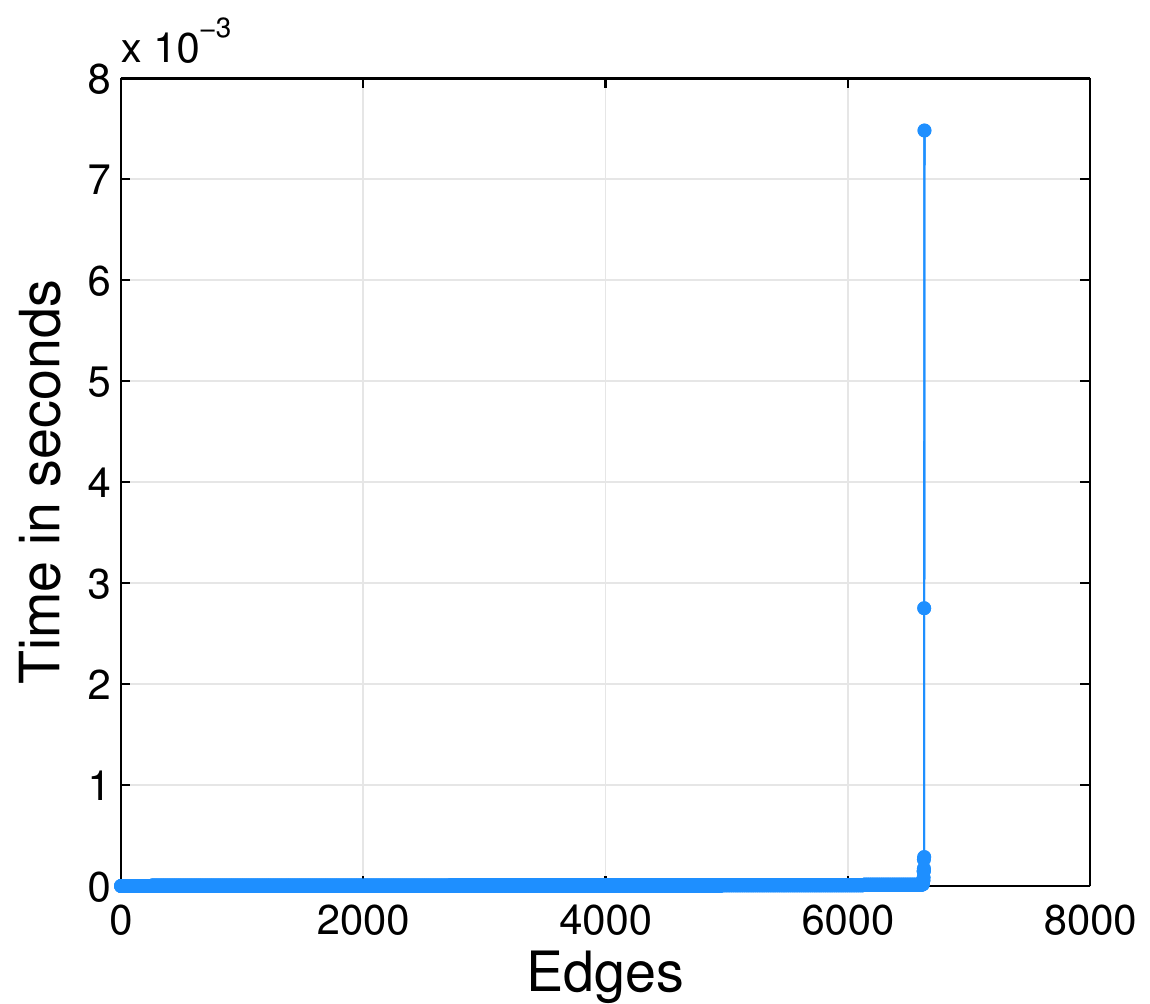}
\vspace{-1.5mm}
\caption{
Power-law relation is observed between the graphlet edge computation time.
The time taken to count $k=\{2,3,4\}$ graphlets for each edge in $\data{tech}{routers\text{-}rf}$ is shown above. 
See text for discussion.
}
\label{fig:graphlet-edge-powerlaw}
\vspace{-0mm}
\end{figure}

\subsubsection{Adaptive estimation}
\label{subsec:est-opt-adaptive-estimation}
Given a set $\Es$ (from the $t$-th iteration), 
the goal is to find the minimum set of edge neighborhoods such that $\Error{\Xs}{\Xs_{t}} \leq \beta$.
At each iteration, how many additional elements $\eta_t = |\Es_{t}| - |\Es_{t-1}|$ should be included in $\Es$?
There are two general approaches.
First, 
the set $\Es$ can be increased by a percent $\phi$ of the remaining edges at each iteration, that is,
$\eta_t = \ceil{ \phi \cdot |\E| - |\Es| }$
\noindent
where $\eta_{1} \geq \cdots \geq \eta_{t-1} \geq \eta_{t} \geq \cdots \geq \eta_{t_{\max}}$. 
Hence, the number of samples to increase $\Es$ by is monotonically decreasing with respect to the iteration $t$.
Clearly, as the sample size increases, the estimation variance decreases.
As a result of the above fact, the growth of $\Es$ is inversely related since the larger $\Es$ becomes, the less variance in estimation.
Thus, $\Es$ should grow at a rate that is inversely proportional to its size. 
Alternatively, $\Es$ may increase by a fixed number of samples at each iteration.

\subsubsection{Complexity} \label{subsec:est-opt-complexity}
The adaptive approach minimizes the relative error between the current best solution $\bX^{}$ and the previous best $\bX_{t-1}$.
Alg~\ref{alg:graphlet-est-optimization} finds an estimate for each $\g_i \in \G$ in: 
\begin{align} \label{eq:adaptive-complexity} 
&&	\O (\K \dmax  (\S_{\max} + T_{\max}) + |\G| t^{\opt} )
\, = \, \O
\bigr(\K \dmax \bigr( \S_{\max} + T_{\max} \bigl)  \bigl)
\nonumber
\end{align}
where $t^{\opt}$ is the number of iterations and $\K$ is the number of selected edge neighborhoods.

\section{Parallel algorithm} \label{sec:parallel}
\noindent
Estimation methods from the framework are parallelized via independent edge-centric graphlet computations over the selected set of edge-induced neighborhoods $\{ \N(e_i),..., \N(e_{\K})\}$.
The parallelization is described such that it could be used for both shared and distributed memory architectures\footnote{In the context of message-passing and distributed memory parallel computing, a node refers to another machine on the network (or bus) with its own set of memory, and multi-core CPUs, etc.}.
The parallel constructs used are a worker task-queue and a global broadcast channel.
Multi-threaded MPI is used for inter-machine communication.
We assume each machine $q$ has a queue and a copy of the graph\footnote{
For implementation on parallel computing architectures with limited memory, 
one only needs to transfer the set of edge-induced neighborhood subgraphs, which can be streamed if needed.} shared among the set of local workers (processing units). 
For macro-level graphlet statistics, the communication cost for a single worker is $O(|\G|)$.

\begin{figure}[h!]
\centering
\includegraphics[width=\linewidth]{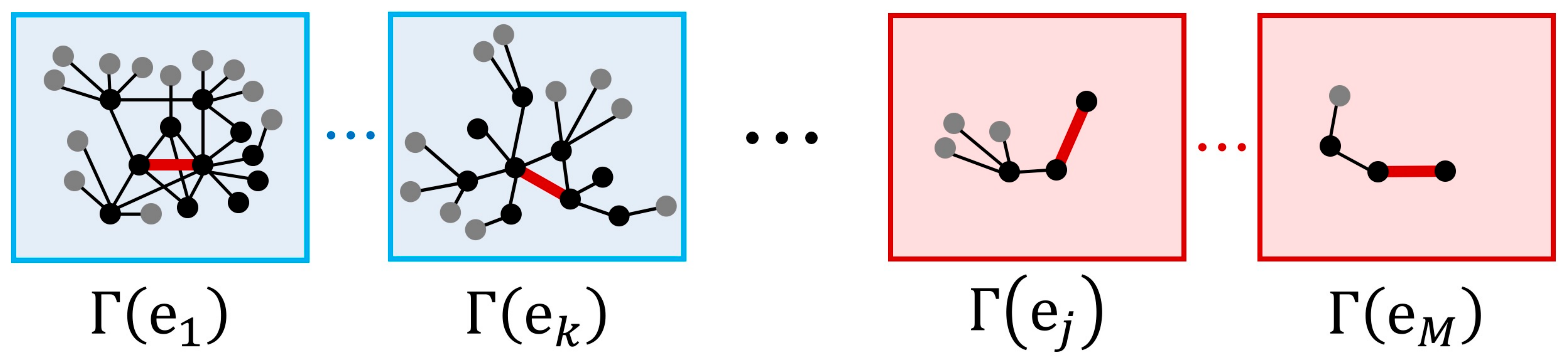}
\caption{Edge neighborhoods are ordered and dynamically partitioned to workers by ``hardness".}
\label{fig:parallel-neighborhood-ordering}
\end{figure}

The main parallel loop can be viewed as a task generator that farms the next $b$ edges out to a worker, which then computes the graphlets centered at each of the $b$ edge neighborhoods. 
Edge neighborhoods are dynamically partitioned to workers by ``hardness'' (Figure~\ref{fig:parallel-neighborhood-ordering}) where the most difficult edge neighborhood is assigned to the first worker, the second most difficult is assigned to the second worker, and so on. 
Furthermore, recall that a handful of edge neighborhoods require a lot of work, whereas the vast majority require only a small amount of work; as observed in Figure~\ref{fig:graphlet-edge-powerlaw}.
This ensures we avoid common problems present in other approaches such as the curse of the last reducer~\cite{suri2011counting}. 
However, notice that computing such a partitioning (Figure~\ref{fig:parallel-neighborhood-ordering}) is computationally intractable and thus we use edge degree (or volume) as an efficient proxy for ``hardness".

\begin{table}[t!]
\vspace{-0mm}
\caption{Estimates of expected value and relative error using $100$K samples.
The graphlet statistic for the full graph is shown in the first column. 
$\lb$ and $\ub$ are $95\perc$ lower and upper bounds, respectively.
Note M=million (mega), B=billion (giga), T=trillion (tera), P=quadrillion (peta).
}
\vspace{-1.0mm}
\label{table:est-rel-error-4-clique}
\centering
\small
\scriptsize
\begin{tabularx}{1.0\linewidth}{@{} c @{} r rrlHrr @{}}
\toprule
\TT \BB
& \textbf{graph}  &  $\Y$  &  $\X$  &  $\frac{|\Y - \X|}{\Y}$   &    &   $\lb$    &     $\ub$ \\
\midrule
\multirow{16}{*}{
\rotatebox[origin=c]{90}{
\mbox{
\includegraphics[scale=0.04]{graphlets/4-clique} \, {  \sc 4-clique }
}}}
& $\datasm{ca}{citeseer}$   &    18.7M    &    18.7M    &    0.0004    &   &    18.3M    &    19M \\ 
& $\datasm{ca}{dblp\text{-}2012}$   &    16.7M    &    16.7M    &    0.0004    &   &    16M    &    17.3M \\ 
& $\datasm{soc}{flickr}$   &    1.7B    &    1.7B    &    0.0003    &   &    1.7B    &    1.7B \\ 
& $\datasm{soc}{friendster}$   &    9B    &    9B    &    0.0038    &   &    8.9B    &    9.1B \\ 
& $\datasm{soc}{gowalla}$   &    6M    &    6M    &    0.0009    &   &    5.9M    &    6.2M \\ 
& $\datasm{soc}{orkut}$   &    3.2B    &    3.2B    &    0.0016    &   &    3.1B    &    3.3B \\ 
& $\datasm{soc}{pokec}$   &    42.9M    &    42.9M    &    0.0002    &   &    41.9M    &    43.9M \\ 
& $\datasm{socfb}{Berkeley13}$   &    26.6M    &    26.6M    &    0.0007    &   &    26.2M    &    27M \\ 
& $\datasm{socfb}{Indiana}$   &    60.1M    &    60.1M    &    0.0004    &   &    59.3M    &    61M \\ 
& $\datasm{socfb}{MIT}$   &    13.6M    &    13.6M    &    0.0004    &   &    13.5M    &    13.8M \\ 
& $\datasm{socfb}{OR}$   &    13.3M    &    13.3M    &    0.0005    &   &    13.1M    &    13.5M \\ 
& $\datasm{socfb}{Texas84}$   &    70.7M    &    70.7M    &    0.0002    &   &    69.6M    &    71.8M \\ 
& $\datasm{socfb}{UCLA}$   &    28.6M    &    28.6M    &    0.0005    &   &    28.2M    &    29M \\ 
& $\datasm{socfb}{UCSB37}$   &    18.1M    &    18.1M    &    $<$10$^{-4}$    &   &    17.9M    &    18.4M \\ 
& $\datasm{socfb}{UF}$   &    97.9M    &    97.9M    &    0.0001    &   &    96.5M    &    99.3M \\ 
& $\datasm{socfb}{UIllinois}$   &    64M    &    63.9M    &    0.0008    &   &    63M    &    64.9M \\ 
& $\datasm{socfb}{Wisconsin87}$   &    23M    &    23M    &    0.0011    &   &    22.7M    &    23.3M \\ 
& $\datasm{web}{wikipedia2009}$   &    1.4M    &    1.4M    &    0.0004    &   &    1.3M    &    1.5M \\ 
\midrule
\multirow{20}{*}{
\rotatebox[origin=c]{90}{
\mbox{
\rotatebox[origin=c]{270}{
\includegraphics[scale=0.04]{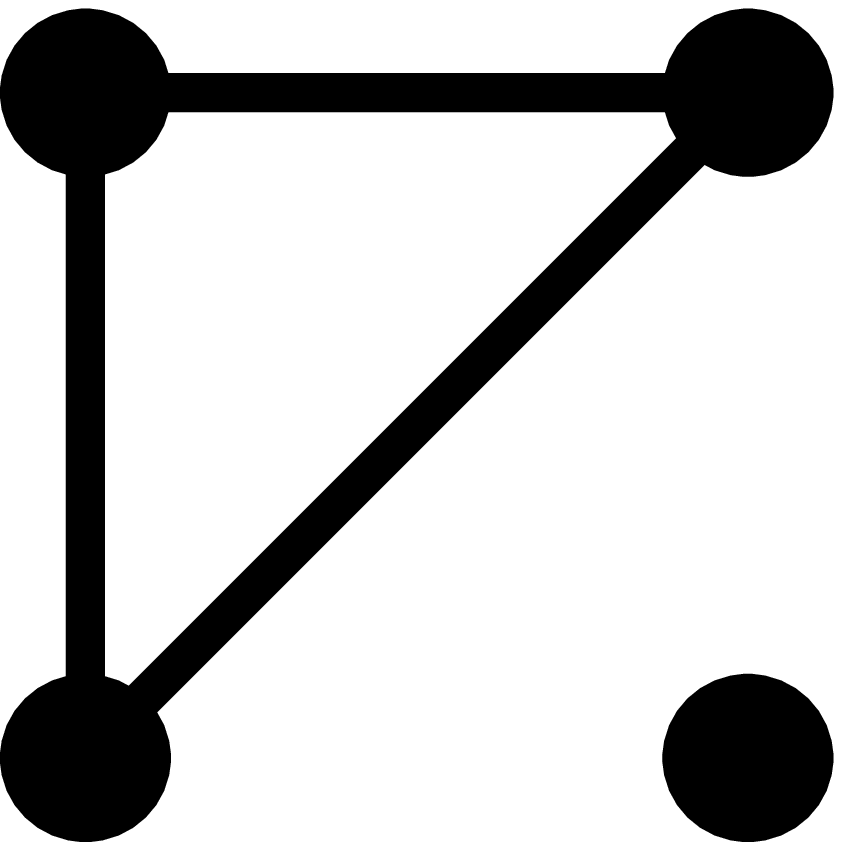}
}
\, {  \sc 4-node-1-tri }
}}}
& $\datasm{ca}{citeseer}$   &    616.6B    &    616.7B    &    0.0003    &   &    611.7B    &    621.8B \\ 
& $\datasm{ca}{dblp\text{-}2012}$   &    705.1B    &    705.1B    &    $<$10$^{-4}$    &   &    696.2B    &    714B \\ 
& $\datasm{soc}{flickr}$   &    30T    &    30T    &    0.0005    &   &    29.7T    &    30.4T \\ 
& $\datasm{soc}{friendster}$   &    273.8P    &    274.4P    &    0.0023    &   &    271.2P    &    277.6P \\ 
& $\datasm{soc}{gowalla}$   &    443.5B    &    443.7B    &    0.0004    &   &    438.1B    &    449.3B \\ 
& $\datasm{soc}{orkut}$   &    1.9P    &    1.9P    &    0.0012    &   &    1.9P    &    1.9P \\ 
& $\datasm{soc}{pokec}$   &    53.1T    &    53.1T    &    0.0003    &   &    52.6T    &    53.7T \\ 
& $\datasm{socfb}{Berkeley13}$   &    119.8B    &    119.8B    &    0.0002    &   &    119B    &    120.6B \\ 
& $\datasm{socfb}{Indiana}$   &    274.6B    &    274.6B    &    $<$10$^{-4}$    &   &    272.7B    &    276.5B \\ 
& $\datasm{socfb}{MIT}$   &    14B    &    14B    &    0.0002    &   &    13.9B    &    14.1B \\ 
& $\datasm{socfb}{OR}$   &    220.8B    &    220.8B    &    $<$10$^{-4}$    &   &    219.2B    &    222.5B \\ 
& $\datasm{socfb}{Texas84}$   &    397.7B    &    397.6B    &    0.0003    &   &    394.8B    &    400.4B \\ 
& $\datasm{socfb}{UCLA}$   &    102.3B    &    102.3B    &    0.0002    &   &    101.6B    &    103B \\ 
& $\datasm{socfb}{UCSB37}$   &    44.7B    &    44.7B    &    $<$10$^{-4}$    &   &    44.4B    &    45B \\ 
& $\datasm{socfb}{UF}$   &    418B    &    418B    &    $<$10$^{-4}$    &   &    415.2B    &    420.9B \\ 
& $\datasm{socfb}{UIllinois}$   &    283.3B    &    283.2B    &    0.0004    &   &    281.3B    &    285B \\ 
& $\datasm{socfb}{Wisconsin87}$   &    113.6B    &    113.6B    &    0.0005    &   &    112.9B    &    114.3B \\ 
& $\datasm{web}{wikipedia2009}$   &    4.1T    &    4.1T    &    $<$10$^{-4}$    &   &    4T    &    4.2T \\ 
\bottomrule
\end{tabularx}
\vspace{-0mm}
\end{table}

\providecommand{\figConfhspace}{-0mm}
\providecommand{\figConfSZ}{0.30\linewidth}
\providecommand{\figsBounds}{graphics/}

\begin{figure*}[t!]
\vspace{-0mm}
\centering

\hspace*{\figConfhspace}\subfigure[$\g_7$ (4-clique)]		{\includegraphics[width=\figConfSZ]{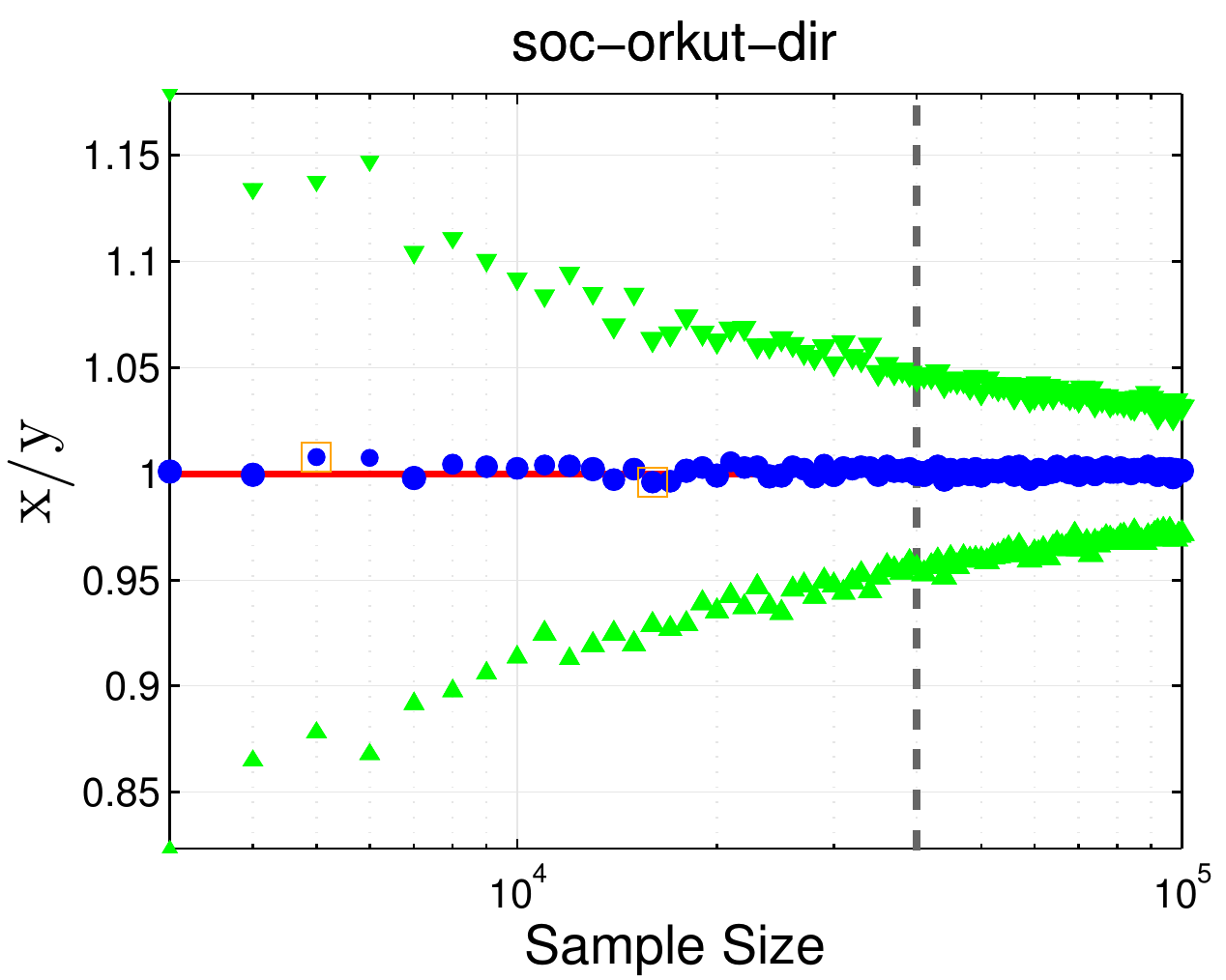}}
\hspace*{\figConfhspace}\subfigure[$\g_{7}$ (4-clique)] 		{\includegraphics[width=\figConfSZ]{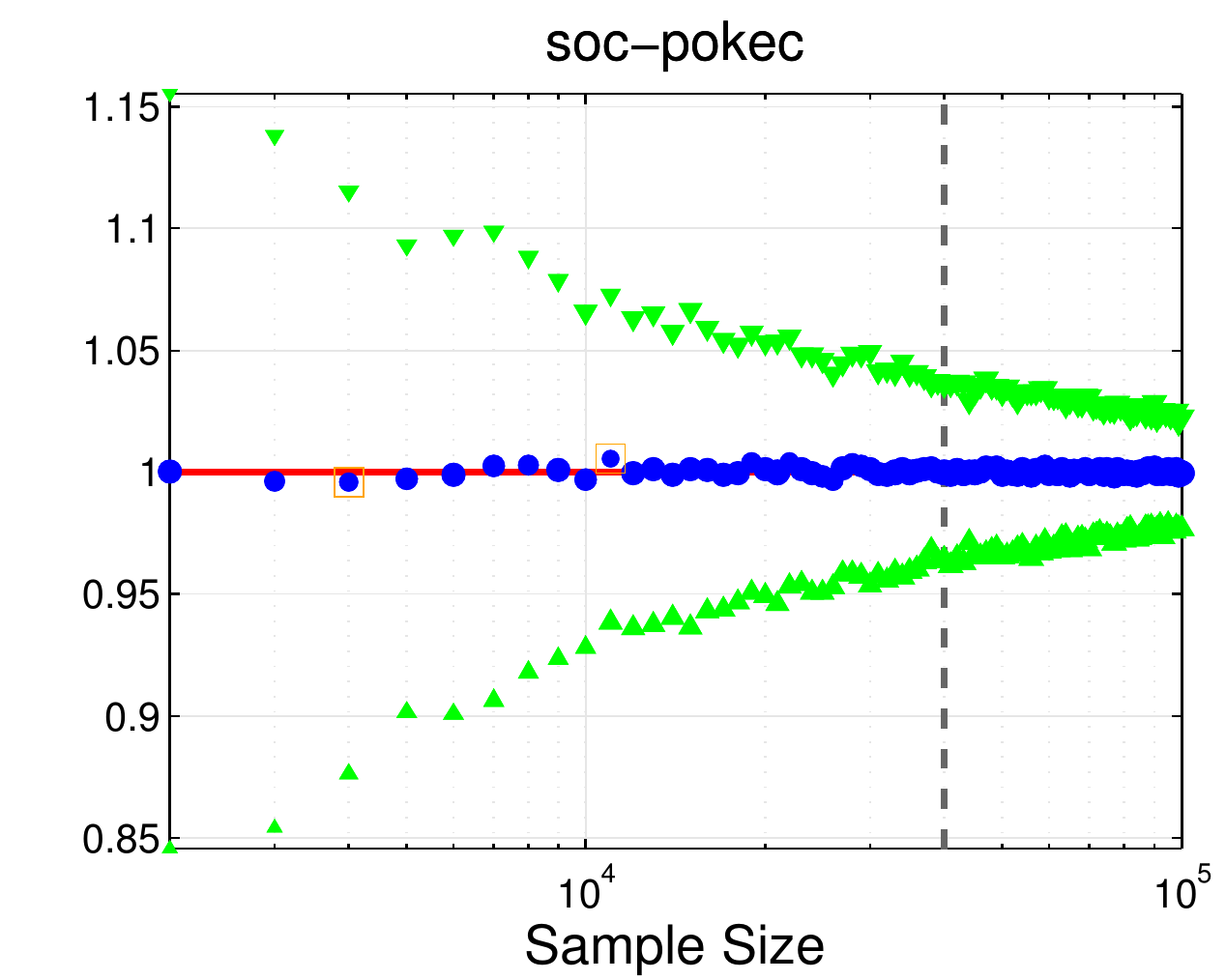}}
\hspace*{\figConfhspace}\subfigure[$\g_{7}$ (4-clique)] 		{\includegraphics[width=\figConfSZ]{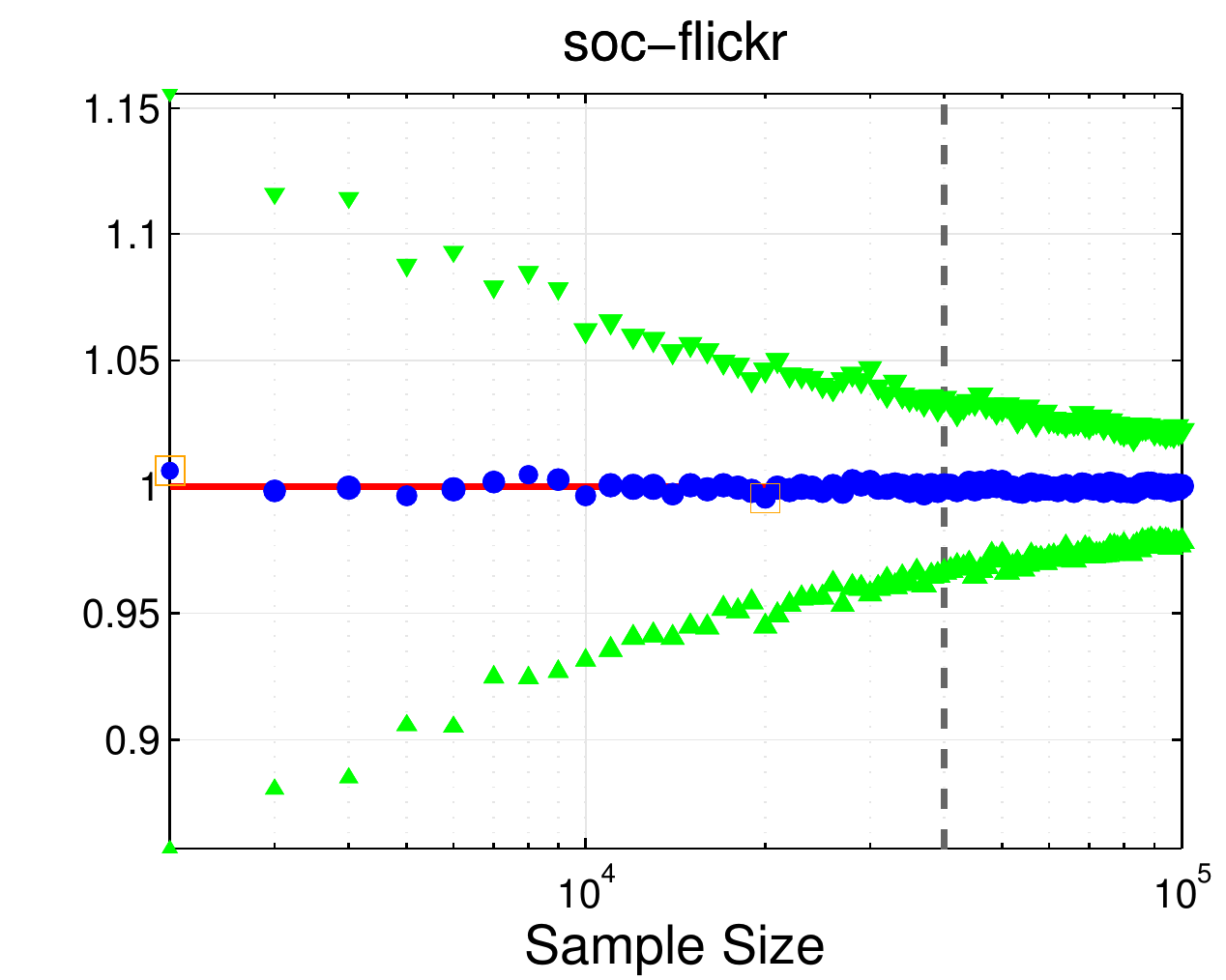}}

\vspace*{-1mm}
\hspace*{\figConfhspace}\subfigure[$\g_{10}$ (4-cycle)] 		{\includegraphics[width=\figConfSZ]{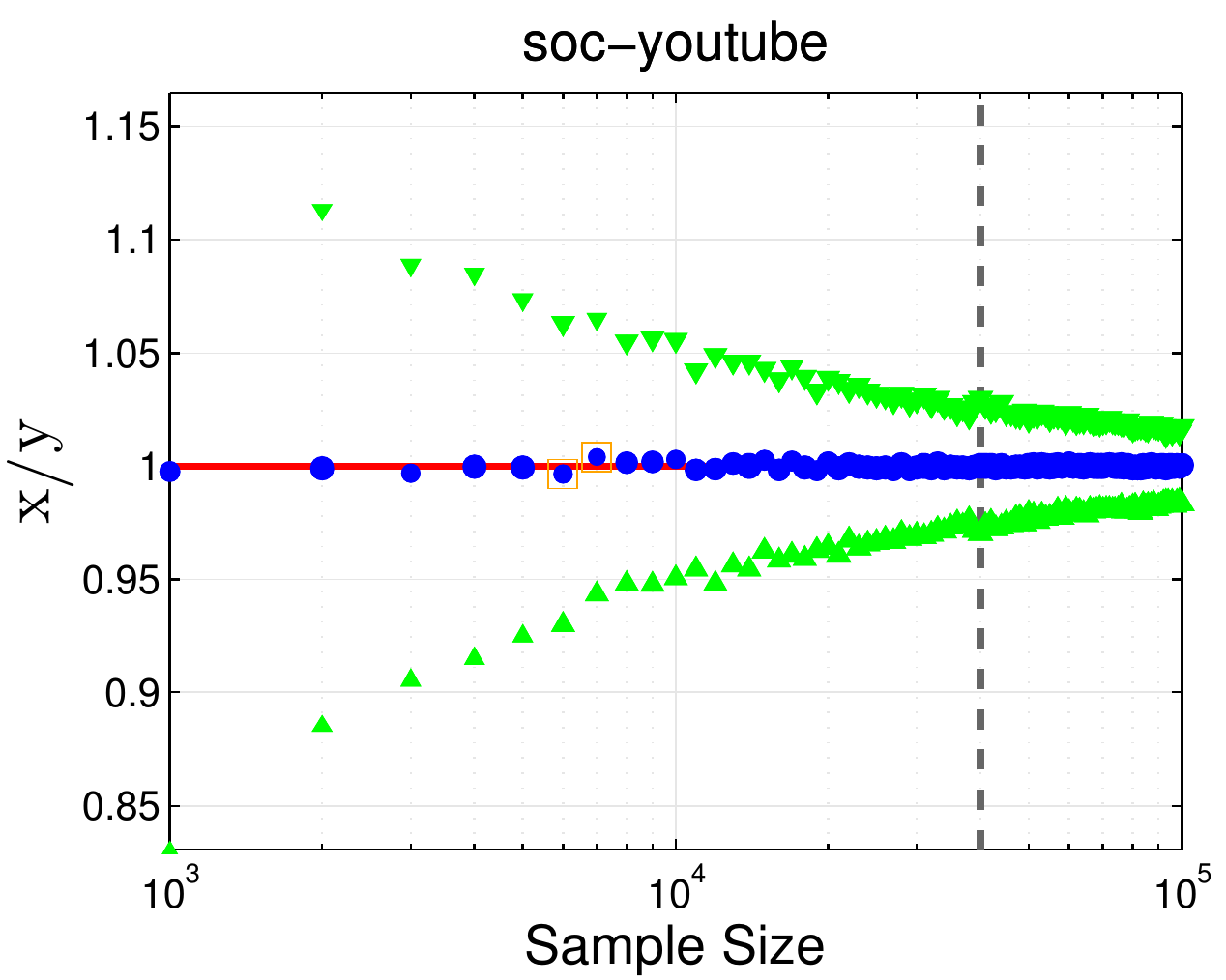}}
\hspace*{\figConfhspace}\subfigure[$\g_{13}$ (4-node-1-tri)] {\includegraphics[width=\figConfSZ]{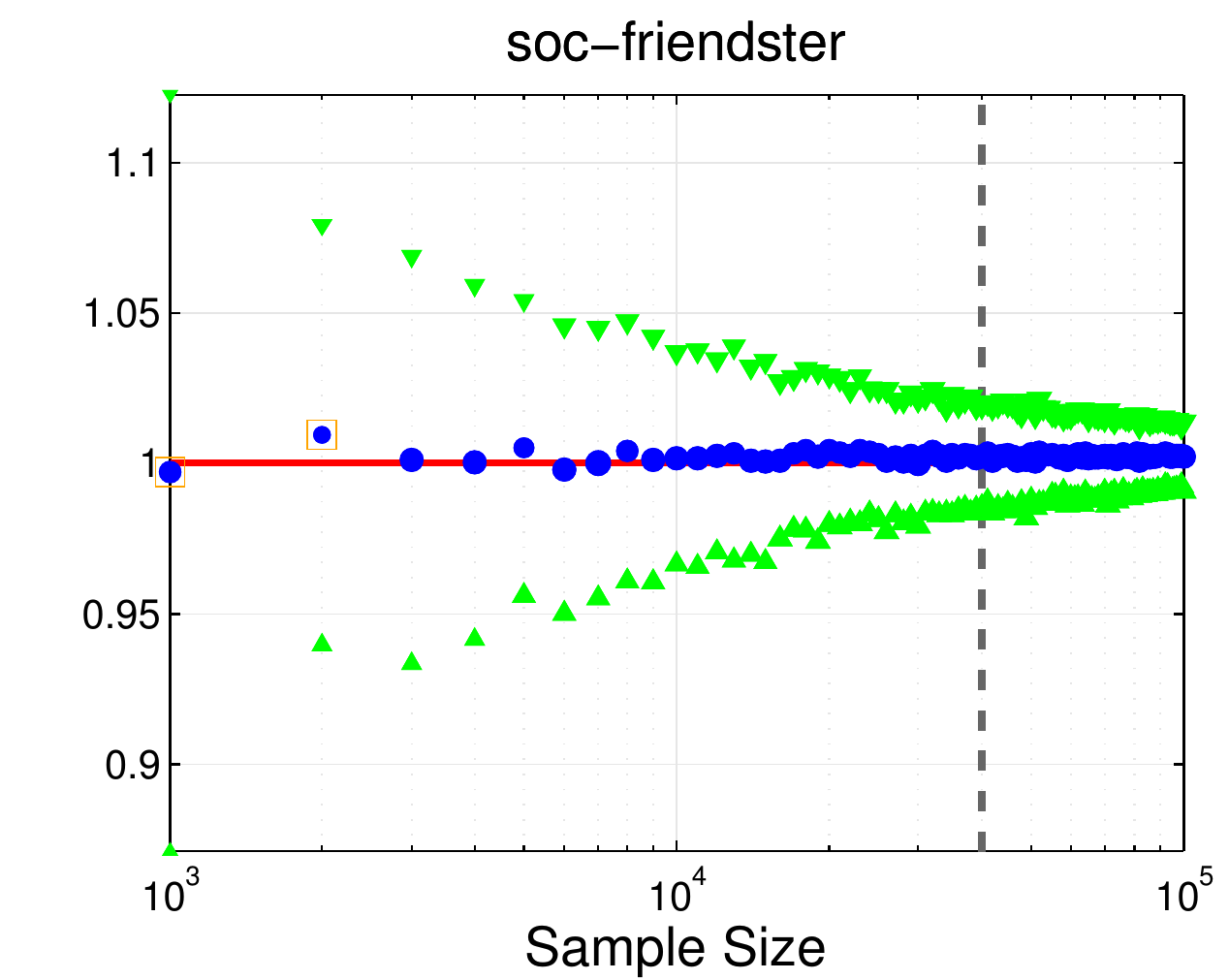}}
\hspace*{\figConfhspace}\subfigure[$\g_{12}$ (4-path)]{\includegraphics[width=\figConfSZ]{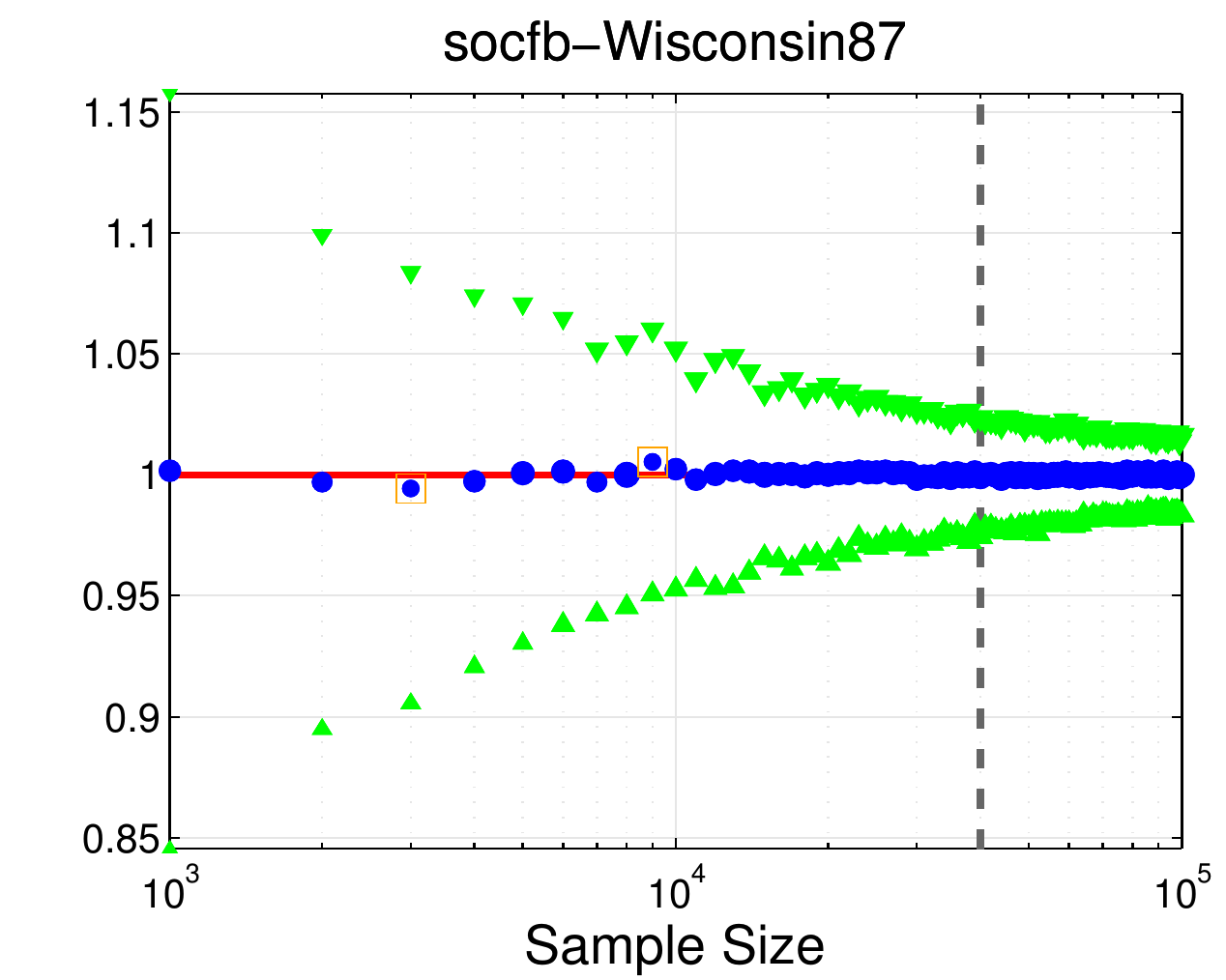}}	

\vspace*{-1mm}
\hspace*{\figConfhspace}\subfigure[$\g_{9}$ (tailed-triangle)]{\includegraphics[width=\figConfSZ]{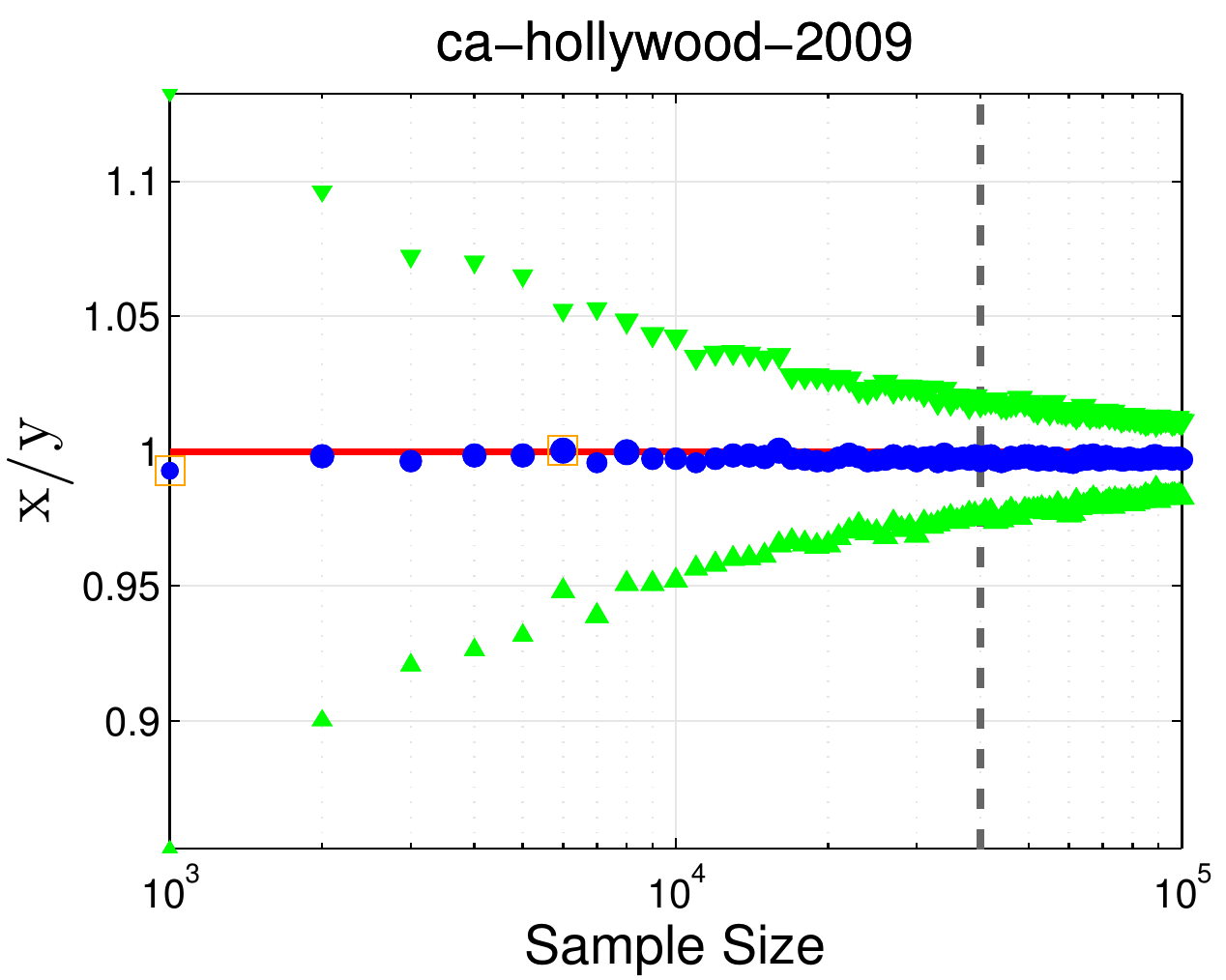}}
\hspace*{\figConfhspace}\subfigure[$\g_{10}$ (4-cycle)]		{\includegraphics[width=\figConfSZ]{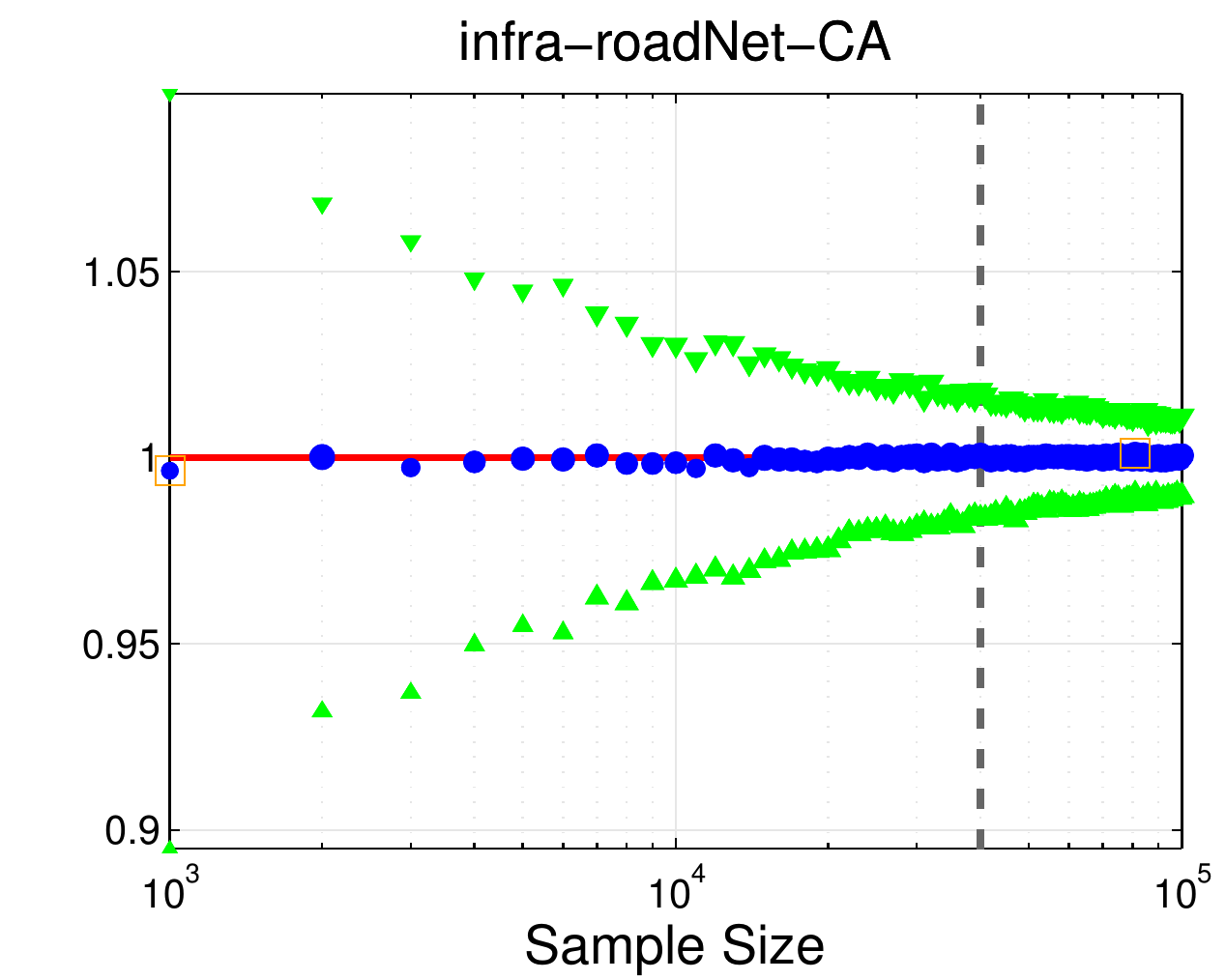}}
\hspace*{\figConfhspace}\subfigure[$\g_{11}$ (3-star)]		{\includegraphics[width=\figConfSZ]{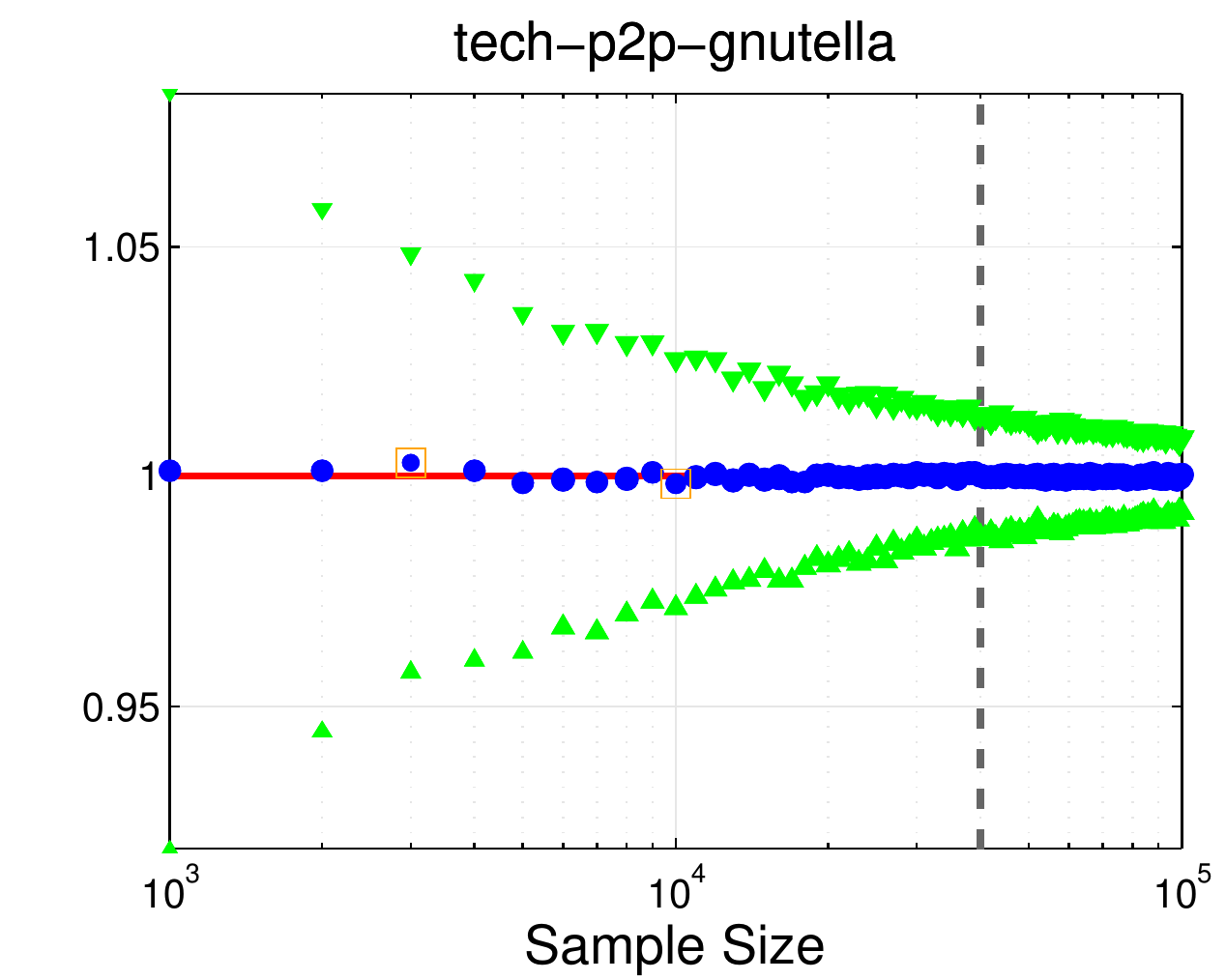}}

\vspace{-2mm}
\caption{
Confidence bounds for connected and disconnected graphlets.
We used graphs from a variety of domains and types.
Note that 4-cliques is understood to be the most difficult to estimate and thus we have dedicated more results for these hard instances.
The properties of the sampling distribution and convergence of the estimates are investigated as the sample size increases.
The circle (blue) represents $\nicefrac{\Xs}{\Y}$ (y-axis) whereas $\vartriangle$ and $\triangledown$ are $\nicefrac{\B_{lb}}{\Y}$ and $\nicefrac{\B_{ub}}{\Y}$, respectively.
The square represents min/max $\nicefrac{\Xs}{\Y}$.
Dashed vertical line (grey) refers to the sample at 40K edges.
Notably, the method has excellent accuracy even at this small sample size.
}
\label{fig:est-confidence-bounds}
\vspace{0mm}
\end{figure*}

The existing state-of-the-art estimation methods are based on sequential algorithms which are inherently slow, difficult to parallelize, and have $t$ dependent parts due to implementation issues, among others.
Furthermore, our edge-centric parallel estimation method provides significantly better load balancing (compared to vertex-based approaches).
It is straightforward to see that if $\n<\m$, then our approach requires significantly less computations per-edge than per-vertex since 
\[ 
\X_i = \sum_{\e \in \E}^{\m} \X_i(\e) = \sum_{v \in V}^{\n} \X_i(v).
\]
Parallelizing via edge-induced neighborhoods provides significantly better load balancing for real-world sparse graphs that follow a power-law.
The time taken to count $k=\{2,3,4\}$ graphlets for each edge is shown in Fig.~\ref{fig:graphlet-edge-powerlaw} and clearly obeys a power-law with only a few edges taking significantly longer than the others.
In addition, each $\N(e)$ graphlet computation may be easily split into $t$ independent tasks, e.g., k-cliques (Alg.~\ref{alg:cliques}), cycles (Alg.~\ref{alg:cycles}), solving the linear system, etc.
Moreover, the edge-centric estimation methods are flexible for situations where one might only be able to retrieve the (induced-) neighborhood of an edge due to privacy or data collection issues, etc. 
In addition, our approach does not require storage, knowledge, and preprocessing of the entire graph (as opposed to existing work).
Other important properties include the neighborhood search order $\Pi$, the batch size $b$, and the dynamic assignment of jobs (for load balancing).
As an aside, there have been a few distributed memory~\cite{ribeiro2012parallel} and shared memory~\cite{aparicio2014parallel,wang2005parallel} exact algorithms.
However, these algorithms are based on older inefficient \emph{exact enumeration} algorithms, whereas this work is focused on \emph{estimation} methods.
In addition, these approaches are all vertex-centric, as opposed to our edge-centric approach, and mainly focus on finding network motifs, \ie, statistically significant subgraph patterns.

\section{Experiments} \label{sec:exp}
\noindent
In this section, we evaluate the empirical error and performance of the methods with extensive experiments.
We use over 300 real-world networks from 20+ domains with different structural characteristics.
All data has been made available at $\nr$~\cite{nr-aaai15}. 

{
\setlength{\tabcolsep}{4.0pt}
\begin{table*}[t!]
\vspace{-0mm}
\caption{
Connected $\gfd$ and disconnected $\gfd$ estimates for a wide variety of sparse graphs.
All estimates have less than $10^{-3}$ relative error and there is no significant difference between the estimate and actual.
Graphlet estimates with relative error less than $10^{-4}$ are highlighted.
}
\vspace{-1mm}
\label{table:exp-sparse-gfd}
\centering
\footnotesize
\scriptsize
\begin{tabularx}{\linewidth}{ r HX XXXXXX  XXXXX rX HH H}
\toprule
& & &
\multicolumn{6}{c}{\bf  Connected $\gfd$} &
\multicolumn{5}{c}{\bf  Disconnected $\gfd$} &
\multicolumn{2}{c}{\bf  KS-Statistic}
\\
\TTT \BBB 
\textbf{Graph} &  & $|E|$ & 
\includegraphics[scale=0.04]{graphlets/4-clique} &
\includegraphics[scale=0.04]{graphlets/chordal-cycle} &
\includegraphics[scale=0.04]{graphlets/tailed-triangle} &
\includegraphics[scale=0.04]{graphlets/4-cycle} &
\includegraphics[scale=0.04]{graphlets/4-star} &
\includegraphics[scale=0.04]{graphlets/4-path} &
\includegraphics[scale=0.04]{graphlets/4-node-triangle} &
\includegraphics[scale=0.04]{graphlets/4-node-star} &
\includegraphics[scale=0.04]{graphlets/4-node-2edges} &
\includegraphics[scale=0.04]{graphlets/4-node-1edge} &
\includegraphics[scale=0.04]{graphlets/4-node-indep} &
\textbf{Conn.} 	&
\textbf{Disconn.} &
& &
\\ \midrule

\textbf{ca-AstroPh}  &  17.9K  &  196.9K  &  0.010\cellcolor{myyellow} & 0.016\cellcolor{myyellow} & 0.193\cellcolor{myyellow} & 0.001\cellcolor{myyellow} & 0.324\cellcolor{myyellow} & 0.455\cellcolor{myyellow} & $<$0.001\cellcolor{myyellow} & $<$0.001\cellcolor{myyellow} & $<$0.001\cellcolor{myyellow} & 0.007\cellcolor{myyellow} & 0.993\cellcolor{myyellow} & $<$10$^{-4}$  &  $<$10$^{-4}$  &&&  \\ 
\textbf{ca-MathSciNet}  &  332.6K  &  820.6K  &  0.001\cellcolor{myyellow} & 0.003 & 0.077\cellcolor{myyellow} & $<$0.001 & 0.461 & 0.457 & $<$0.001\cellcolor{myyellow} & $<$0.001 & $<$0.001\cellcolor{myyellow} & $<$0.001 & 0.999 & $<$10$^{-4}$  &  $<$10$^{-4}$  &  &&  \\ 
\midrule 
\textbf{ia-email-EU}  &  32.4K  &  54.3K  &  $<$0.001\cellcolor{myyellow} & 0.001\cellcolor{myyellow} & 0.031\cellcolor{myyellow} & $<$0.001\cellcolor{myyellow} & 0.715\cellcolor{myyellow} & 0.252\cellcolor{myyellow} & $<$0.001\cellcolor{myyellow} & $<$0.001\cellcolor{myyellow} & $<$0.001\cellcolor{myyellow} & $<$0.001\cellcolor{myyellow} & 0.999\cellcolor{myyellow} & 0.0005  &  $<$10$^{-4}$  &  &&  \\ 
\textbf{ia-enron-large}  &  33.6K  &  180.8K  &  $<$0.001\cellcolor{myyellow} & 0.004 & 0.060\cellcolor{myyellow} & 0.001\cellcolor{myyellow} & 0.716\cellcolor{myyellow} & 0.219\cellcolor{myyellow} & $<$0.001\cellcolor{myyellow} & $<$0.001 & $<$0.001\cellcolor{myyellow} & 0.002\cellcolor{myyellow} & 0.998\cellcolor{myyellow} & $<$10$^{-4}$  &  $<$10$^{-4}$  &  &&  \\ 
\midrule 

\textbf{rt-retweet-crawl}  &  1.1M  &  2.2M  &  $<$0.001 & $<$0.001\cellcolor{myyellow} & $<$0.001 & $<$0.001\cellcolor{myyellow} & 0.898\cellcolor{myyellow} & 0.101\cellcolor{myyellow} & $<$0.001 & $<$0.001\cellcolor{myyellow} & $<$0.001 & $<$0.001\cellcolor{myyellow} & 0.999\cellcolor{myyellow} & $<$10$^{-4}$  &  $<$10$^{-4}$  & &&  \\ 
\midrule 

\textbf{soc-douban}  &  154.9K  &  327.1K  &  $<$0.001\cellcolor{myyellow} & $<$0.001 & 0.012\cellcolor{myyellow} & $<$0.001\cellcolor{myyellow} & 0.436\cellcolor{myyellow} & 0.552\cellcolor{myyellow} & $<$0.001\cellcolor{myyellow} & $<$0.001 & $<$0.001\cellcolor{myyellow} & $<$0.001\cellcolor{myyellow} & 0.999\cellcolor{myyellow} & 0.0005  &  $<$10$^{-4}$  & &&  \\ 

\textbf{soc-youtube-s}  &  1.1M  &  2.9M  &  $<$0.001 & $<$0.001 & 0.002 & $<$0.001\cellcolor{myyellow} & 0.982\cellcolor{myyellow} & 0.016 & $<$0.001 & $<$0.001 & $<$0.001 & $<$0.001\cellcolor{myyellow} & 0.999\cellcolor{myyellow} & 0.0003  &  $<$10$^{-4}$  &  &&  \\ 

\textbf{soc-flickr}  &  513.9K  &  3.1M  &  0.003\cellcolor{myyellow} & 0.020\cellcolor{myyellow} & 0.132\cellcolor{myyellow} & 0.010\cellcolor{myyellow} & 0.477\cellcolor{myyellow} & 0.358\cellcolor{myyellow} & $<$0.001\cellcolor{myyellow} & $<$0.001\cellcolor{myyellow} & $<$0.001\cellcolor{myyellow} & $<$0.001\cellcolor{myyellow} & 0.999\cellcolor{myyellow} & 0.0007  &  $<$10$^{-4}$  &  &&  \\ 

\textbf{soc-twitter-higgs}  &  456.6K  &  14.8M  &  $<$0.001\cellcolor{myyellow} & $<$0.001 & 0.003\cellcolor{myyellow} & $<$0.001\cellcolor{myyellow} & 0.972\cellcolor{myyellow} & 0.024\cellcolor{myyellow} & $<$0.001\cellcolor{myyellow} & $<$0.001 & $<$0.001\cellcolor{myyellow} & $<$0.001\cellcolor{myyellow} & 0.999\cellcolor{myyellow} & $<$10$^{-4}$  &  $<$10$^{-4}$  & &&  \\ 

\textbf{soc-friendster}  &  65.6M  &  1.8T  &  $<$0.001\cellcolor{myyellow} & $<$0.001 & 0.009\cellcolor{myyellow} & $<$0.001 & 0.400\cellcolor{myyellow} & 0.590\cellcolor{myyellow} & $<$0.001\cellcolor{myyellow} & $<$0.001 & $<$0.001\cellcolor{myyellow} & $<$0.001 & 0.999\cellcolor{myyellow} & $<$10$^{-4}$  &  $<$10$^{-4}$  &  &&  \\ 
\midrule 

\textbf{socfb-UIllinois}  &  30.7K  &  1.2M  &  0.001\cellcolor{myyellow} & 0.005\cellcolor{myyellow} & 0.071\cellcolor{myyellow} & 0.002\cellcolor{myyellow} & 0.499\cellcolor{myyellow} & 0.422\cellcolor{myyellow} & $<$0.001\cellcolor{myyellow} & $<$0.001\cellcolor{myyellow} & $<$0.001\cellcolor{myyellow} & 0.016\cellcolor{myyellow} & 0.984\cellcolor{myyellow} & 0.0001  &  $<$10$^{-4}$  &  && \\ 
\textbf{socfb-Indiana}  &  29.7K  &  1.3M  &  0.001\cellcolor{myyellow} & 0.006\cellcolor{myyellow} & 0.089\cellcolor{myyellow} & 0.003\cellcolor{myyellow} & 0.300\cellcolor{myyellow} & 0.600\cellcolor{myyellow} & $<$0.001\cellcolor{myyellow} & $<$0.001\cellcolor{myyellow} & $<$0.001\cellcolor{myyellow} & 0.017\cellcolor{myyellow} & 0.982\cellcolor{myyellow} & $<$10$^{-4}$  &  $<$10$^{-4}$  & &&  \\ 
\textbf{socfb-Penn94}  &  41.5K  &  1.3M  &  $<$0.001\cellcolor{myyellow} & 0.002\cellcolor{myyellow} & 0.039\cellcolor{myyellow} & 0.001\cellcolor{myyellow} & 0.652\cellcolor{myyellow} & 0.304\cellcolor{myyellow} & $<$0.001\cellcolor{myyellow} & $<$0.001\cellcolor{myyellow} & $<$0.001\cellcolor{myyellow} & 0.009\cellcolor{myyellow} & 0.991\cellcolor{myyellow} & $<$10$^{-4}$  &  $<$10$^{-4}$  &  && \\ 
\textbf{socfb-Texas84}  &  36.3K  &  1.5M  &  $<$0.001\cellcolor{myyellow} & 0.002\cellcolor{myyellow} & 0.043\cellcolor{myyellow} & 0.001\cellcolor{myyellow} & 0.667 & 0.287\cellcolor{myyellow} & $<$0.001\cellcolor{myyellow} & $<$0.001\cellcolor{myyellow} & $<$0.001\cellcolor{myyellow} & 0.014\cellcolor{myyellow} & 0.986 & 0.0007  &  $<$10$^{-4}$  & &&  \\ 
\midrule 
\textbf{tech-internet-as}  &  40.1K  &  85.1K  &  $<$0.001\cellcolor{myyellow} & $<$0.001 & 0.005 & $<$0.001\cellcolor{myyellow} & 0.963 & $<$0.001\cellcolor{myyellow} & 0.000\cellcolor{myyellow} & $<$0.001 & $<$0.001 & $<$0.001\cellcolor{myyellow} & 0.999 & 0.0003  &  $<$10$^{-4}$  & &&  \\ 
\bottomrule
\end{tabularx}
\vspace{0mm}
\end{table*}
}

\subsection{Estimating Macro Graphlet Statistics} \label{sec:est-frequency}
\noindent
We proceed by first demonstrating the effectiveness of the proposed methods for estimating the frequency of both connected and disconnected graphlets up to size $k=4$.
Given an estimated statistic $\Xs_i$ of an arbitrary graphlet $\g_i \in \G$, 
we consider the relative error:
{
\vspace{-1.9mm}
\begin{equation} \nonumber \label{eq:abs-rel-error}
\Error{\X_i}{\Y_i} = \frac{|\X_i - \Y_i |}{\Y_i} 
\end{equation}
\noindent
where $\Y_i$ is the actual statistic (e.g., frequency) of $\g_i$.
Thus, this is a measure of how far the estimated statistic is from the actual graphlet statistic of interest, where $\X_i$ is the mean estimated value across $100$ independent runs.
The relative error indicates the quality of an estimated graphlet statistic relative to the magnitude of the exact statistic.
}
Results for both connected and disconnected graphlets are provided in Table~\ref{table:est-rel-error-4-clique} for a wide range of graphs from various domains.
Overall, the results demonstrate the effectiveness of the estimation methods as they have excellent empirical accuracy.
Further, the estimation error for the disconnected graphlets is considerably smaller than the error for connected graphlets.

\begin{figure}[h!]
\centering
\includegraphics[width=0.8\linewidth]{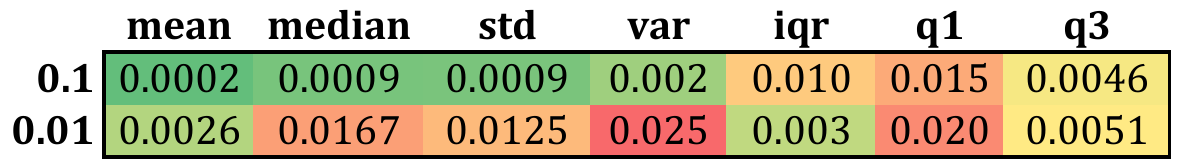}
\vspace{-1.5mm}
\caption{Estimation error for a variety of univariate statistics for the local 4-clique graphlet distribution.
These results are from $\datasm{socfb}{MIT}$ and thus even a sample size of $1\%$ is small.   
}
\label{fig:univariate-stats}
\vspace{-0mm}
\end{figure}

We also estimated univariate graphlet statistics beyond simple macro-level global counts such as the median, standard deviation, variance, irq, Q1, Q3, and others.
Overall, the methods are found to be accurate for many of the new graphlet statistics as shown in Figure~\ref{fig:univariate-stats}.
As an aside, for estimating the max 4-cliques, we found that selecting edges via the k-core distribution resulted in high accuracy at very low sample rates.

\subsection{Confidence Bounds} \label{sec:exp-confidence-bounds}
\noindent
Given an arbitrary graphlet $\g_i \in \G$, we compute $\Xs_i$ using the estimators from the framework derived in Section~\ref{sec:framework} and construct confidence bounds for the unknown $\Y_i$.
Using the large sampling distribution, we derive lower and upper bounds such that
{
\vspace{-2.5mm}
\begin{equation} \label{eq:bounds}
\lb \leq \Y_i \leq \ub
\end{equation}
\noindent
where 
}
{
\vspace{-2.5mm}
\begin{equation} \label{eq:lb}
\ub = \Xs_i - z_{\alpha/2} \cdot \sqrt{\Var[\Xs_i]}
\end{equation}
\noindent
and
}
{
\vspace{-2.5mm}
\begin{equation} \label{eq:ub}
\lb = \Xs_i + z_{\alpha/2} \cdot \sqrt{\Var[\Xs_i]}
\end{equation}
\noindent
}
The estimates $\Xs_i$ and $\Var(\Xs_i)$ are computed using the equations of the unbiased estimators of counts and their variance.
Thus, $\alpha=0.05$ and $z_{\alpha/2}=z_{0.025}=1.96$ for a $95\%$ confidence interval for the unknown $\Y_i$.
This gives
\begin{equation} \label{eq:bounds-95}
\Xs_i - 1.96 \sqrt{\Var[\Xs_i]} \;\; \leq\; \Y_i \; \leq \;\; \Xs_i + 1.96 \sqrt{\Var[\Xs_i]}
\end{equation}
Further, the sample size needed is
$\textstyle K = (\nicefrac{z_{\alpha/2} \cdot \sqrt{\Var[\Xs_i]}} {\alpha/2})^2$.

\noindent
The $95$\% upper and lower bounds (i.e., $\ub$ and $\lb$) for the 4-clique (connected graphlet) and 4-node-1-triangle (disconnected graphlet) are shown in Table~\ref{table:est-rel-error-4-clique} (other graphlet results were removed due to space). 
In all cases, the actual graphlet statistics lie inside the error bounds, $\lb \leq \Y_i \leq \ub$. 
Figure~\ref{fig:est-confidence-bounds} investigates the properties of the sampling distribution as the sample size increases. 
The circle (blue) in Figure~\ref{fig:est-confidence-bounds} represents the fraction $\nicefrac{\Xs_i}{\Y_i}$.
Further, $\nicefrac{\B_{lb}}{\Y_i}$ and $\nicefrac{\B_{ub}}{\Y_i}$ are represented in Figure~\ref{fig:est-confidence-bounds} by $\vartriangle$ and $\triangledown$, respectively. 

\smallskip
\noindent
The key findings are summarized below.
\begin{itemize}[{\small$\bullet$}]
\setlength{\parskip}{0.2em}
\item The sampling distribution is centered and balanced over the actual graph statistic (represented by the red line).
\item Upper and lower bounds always contain the actual value.
\item As the sample size increases, the bounds \emph{converge} to the actual value of the graphlet statistic
The \emph{estimated} variance decreases as $k$ grows larger.
\item Confidence bounds are within $5\%$ of the actual for all graphs and subgraph patterns.
\item Thus, the sampling distribution of the estimation framework has many attractive properties including unbiased estimates for all subgraph patterns and low variance even for very small sample sizes (and variance decreases as a function of the sample size).
\end{itemize}
\medskip
\noindent
Let $\Pr(\lb \leq \Y \leq \ub)$ be the exact coverage probability of our bounds.
We observe that the confidence bounds are tight (for all subgraph patterns) and holds to a good approximation that is within $5\%\pm$ of the actual value for all 300+ graphs.

\begin{figure}[b!]
\centering
\hspace*{-4mm}
\includegraphics[width=1.04\linewidth]{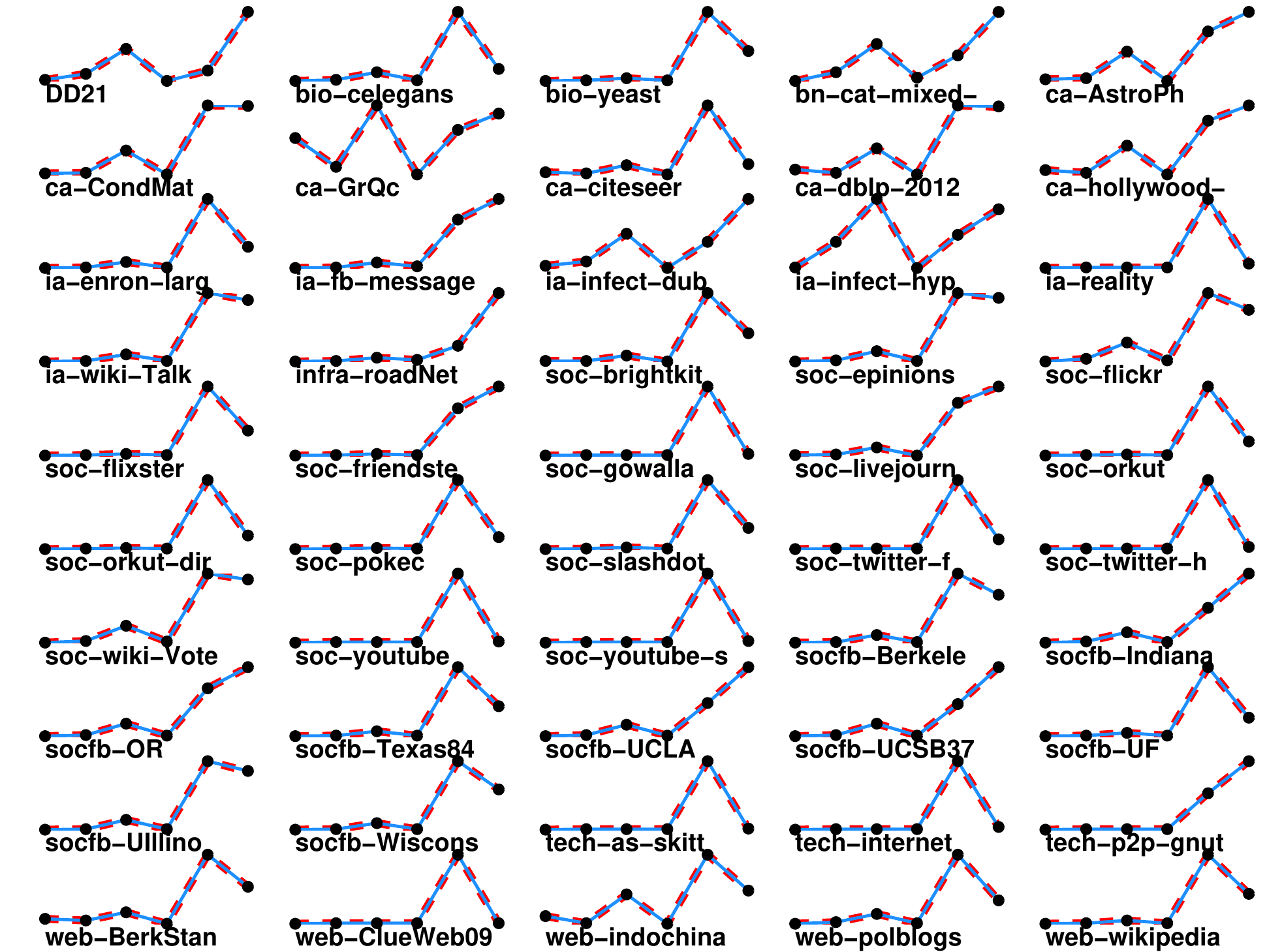}
\vspace{-4mm}
\caption{Estimated $\gfd$ is indistinguishable from the actual (larger dotted red line), even across a wide variety of graphs with fundamentally different structural characteristics.
The y-axis is the normalized 4-vertex connected graphlet counts $\vx^{\prime} = \vx-\min(\vx) / \max (\vx) -\min (\vx)$ where $\vx$ is the vector of graphlet counts.
Nevertheless, similar results were found for other graphlet sizes and $\gfd$ variants such as the disconnected $\gfd$ and $\gfd$ consisting of both connected and disconnected graphlets. 
}
\label{fig:gfd-conn-memberships-est-vs-exact}
\end{figure}

\subsection{Graphlet Frequency Distribution ($\gfd$)} 
\label{sec:est-gfd}
\noindent
We investigate the methods for approximating three different distributions: connected $\gfd$, disconnected $\gfd$, and the combined $\gfd$ consisting of both connected and disconnected graphlets.
Strikingly, the estimated $\gfd$ from our approach almost perfectly matches the actual $\gfd$ (Figure~\ref{fig:gfd-conn-memberships-est-vs-exact}).
Observe that the methods are evaluated by how well they estimate the entire $\gfd$ and thus Figure~\ref{fig:gfd-conn-memberships-est-vs-exact} indicates that the proposed methods estimate all such induced subgraphs from Table~\ref{table:graphlet_notation} with excellent accuracy (matching the actual $\gfd$ in all cases).
Results for sparse graphs are shown in Table~\ref{table:exp-sparse-gfd} and dense graphs are shown in Table~\ref{table:exp-dense-dimacs-gfd}.
The KS-Statistic for both the connected $\gfd$ and disconnected $\gfd$ is very small for all graphs.

\begin{figure}[h!]
\centering
\includegraphics[width=0.7\linewidth]{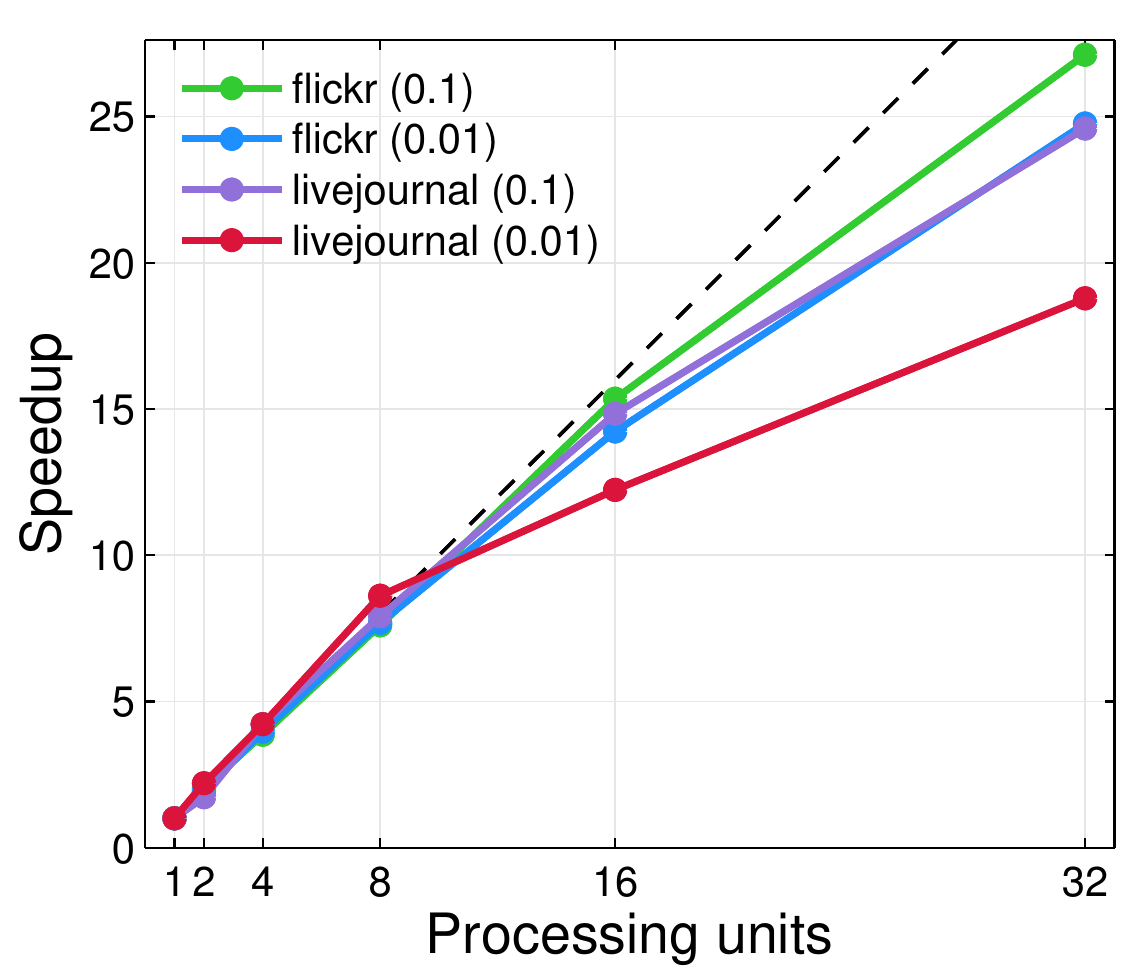}
\vspace{-1mm}
\caption{Strong scaling results for various graphlet estimation problems.
See text for discussion.
}
\label{fig:parallel-scaling}
\end{figure}

\subsection{Scalability}
\label{sec:parallel-scaling}
\noindent
This section investigates the scalability of the \emph{parallel graphlet estimation methods}.
We use speedup to evaluate the effectiveness of the parallel algorithm.
Speedup is simply $S_p = \frac{T_1}{T_p}$ where $T_1$ is the execution time of the sequential algorithm, and $T_p$ is the execution time of the parallel algorithm with $p$ processing units.
For the results in Figure~\ref{fig:parallel-scaling}, we used a 4-processor Intel Xeon E5-4627 v2 3.3GHz CPU.
Overall, the methods show strong scaling (See Figure~\ref{fig:parallel-scaling}).
Similar results were found for other graphs and sample sizes.

\subsection{Runtime Comparison} \label{sec:perf-speedup}
\noindent
This section investigates the performance of the proposed class of localized graphlet estimation methods.
\begin{itemize}[{\small$-$}]
\setlength{\parskip}{0.2em}

\item \smallskip\noindent\textit{Small and medium sized graphs:\;}
For hundreds of small and medium sized graphs, our method is on average $2895$x faster than other existing \emph{exact} approaches~\cite{rage,fanmod,orca,pgd}.

\item \smallskip \noindent\textit{Large networks:\;} 
For larger networks with hundreds of millions of edges, our method is over $200K$ times.
We observe that the speedup (relative to existing methods) increases with the size of the network.
Nevertheless, if we exclude $\pgd$, the difference in runtime between other approaches~\cite{orca,rage,fanmod} is even larger.
In many instances, these methods never finished and/or crashed after exceeding a day (even for relatively small graphs), whereas for these same graphs, our method takes less than a second to obtain accurate estimates $\leq 0.1\%$ for each graphlet $\g_i \in \G$.
Nevertheless, in all cases our approach is significantly faster, and most importantly, our approach is capable of computing graphlets on massive networks with more than a billion edges.
\end{itemize}

\begin{figure}[h!]
\vspace{-2mm}
\centering
\includegraphics[width=0.59\linewidth]{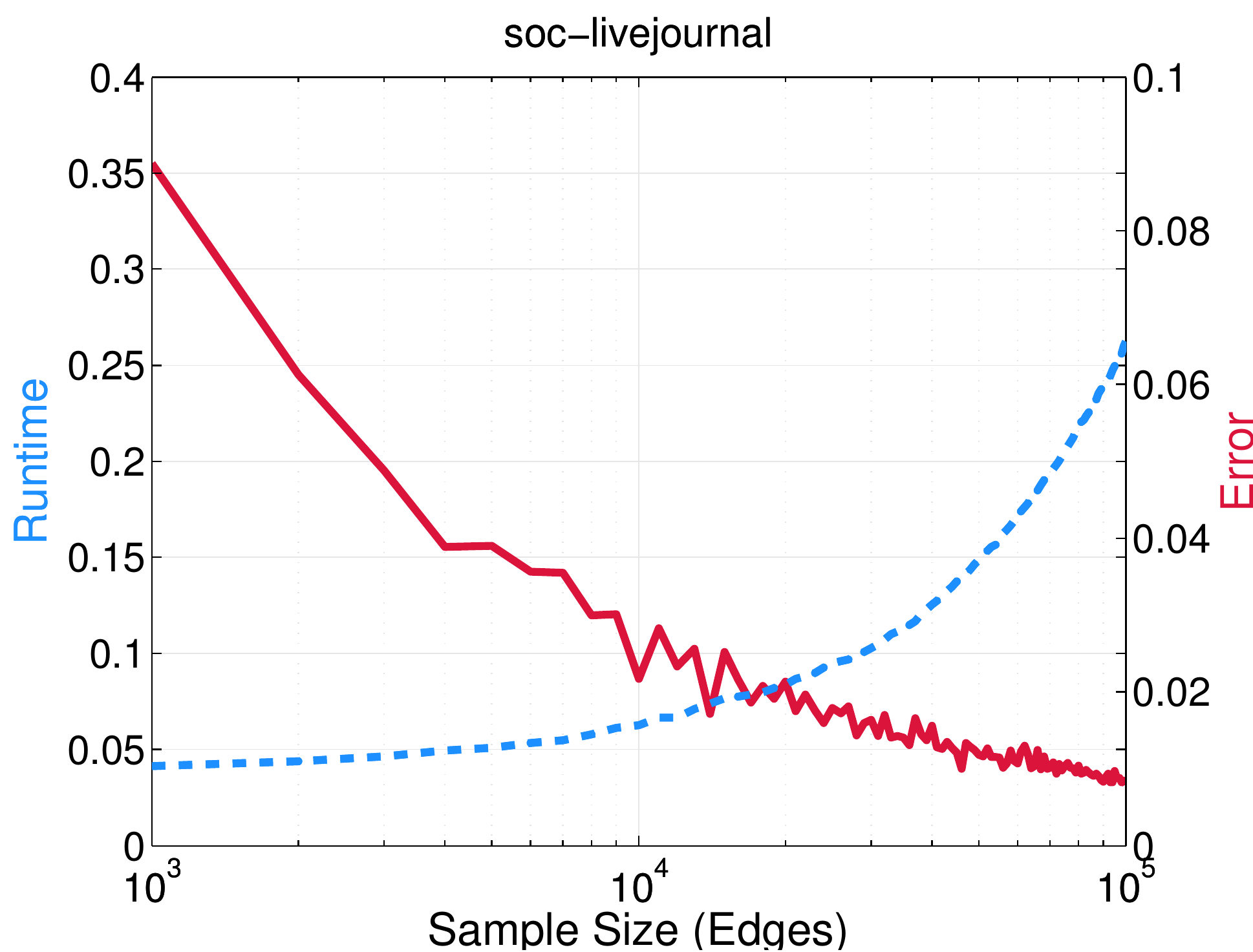}
\vspace{-1mm}
\caption{Effectively balancing speed and accuracy (4-cycles). 
}
\label{fig:est-opt-balance-speed-and-accuracy}
\vspace{-2mm}
\end{figure}

{
\setlength{\tabcolsep}{4.0pt}
\begin{table*}[t!]
\vspace{1mm}
\caption{
Connected and disconnected $\gfd$ estimates for graphs that are significantly more dense.
All estimates have less than $10^{-3}$ relative error and there is no significant difference between the estimate and actual.
Graphlet estimates with relative error <$10^{-4}$ are highlighted.
}
\vspace{-1mm}
\label{table:exp-dense-dimacs-gfd}
\centering
\scriptsize
\begin{tabularx}{\linewidth}{ r HX XXXXXX  XXXXX XX HH H}
\toprule
& & &
\multicolumn{6}{c}{\bf  Connected $\gfd$} &
\multicolumn{5}{c}{\bf  Disconnected $\gfd$} &
\multicolumn{2}{c}{\bf  KS-Statistic}
\\
\TTT \BBB 
\textbf{Graph} & $|V|$ & $|E|$ & 
\includegraphics[scale=0.04]{graphlets/4-clique} &
\includegraphics[scale=0.04]{graphlets/chordal-cycle} &
\includegraphics[scale=0.04]{graphlets/tailed-triangle} &
\includegraphics[scale=0.04]{graphlets/4-cycle} &
\includegraphics[scale=0.04]{graphlets/4-star} &
\includegraphics[scale=0.04]{graphlets/4-path} &
\includegraphics[scale=0.04]{graphlets/4-node-triangle} &
\includegraphics[scale=0.04]{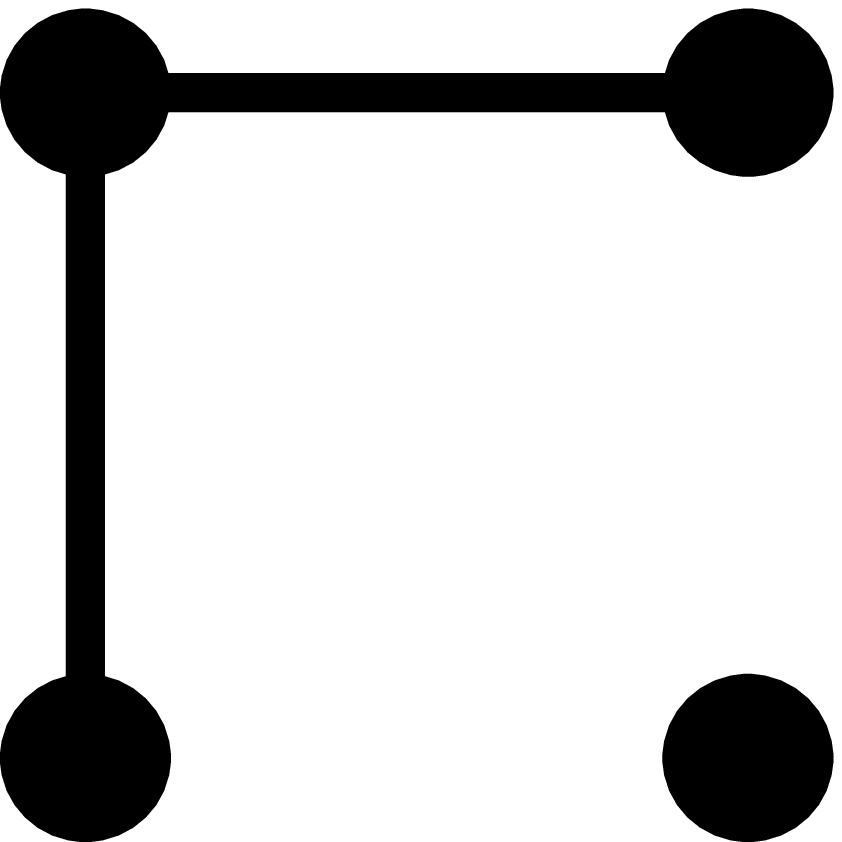} &
\includegraphics[scale=0.04]{graphlets/4-node-2edges} &
\includegraphics[scale=0.04]{graphlets/4-node-1edge} &
\includegraphics[scale=0.04]{graphlets/4-node-indep} &
\textbf{Conn.} 	&
\textbf{Disconn.} &
\\ \midrule

\textbf{johnson32-2-4}  &  496  &  107.8K  &  0.446\cellcolor{myyellow} & 0.428\cellcolor{myyellow} & 0.066\cellcolor{myyellow} & 0.033\cellcolor{myyellow} & 0.023\cellcolor{myyellow} & 0.005\cellcolor{myyellow} & $<$0.001\cellcolor{myyellow} & 0.887\cellcolor{myyellow} & 0.008\cellcolor{myyellow} & 0.032\cellcolor{myyellow} & 0.074\cellcolor{myyellow} & $<$10$^{-4}$  &  $<$10$^{-4}$  &  &&  \\ 

\textbf{brock800-3}  &  800  &  207.3K  &  0.089\cellcolor{myyellow} & 0.290\cellcolor{myyellow} & 0.314\cellcolor{myyellow} & 0.079\cellcolor{myyellow} & 0.057\cellcolor{myyellow} & 0.170\cellcolor{myyellow} & 0.285\cellcolor{myyellow} & 0.463\cellcolor{myyellow} & 0.116\cellcolor{myyellow} & 0.125\cellcolor{myyellow} & 0.011\cellcolor{myyellow} & $<$10$^{-4}$  &  0.0005  &  &&  \\ 
\textbf{brock800-1}  &  800  &  207.5K  &  0.090\cellcolor{myyellow} & 0.291\cellcolor{myyellow} & 0.314\cellcolor{myyellow} & 0.079\cellcolor{myyellow} & 0.057\cellcolor{myyellow} & 0.170\cellcolor{myyellow} & 0.285\cellcolor{myyellow} & 0.463\cellcolor{myyellow} & 0.116\cellcolor{myyellow} & 0.125\cellcolor{myyellow} & 0.011\cellcolor{myyellow} & $<$10$^{-4}$  &  0.0003  &  &&  \\ 

\textbf{san1000}  &  1K  &  250.5K  &  0.120\cellcolor{myyellow} & 0.192\cellcolor{myyellow} & 0.274\cellcolor{myyellow} & 0.037\cellcolor{myyellow} & 0.063\cellcolor{myyellow} & 0.315\cellcolor{myyellow} & 0.367\cellcolor{myyellow} & 0.247\cellcolor{myyellow} & 0.277\cellcolor{myyellow} & 0.093\cellcolor{myyellow} & 0.017\cellcolor{myyellow} & $<$10$^{-4}$  &  0.0013  &  &&  \\ 
\textbf{p-hat1500-1}  &  1.5K  &  284.9K  &  0.004\cellcolor{myyellow} & 0.047\cellcolor{myyellow} & 0.218\cellcolor{myyellow} & 0.048\cellcolor{myyellow} & 0.190\cellcolor{myyellow} & 0.494\cellcolor{myyellow} & 0.036\cellcolor{myyellow} & 0.275\cellcolor{myyellow} & 0.058\cellcolor{myyellow} & 0.401\cellcolor{myyellow} & 0.230\cellcolor{myyellow} & 0.0010  &  0.0003  &  &&  \\ 
\textbf{C2000-5}  &  2K  &  999.8K  &  0.026\cellcolor{myyellow} & 0.158\cellcolor{myyellow} & 0.316\cellcolor{myyellow} & 0.079\cellcolor{myyellow} & 0.105\cellcolor{myyellow} & 0.316\cellcolor{myyellow} & 0.154\cellcolor{myyellow} & 0.462\cellcolor{myyellow} & 0.115\cellcolor{myyellow} & 0.231\cellcolor{myyellow} & 0.038\cellcolor{myyellow} & $<$10$^{-4}$  &  0.0001  &  && \\ 
\textbf{C4000-5}  &  4K  &  4M  &  0.026\cellcolor{myyellow} & 0.158\cellcolor{myyellow} & 0.316\cellcolor{myyellow} & 0.079\cellcolor{myyellow} & 0.105\cellcolor{myyellow} & 0.316\cellcolor{myyellow} & 0.154\cellcolor{myyellow} & 0.462\cellcolor{myyellow} & 0.115\cellcolor{myyellow} & 0.231\cellcolor{myyellow} & 0.038\cellcolor{myyellow} & $<$10$^{-4}$  &  $<$10$^{-4}$  &  &&  \\ 
\bottomrule
\end{tabularx}
\vspace{-0mm}
\end{table*}
}

\subsection{Effectiveness of Adaptive Approach}
\label{sec:est-opt-results}
\noindent
Given an error bound (which may be specified by the user), the proposed method from Section~\ref{sec:est-opt} automatically finds estimates for all graphlets $\g_i \in \G$ such that $\Error{\hat{\vx}}{\vy} < \B$ where $\B$ is usually small (e.g., $\B=10^{-4}$) but can be adjusted by the user to balance the trade-off between accuracy and time.
For instance, many applications require fast methods that operate in real-time (with interactive rates).
To achieve such rates, our approach trades off accuracy for time.
Results are shown in Table~\ref{table:est-opt} and Figure~\ref{fig:est-opt-balance-speed-and-accuracy}.
Overall, the methods are fast, scalable (nearly linear scaling), and accurate with a very small KS and KL-divergence <$10^{-4}$ from the actual.
As expected, we find that the relative error between the actual graphlet statistics and the final estimates returned by the method are within the desired error bound (\eg, user-specified).

\begin{table}[h!]
\vspace{-0mm}
\setlength{\tabcolsep}{5.0pt}
\scriptsize
\caption{
Adaptive estimation results for a variety of networks. 
The methods have excellent accuracy (very small KS/KL-div.).
In all cases, the maximum relative error is <$0.001$ and usually much less.
This method has been shown to be effective for both large sparse and dense networks that arise in many real-world applications.
Recall that $\phi$ is the fraction of edge neighborhoods used (converged), 
$t^{\opt}$ is the total number of steps from Alg~\ref{alg:graphlet-est-optimization}, and 
$\delta^{\opt}$ is the converged objective.
The KS-stat. and KL-div. below is shown for connected graphlets, since it is even smaller for disconnected graphlets.
}
\tiny
\vspace{-2mm}
\label{table:est-opt}
\centering\small\scriptsize
\begin{tabularx}{\linewidth}{  r Hr HHHHHH HHHHH cHcc HHH H XH HXH HH}
\toprule
\BB
& & $|E|$ & 
& & & & & & & & & & &
$\phi$  &  
 &  
$t^{\opt}$  &  
$\delta^{\opt}$ &  
&&&& KS  &  && KL  &&
\\ 
\midrule

\textbf{C4000-5}  &   &  4M  &  
& & & & & & & & & & &
0.0003  &    &  4  &  0.0003  &  
& & & &
<$10^{-4}$  &&&  <$10^{-4}$ 
\\ 

\textbf{soc-douban}  &   &  327.1K  &  
& & & & & & & & & & &
<$10^{-6}$  &    &  150  &  0.0007  &  
&&&&
0.0005  &&&  <$10^{-4}$ 
\\ 

\textbf{soc-friendster}  &  &  1.8B  &  
& & & & & & & & & & &
<$10^{-6}$ &    &  50  &  0.0006  &  
&&&&
<$10^{-4}$  &&&  <$10^{-4}$  
\\ 

\textbf{soc-gowalla}  &   &  950.3K  &  
& & & & & & & & & & &
0.0283  &   &  287  &  0.0007  &  
&&&&
0.0002  &&& <$10^{-4}$ 
\\ 

\textbf{soc-twitter-higgs}  &   &  14.8M  &  
& & & & & & & & & & &
0.0000  &   &  161  &  0.0007  &  
&&&&
<$10^{-4}$  &&&  <$10^{-4}$  
\\ 

\textbf{socfb-Indiana}  &   &  1.3M  &  
& & & & & & & & & & &
0.0080  &    &  81  &  0.0009  &  
&&&&
<$10^{-4}$  &&&  <$10^{-4}$  
\\ 

\textbf{socfb-Penn94}  &   &  1.3M  &  
& & & & & & & & & & &
0.0175  &    &  177  &  0.0007  &  
&&&&
<$10^{-4}$  &&&  <$10^{-4}$  
\\ 
\bottomrule
\end{tabularx} 
\vspace{-0mm}
\end{table}

{
\setlength{\tabcolsep}{2.0pt}
\begin{table}[b!]
\vspace{1mm}
\caption{Micro graphlet estimation experiments. 
For each graph problem, we report the relative error averaged over 500 randomly selected edges.
These experiments use $\pe_{\e}=0.001$ (See Section~\ref{sec:local-graphlet-est} for more details).
In addition to the high accuracy, 
the micro graphlet estimation methods are between 900-1000K times faster, and thus fast and highly scalable.
}
\vspace{-2mm}
\label{table:exp-local-graphlet-est}
\centering
\scriptsize
\begin{tabularx}{\linewidth}{@{} r r HHH XXXXXX HXX HHHH H @{}}
\toprule
& & & & & \multicolumn{6}{c}{\textsc{relative error}} \BBB 
\\

\textbf{graph} && &&&
\includegraphics[scale=0.04]{graphlets/4-clique} &
\includegraphics[scale=0.04]{graphlets/chordal-cycle} &
\includegraphics[scale=0.04]{graphlets/tailed-triangle} &
\includegraphics[scale=0.04]{graphlets/4-cycle} &
\includegraphics[scale=0.04]{graphlets/4-star} &
\includegraphics[scale=0.04]{graphlets/4-path} &
& \textbf{KL} & \textbf{L1} & 
&&&&
\\
\midrule

$\mathsf{soc}$-$\mathsf{flickr}$ && &&&
{ 0.001}  & { 0.001}  & { 0.001\cellcolor{white}} & { 0.001}  & { 0.001\cellcolor{white}} & { 0.001\cellcolor{white}} & 
& { 0.0001\cellcolor{white}} & { <$10^{-4}$ \cellcolor{white}} & 
&&&& \\

$\mathsf{bio}$-$\mathsf{human}$-$\mathsf{gene1}$ && &&&
{ 0.002\cellcolor{white}} & { 0.002}  & { 0.001\cellcolor{white}} & { 0.001}  & { 0.001\cellcolor{white}} & { 0.001\cellcolor{white}} & 
& { 0.0004\cellcolor{white}} & { <$10^{-4}$ \cellcolor{white}} & 
&&&& \\

$\mathsf{tech}$-$\mathsf{internet}$-$\mathsf{as}$ && &&&
{ 0.0001\cellcolor{white}} & { 0.0001\cellcolor{white}} & { 0.0012}  & { 0.0002}  & { 0.001\cellcolor{white}} & { 0.0002\cellcolor{white}} & 
& { 0.001\cellcolor{white}} & { <$10^{-4}$ \cellcolor{white}} & 
&&&& \\

$\mathsf{sc}$-$\mathsf{nasasrb}$ && &&&
{ 0.004 \cellcolor{white}} & { 0.004}  & { 0.001}  & { 0.002}  & { 0.003 \cellcolor{white}} & 0.002  & 
& { 0.004} { \cellcolor{white}} & 0.001 { \cellcolor{white}} & 
&&&& \\
\bottomrule
\end{tabularx}
\vspace{-0mm}
\end{table}
}

\subsection{Micro Graphlet Estimation Experiments} \label{sec:exp-local-graphlet-est}
\noindent
This section investigates the accuracy, runtime, and scalability of the computational framework presented in Section~\ref{sec:local-graphlet-est} for estimating micro graphlet statistics and distributions of individual graph elements such as an edge (or node, path, or subgraph) as opposed to estimating macro-level graphlet statistics over the entire graph $G$.
Results are shown in Table~\ref{table:exp-local-graphlet-est}.
Note that for simplicity, nodes are selected uniformly at random, thus $\Fdist$ in Alg~\ref{alg:lge-local} represents a uniform distribution over the neighbors.

{
\setlength{\tabcolsep}{6.0pt}
\begin{table}[b!]
\vspace{-0mm}
\caption{Results for two of our proposed techniques for estimating the maximum frequency of an arbitrary induced subgraph centered at an edge in $G$.
The results below use $\p_i=0.005$ and are for $\data{socfb}{Texas}$.
Similar results were found with different graphs and sampling probabilities, and thus, removed for brevity.
Note that the runtime is the total time taken to estimate all graphlet statistics.
Clearly, selecting edge neighborhoods using the weighted probability distribution based on k-core numbers gives significantly better estimates for the vast majority of statistics below.
In particular, at $\p_i=0.005$, uniform does better only for estimating the maximum 3-stars centered at any edge in G.
Nevertheless, both are orders of magnitude faster than the exact method.
For instance, the k-core approach is 157x faster than the exact method (on average using $\p=0.005$), whereas the uniform method is 185x faster.
Note that the best result among the estimation methods is bold, whereas $^{*}$ indicates that the estimate returned by the method is optimal (that is, it matches the actual maximum returned by the exact algorithm).
}
\vspace{-0mm}
\label{table:exp-max-edge-graphlet-stats}
\centering
\scriptsize
\begin{tabularx}{\linewidth}{@{} r  H X XXXXXX H @{}}
\toprule
& & &
\multicolumn{6}{c}{\bf  Maximum connected graphlet counts} \BB
&
\\
\textbf{Method}
\TTT \BBB & 
& 
\rotatebox{35}{\textbf{Speedup}} &
\includegraphics[scale=0.04]{graphlets/4-clique} &
\includegraphics[scale=0.04]{graphlets/chordal-cycle} &
\includegraphics[scale=0.04]{graphlets/tailed-triangle} &
\includegraphics[scale=0.04]{graphlets/4-cycle} &
\includegraphics[scale=0.04]{graphlets/4-star} &
\includegraphics[scale=0.04]{graphlets/4-path} &
\\ 
\midrule
\textsc{kcore}       &	 & $157$x		&	 \textbf{45650}$^{*}$ 	&	 \textbf{3.85M}$^{*}$ 	&	 \textbf{26509}$^{*}$ &	 \textbf{50351}$^{*}$ &	 19.51M 
&	 \textbf{11.01M}$^{*}$ &	 \\ 
\textsc{uniform}     &	  & $185$x		&	 8172 					&	 22180 					&	 12112 				 &	 24429 				 &	 \textbf{19.89M} 
&	 3.35M &	 \\
\midrule 
$\mathsf{Exact}$ 	&	 & $-$	 
&	 45650 					&	 3.85M 				&	 26509 				&	 50351 				 &	 19.91M 
&	 11.01M &	 \\
\bottomrule
\end{tabularx}
\end{table}
}

\subsection{Extremal Graphlet Estimation} \label{sec:extremal-max-graphlet-problem}
\noindent

{
\noindent
\textbf{Problem.}\; ({\sc Max Graphlet Estimation})
\label{sec:max-local-graphlet-counts}
Given a graph $G$, 
and a graphlet pattern $\g_j$ of size $k$, find
\begin{equation} 
Z_j = \max_{e_i \in \{e_1,...,e_m\}} \, \Big[ X_{j}(\e_i) \Big]
\end{equation}
\noindent
where $Z_j$ is the maximum number of times graphlet $\g_j$ occurs at any edge $e_i \in E$ in $G$.
\smallskip
}

The aim is to compute the maximum frequency that graphlet $\g_j$ occurs at any edge $e_i \in E$ in $G$. 
For this problem, we leverage the proposed $\lge$ framework from Section~\ref{sec:framework} and bias the estimation method towards selecting a small set of edge $\Es$ where $\g_j$ is most likely to appear at larger frequencies.
The set of edges $\Es$ are sampled via a graph parameter/distribution that appropriately biases selection of edges that are most likely to induce large quantities of the graphlet $\g_j$.
For relatively dense graphlets such as the $k$-clique (chordal-cycle/diamond, etc.), we investigated sampling edges from the largest k-core subgraphs.
More specifically, 
instead of selecting edge neighborhoods via a uniform distribution $\Fdist$, our approach replaces $\Fdist$ in Line~\ref{algline:select-edge-neighborhood-via-F} of Alg~\ref{alg:graphlet-est} with a weighted distribution that biases the selection of edge neighborhoods towards those in large k-core subgraphs (i.e., edge neighborhoods centered at edges with large k-core numbers).
Similarly, one may also use the triangle-core subgraphs if computed to obtain an estimate with lower error.
Results demonstrate the effectiveness of this approach in Table~\ref{table:exp-max-edge-graphlet-stats}.
Strikingly, the above approach finds the optimal solution (while taking only a fraction of the time) for many graphs as well as many of the $k$-vertex induced subgraphs.

\begin{table}[h!]
\vspace{-0mm}
\ra{1.2}
\caption{
Results for counting connected graphlets for four \emph{massive networks} and one smaller graph (see text for discussion).
For each method, 
we report the time required until the relative error is less than $\B=0.01$.
A hyphen ($-$) indicates that the method did not terminate within 12 hours.
The best time for each problem instance is bolded.
}
\vspace{-2mm}
\label{table:conn-counts-3path-sampling}
\centering
\small
\scriptsize
\begin{tabularx}{1.0\linewidth}{@{}lHX l lXXc@{}}
\toprule
& 		& &	\multicolumn{5}{c}{\bf Time in seconds} 
\\ 
\textbf{graph}  					& 		& 		$|\E|$ 			&  $\lge$  &   {\tiny 3}\textsc{-path}  &   \textsc{guise}   &   \textsc{graft} & \textsc{pgd} (exact) \\ 
\midrule
$\datasm{soc}{sinaweibo}$   		&   	&  $261$M 			&	\textbf{12.3}			&	$-$			&	$-$			& $-$			&  			$33359$		\\ 
$\datasm{web}{ClueWeb09}$			& 		&  $7.81$B			&	\textbf{65.6}			&	$-$			&	$-$			& $-$			&  			$-$	 	\\ 
$\datasm{soc}{friendster}$   		& 		&  $1.81$B 			&	\textbf{44.1}			&	$-$			&	$-$			& $-$			&  			$-$ 		\\ 	
$\datasm{soc}{twitter}$   		& 		&  $1.20$B 			&	\textbf{341.2}			&	$-$			&	$-$			& $-$			&  			$-$ 		\\ 	
$\datasm{wiki}{Talk}$   			&  				&  $4.6$M	 		&	\textbf{0.0007}		&	$1.04$			&	$-$			& $-$			&  			$0.14$ 	\\ 
\bottomrule
\end{tabularx}
\end{table}

\subsection{Comparison to Previous Work}
\noindent
We compare to recent work done on approximating simple counts of a few connected graphlets.
Results are provided in Table~\ref{table:conn-counts-3path-sampling}.
As an aside, it is worth mentioning that existing work is fundamentally different than ours, both in techniques, as well as in the estimation problems themselves.
For instance, these methods estimate only simple macro-level counts of connected graphlets, whereas 
the proposed class of $\lge$ methods accurately estimate a wide variety of macro and micro-level statistics (including simple counts) and distributions for both connected and disconnected graphlets.
See Table~\ref{table:taxonomy-and-overview} for a summary of the differences.
Note that the 3-path sampling heuristic by Jha~\etal~\cite{path-sampling} requires significantly more samples to obtain estimates with similar accuracy.
In addition, that approach requires two different methods for estimating connected graphlets counts of size $4$, and thus requires 2x the samples.
In particular, we find that 3-path sampling, GUISE, and GRAFT are unable to obtain accurate estimates within a reasonable amount of time\footnote{Furthermore, GUISE and GRAFT did not converge, even despite using millions of samples, which is consistent with recent findings~\cite{path-sampling}, and especially true for the massive networks used in this work.}. 
See Table~\ref{table:conn-counts-3path-sampling}.
In some cases, the runtime of these methods even exceeded an exact graphlet algorithm, and thus not useful in practice.
Notably, our method is not only more accurate at lower sampling rates, but significantly faster than these methods.
For instance, on $\data{soc}{flickr}$ we are $8047$x faster than the path-sampling heuristic.
In some cases, we even find that our exact method is significantly faster than the 3-path heuristic (for instance, on $\data{wiki}{talk}$ and others).
We also investigated selecting node-centric neighborhoods and other methods based on sampling graphlets directly, though, the accuracy was worse in all cases, and thus removed for brevity.

\begin{figure}[h!]
\centering
\includegraphics[width=0.78\linewidth]{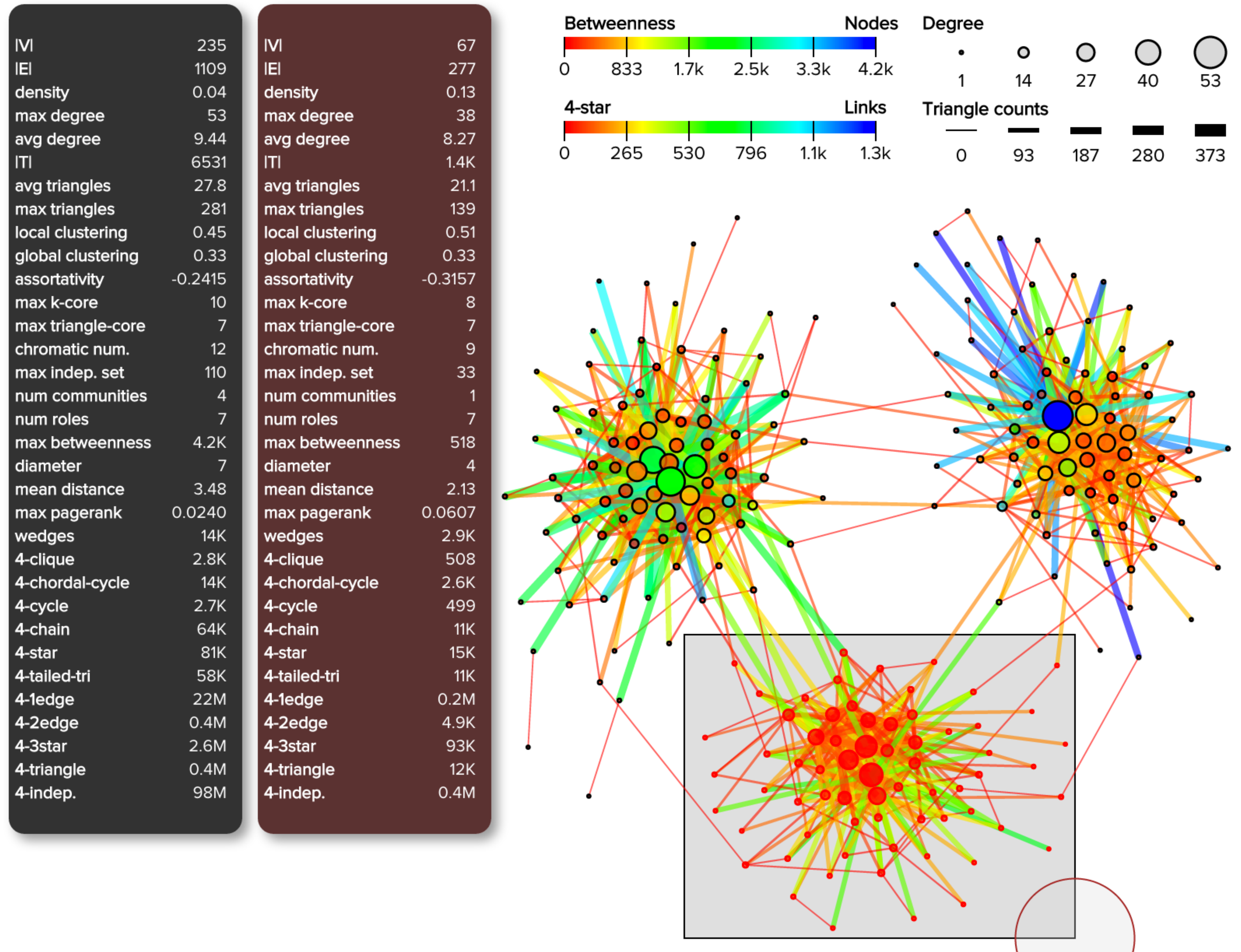}
\vspace{-2.5mm}
\caption{
Application of the fast and accurate approximation methods for real-time interactive graph mining and predictive modeling tasks (\eg, node classification).
}
\label{fig:application-real-time}
\end{figure}

\section{Applications} \label{sec:applications}
\noindent
Due to the volume and the velocity of big data, approximate results are often a necessity.
Graphlet estimators are implemented in a web-based visual graph analytics platform (Figure~\ref{fig:application-real-time}).
Graphlet estimation methods (from the proposed estimation framework in Section~\ref{sec:framework}) are implemented in a recent web-based visual graph analytics platform~\cite{gvis-icwsm15} called $\mathsf{graph}{\sc vis}$ (Figure~\ref{fig:application-real-time}).
Across all experiments, the graphlet methods are fast and scalable taking <$1$ ms 
for $99\%$ of the interactive queries and graphs,
while also accurate (no significant difference).
Thus, the graphlet estimation methods are able to support \emph{real-time} interactive queries for visual graph mining, exploration, and predictive modeling tasks (such as relational classification).
Other applications were removed for brevity.

\section{Conclusion} \label{sec:conc}
\noindent
We have shown that even when dealing with massive networks with more than a billion edges, one can compute graphlets fast and with exceptional accuracy. 
The newly introduced family of graphlet estimators significantly improves the scalability, flexibility, and utility of graphlets.
In addition, this paper studied and proposed estimators for new graphlet problems and statistics including methods for both connected and disconnected graphlets, as well as estimating a number of novel macro and micro-level graphlet statistics.
Moreover, we proposed a fast and scalable parallel scheme that generalizes for the family of edge-centric estimation methods in the framework.
In addition, an optimization method that automatically finds an estimate within a user-defined 
level of accuracy without requiring the user to input the sample size.
Finally, the methods give rise to new opportunities and applications for graphlets (as shown in Section~\ref{sec:applications}).

\ifCLASSOPTIONcaptionsoff
\newpage
\fi

{
\fontsize{7}{8}\selectfont
\balance
\bibliographystyle{IEEEtran}
\bibliography{rossi}
}

\end{document}